\documentclass[prd,showkeys,preprintnumbers,floatfix,
nofootinbib,superscriptaddress]{revtex4}
\usepackage{float}
\usepackage{nicefrac}
\usepackage{slashed}
\usepackage{mathtools}
\usepackage{amsfonts} 
\usepackage{amssymb} 
\usepackage{amsmath} 
\usepackage{graphicx} 
\usepackage{subfigure} 
\usepackage{array} 
\usepackage{dcolumn} 
\usepackage{bm} 
\usepackage{latexsym} 
\usepackage{longtable} 
\usepackage{hyperref} 
\usepackage{verbatim}
\usepackage{epsfig}
\usepackage{color}
\newcommand{\T}[0]{{\mathcal T}}
\newcommand{\df}[0]{\mathrm{df}}

\newcommand{\nn}[0]{\nonumber}

\newcommand{\MLL}[0]{   \mathcal M_{L} }
\newcommand{\ML}[1]{   \mathcal M_{L,#1} }

\newcommand{\cM}[0]{\mathcal M}
\newcommand{\cF}[0]{\mathcal F}
\newcommand{\cK}[0]{\mathcal K}
\newcommand{\cD}[0]{\mathcal D}
\newcommand{\cS}[0]{\mathcal S}
\newcommand{\cC}[0]{\mathcal C}
\newcommand{\cL}[0]{\mathcal L}
\newcommand{\cR}[0]{\mathcal R}
\newcommand{\cZ}[0]{\mathcal Z}
\newcommand{\CH}[0]{\mathcal C^H}
\newcommand{\Kdf}[0]{\mathcal K_{\rm df,3}}

\newcommand{\KD}[0]{\mathcal K_{\rm df,D}}
\newcommand{\hsp}[1]{\hspace{#1 pt}}
\newcommand{\tI}[0]{\widetilde I}
\newcommand{\tM}[0]{B}

\newcommand{\tA}[0]{\widetilde A}

\begin{document}
 \preprint{\vbox{\hbox{JLAB-THY-17-2400} }}
\title{Relating the finite-volume spectrum and the two-and-three-particle $S$ matrix\\ for relativistic systems of identical scalar particles}
\author{Ra\'ul A. Brice\~no}
\email[e-mail: ]{rbriceno@jlab.org}
\affiliation{
Thomas Jefferson National Accelerator Facility, 12000 Jefferson Avenue, Newport News, VA 23606, USA\\
}

\author{Maxwell T. Hansen}
\email[e-mail: ]{hansen@kph.uni-mainz.de}
\affiliation{
Institut f\"ur Kernphysik and Helmholtz Institute Mainz, Johannes Gutenberg-Universit\"at Mainz,
55099 Mainz, Germany\\
}
\author{Stephen R. Sharpe}
\email[e-mail: ]{srsharpe@uw.edu}
\affiliation{
Physics Department, University of Washington, Seattle, WA 98195-1560, USA\\
}

\date{\today}
\begin{abstract}
Working in relativistic quantum field theory, we derive the quantization condition satisfied by coupled two- and three-particle systems of identical scalar particles confined to a cubic spatial volume with periodicity $L$. This gives the relation between the finite-volume spectrum and the infinite-volume $\textbf 2 \to \textbf 2$, $\textbf 2 \to \textbf 3$ and $\textbf 3 \to \textbf 3$ scattering amplitudes for such theories. The result holds for relativistic systems composed of scalar particles with nonzero mass $m$, whose center of mass energy lies below the four-particle threshold, and for which the two-particle $K$ matrix has no singularities below the three-particle threshold. The quantization condition is exact up to corrections of the order $\mathcal{O}(e^{-mL})$ and holds for any choice of total momenta satisfying the boundary conditions.  

\end{abstract}

\maketitle

\section{Introduction}
Over the past few decades, enormous progress has been made in determining the properties of hadrons  
directly from the fundamental theory of the strong force,   
quantum chromodynamics (QCD). 
A key tool in such investigations is lattice QCD (LQCD), which can be used to numerically
calculate correlation functions defined on a discretized, finite, Euclidean spacetime.
State-of-the-art LQCD
calculations of stable hadronic states use dynamical up, down, strange, and even charm quarks, with physical quark masses, and include isospin breaking both from the mass difference of the up and down quarks and from the effects of quantum electrodynamics (QED). 
For recent reviews, see Refs.~\cite{Prelovsek:2014zga,Portelli:2015wna,Liu:2016kbb}.
  
Using LQCD to investigate hadronic resonances that decay via the strong force is significantly more challenging. Resonances do not correspond to eigenstates of the QCD Hamiltonian and thus cannot be studied by directly interpolating a state with the desired quantum numbers. Instead, resonance properties are encoded in scattering and transition amplitudes, and only by extracting these observables can one make systematic, quantitative statements. In fact, it is not {\em a priori} clear that one can extract such observables using LQCD. Confining the system to a finite volume obscures the meaning of asymptotic states and restricting to Euclidean momenta prevents one from directly applying the standard approach of Lehmann-Symanzik-Zimmermann reduction. In addition, since one can only access numerically determined Euclidean correlators with nonvanishing noise, analytic continuation to Minkowski momenta is, in general, an
ill-posed problem.
  
For two-particle states, it is by now well known that scattering amplitudes
can be constrained indirectly, by first extracting the discrete finite-volume energy spectrum. 
The approach follows from seminal work by L\"uscher~\cite{Luscher:1986pf, Luscher:1990ux} who derived a relation between the finite-volume energies and the elastic two-particle scattering amplitude for a system of identical scalar particles. 
Since then, this relation has been generalized to accommodate non zero spatial momentum in the finite-volume frame and also to describe more complicated two-particle systems, including nonidentical and nondegenerate particles as well as particles with intrinsic spin~\cite{Rummukainen:1995vs, He:2005ey, Kim:2005gf,
  Christ:2005gi, Lage:2009zv, Bernard:2010fp, Hansen:2012tf,
  Briceno:2012yi, Briceno:2014oea}. This formalism has been applied in many numerical LQCD calculations to determine the properties of low-lying resonances that decay into a single two-particle channel~\cite{Dudek:2012xn, Lang:2015hza,
  Lang:2014yfa, Feng:2010es, Prelovsek:2013ela, Aoki:2007rd},
including most recently the first study of the lightest hadronic
resonance, the $\sigma/f_0(500)$~\cite{Briceno:2016mjc}. The extension
to systems with multiple coupled two-particle channels~\cite{He:2005ey,Lage:2009zv, Bernard:2010fp,Hansen:2012tf, Briceno:2012yi}, has led to the first LQCD results for resonances at higher energies, where more than one decay channel is open~\cite{Moir:2016srx, Dudek:2016cru, Wilson:2015dqa,
  Wilson:2014cna, Dudek:2014qha}.

Thus far, however, no LQCD calculations have been performed for resonances that have a significant branching
fraction into three or more particles. This is largely because the formalism needed to do so, the three-particle 
extension of the relations summarized above, is still under construction. 
Early work in this direction includes the nonrelativistic studies presented 
in Refs.~\cite{Polejaeva:2012ut, Briceno:2012rv}. 
More recently, in Refs.~\cite{Hansen:2014eka, Hansen:2015zga}, 
two of the present authors  derived a three-particle quantization condition for identical scalar particles
using a generic relativistic quantum field theory (subject to some restrictions described below).
Since these articles are the starting point for the present work, 
we briefly summarize their methodology.\footnote{%
We also note that additional checks of the quantization condition have been given in
Refs.~\cite{Hansen:2016fzj, Hansen:2016ync}. }

Reference~\cite{Hansen:2014eka} studied a three-particle finite-volume correlator and determined its pole positions,
which correspond to the finite-volume energies, in terms of an infinite-volume scattering quantity. This was done by deriving a skeleton expansion, expressing each finite-volume Feynman diagram in terms of its infinite-volume counterpart plus a finite-volume residue, summing the result into a closed form and then identifying the pole locations. The resulting expression for the finite-volume energies depends on a nonstandard infinite-volume scattering 
quantity---the divergence-free $K$ matrix, denoted $\Kdf$. 
A drawback of this result is that $\Kdf$, as well as other quantities in the quantization condition, 
depends on a smooth cutoff function (denoted $H_3$ below), 
although the energies themselves are independent of this cutoff.
Thus the relation to the infinite-volume scattering amplitude is not explicit.

The second publication, Ref.~\cite{Hansen:2015zga}, resolved this issue  by deriving 
the relation between $\Kdf$ and the standard infinite-volume three-to-three scattering 
amplitude $\mathcal M_3$. We comment that, like the two-to-two scattering amplitude, $\mathcal M_2$, the three-particle scattering amplitude must satisfy constraints relating its real and imaginary parts that are dictated by unitarity. These constraints are built into quantum field theory, and can be recovered order by order in a diagrammatic 
expansion. In the two-particle case, both the definition of the $S$ matrix and the diagrammatic analysis can be 
used to show that $[\mathcal M_2]^{-1} \propto  \cot \delta - i $ where the scattering phase shift $\delta$ 
(and the proportionality constant) is real. 
In the three-particle sector, unitarity takes a much more complicated form but enters our result through 
the condition that $\Kdf$ is a real function on a three-particle phase space. 
The relation to $\mathcal M_3$ then automatically produces the required unitarity properties, 
in addition to removing the scheme dependence.

As mentioned above, the results of Refs.~\cite{Hansen:2014eka,Hansen:2015zga}
were obtained under some restrictions.
The finite spatial volume was taken to be cubic (with linear extent $L$),
with periodic boundary conditions on the fields,
and the particles were assumed to be spinless and identical (with mass $m$).
The more important restrictions concerned the class of interactions considered.
These were assumed to satisfy the following two properties:
\begin{enumerate}
\item They have a $\mathbb Z_2$ symmetry such that 
$\textbf 2 \leftrightarrow \textbf 3$ transitions are forbidden;
 i.e.~only even-legged vertices are allowed.
\item They are such that the two-particle $K$ matrix, appearing due to subprocesses in which two particles scatter while the third spectates, is smooth in the kinematically available energy range. 
\end{enumerate}
The relation between the three-particle finite-volume energies and the three-to-three scattering amplitude, summarized above, holds for any system satisfying these restrictions. The relation is valid up to exponentially suppressed corrections scaling as $e^{- m L}$, which we assume are also negligible here, and holds for any allowed
value of the total three-momentum in the finite-volume frame.

In this work we remove the first of the two major restrictions; i.e.~we consider theories without a $\mathbb Z_2$ symmetry, so that all vertices are allowed
in the field theory.
We continue to impose the second restriction.
This leads to a relativistic, model-independent quantization condition that can be used to 
extract coupled two- and three-particle scattering amplitudes from LQCD. 
We otherwise use the setup of the previous studies. In particular, we assume a theory of 
identical scalar particles in a periodic, cubic box. 
Given past experience in the two-particle sector,  we expect that these restrictions on particle content
will be straightforward to remove. We also expect that the generalization to multiple two- and three-body
channels will be straightforward.
We defer consideration of these cases until a later publication.

The generalization that we derive here is a necessary step toward using LQCD to study resonances that decay into both two- and three-particle states. A prominent example is the Roper resonance, $N(1440)$, the lowest lying excitation of the nucleon. This state is counterintuitive from the perspective of quark models, as it lies below the 
first negative parity excited state.
The Roper resonance  is estimated to decay to $N\pi$ with a branching fraction of $55\%-75\%$ 
and otherwise to $N\pi\pi$, with other open channels highly suppressed.
Similarly, nearly all of the recently discovered $XYZ$ states  
have significant branching fractions into both two- and three-particle final states (see Refs.~\cite{Liu:2013waa, Chen:2016qju} for recent reviews). 
These states exhibit the rich phenomenology of nonperturbative QCD and 
it is thus highly desirable to have theoretical methods to extract their
 properties directly from the underlying theory.

\bigskip
  
 This article derives  two main results: 
 The relation between the discrete finite-volume spectrum and the generalized divergence-free $K$ matrix,
  given in Eq.~(\ref{eq:QC}), and the relation between the $K$ matrix and the coupled 
  two- and three-particle scattering amplitudes, given compactly in Eq.~(\ref{eq:compactKtoM}) and more explicitly throughout Sec.~\ref{sec:KtoM}. 
  These results generalize those of Refs.~\cite{Hansen:2014eka} and \cite{Hansen:2015zga}, respectively.
   The first, Eq.~(\ref{eq:QC}), has a form reminiscent of the coupled two-particle result~\cite{He:2005ey,Lage:2009zv, Bernard:2010fp,Hansen:2012tf, Briceno:2012yi}. The finite-volume effects are contained in a diagonal two-by-two matrix with entries $F_2$ in the two-particle sector and $F_3$ in the three-particle sector. 
   Aside from minor technical changes, these are the same finite-volume quantities that arise in the previously derived two- and three-particle quantization conditions~\cite{Luscher:1986pf, Luscher:1990ux, Kim:2005gf, Hansen:2012tf,Briceno:2012yi, Briceno:2014oea,Hansen:2014eka,Hansen:2015zga}. The coupling between channels is captured by the generalized divergence-free $K$ matrix. This contains diagonal elements, mediating two-to-two and three-to-three transitions, as well as off-diagonal elements that encode the two-to-three transitions.

To obtain both the quantization condition and the relation to the scattering amplitude from a single calculation, we use a matrix of finite-volume
correlators, $\MLL$, chosen so that it goes over to the corresponding matrix of 
infinite-volume scattering amplitudes when the $L\to\infty$ limit is taken appropriately. 
This differs from the type of correlator used in Ref.~\cite{Hansen:2014eka},
but is the direct generalization of that considered in Ref.~\cite{Hansen:2015zga}.

\begin{figure}[t]
\centering \includegraphics[width = \textwidth]{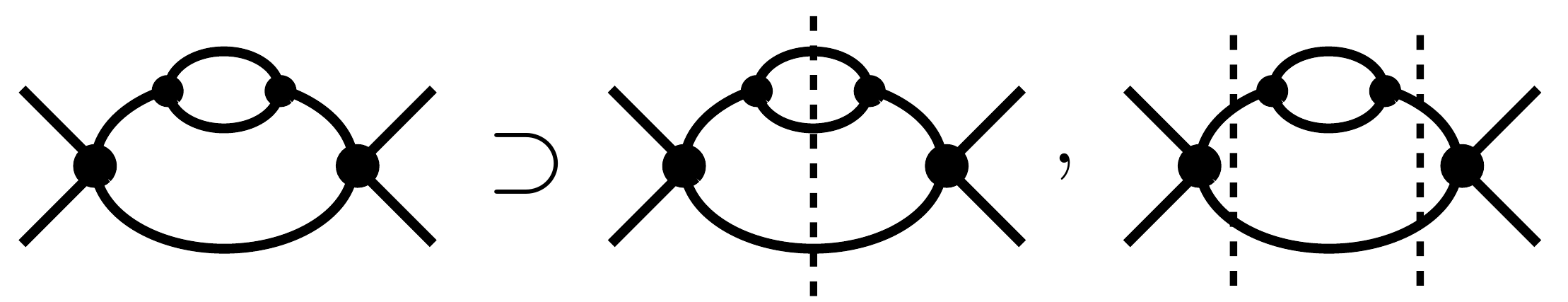}
\caption{An example of a Feynman diagram contributing to the finite-volume correlator. Above the three-particle threshold, this  diagram has cuts due to both two- and three-particle states, as shown on the right-hand side. 
The three-particle cut runs through a self-energy bubble, meaning that such diagrams must be explicitly displayed 
(rather than subsumed into a dressed propagator) in order to properly identify all finite-volume effects.}
\label{fig:examples_cuts} 
\end{figure}

The results of this work, like those given in Refs.~\cite{Luscher:1986pf, Luscher:1990ux, Kim:2005gf, Hansen:2012tf,Briceno:2012yi, Briceno:2014oea,Hansen:2014eka,Hansen:2015zga}, are derived by analyzing an infinite set of finite-volume Feynman diagrams and identifying the power-law finite-volume effects. 
The central complication new to the present derivation comes from diagrams such as that of
Fig.~\ref{fig:examples_cuts}, in which a two-to-three transition is mediated by a one-to-two transition 
together with a spectator particle. The cuts on the right-hand side of the figure indicate that this diagram 
gives rise to finite-volume effects from both two- and three-particle states. As we describe in detail below, 
a consequence of such diagrams is that we cannot use standard fully dressed propagators in two-particle loops, 
but instead need to introduce modified propagators built from two-particle-irreducible (2PI) self-energy diagrams.
 In addition, we must keep track of the fact that the two- and three-particle 
 states in these diagrams share a common coordinate. This makes it more 
 challenging to separate the finite-volume effects arising from the two- and three-particle states
 in diagrams such as that of  Fig.~\ref{fig:examples_cuts}.

To address this complication, and other technical issues that arise, we use here an approach
for studying the finite-volume correlator that differs from the skeleton-expansion-based methods of 
Refs.~\cite{Luscher:1986pf, Luscher:1990ux, Kim:2005gf, Hansen:2012tf,Briceno:2012yi, Briceno:2014oea,Hansen:2014eka,Hansen:2015zga}. 
In particular, we construct an expansion using a mix of fully dressed and
modified two- and three-particle irreducible propagators, which are connected via the local interactions of the
general quantum field theory. 
We then identify all power-law finite-volume effects using time-ordered perturbation theory (TOPT).
We also introduce smooth cutoff functions, $H_2$ and $H_3$, that only have support in the vicinity of the two- and three-particle poles, respectively. A key simplification of this construction is that,
 in disconnected two-to-three transitions such as that shown in Fig.~\ref{fig:examples_cuts}, 
 the two- and three-particle poles do not contribute simultaneously. 
 This is an extension of the result that an on-shell one-to-two transition 
 is kinematically forbidden for stable particles.

After eliminating such disconnected two-to-three transitions we are left with a series of terms built from two- and three-particle poles, summed over the spatial momenta allowed in the periodic box, and with all two-to-three transitions mediated by smooth functions. To further reduce these expressions, we apply the results of Refs.~\cite{Luscher:1986pf, Luscher:1990ux, Kim:2005gf,Hansen:2014eka,Hansen:2015zga}, to express the sums over poles as products of infinite-volume quantities and finite-volume functions. The modifications that we make to accommodate two-to-three transitions affect the exact forms of these poles, so that some effort is required to extend the previous results to rigorously apply here. With these modified relations we are able to derive a closed form for the finite-volume correlator and to express its pole positions in terms of a quantization condition.

\bigskip

The remainder of this work is organized as follows. In the following section we derive the quantization condition relating the discrete finite-volume spectrum to the generalized divergence-free $K$ matrix. After giving the precise definition of the finite-volume correlator, $\mathcal M_L$, and introducing various kinematic variables, we divide the bulk of the derivation into four subsections.  In Sec.~\ref{sec:naive} we apply standard TOPT to identify all of the two- and three-particle states that lead to important finite-volume effects. However, because of technical issues, the form reached via the standard approach is not useful for the subsequent derivation. Thus, in Sec.~\ref{sec:correctder}, we provide an alternative procedure that displays the same finite-volume effects in a more useful form. This improved derivation is highly involved and we relegate the technical details to Appendix~\ref{app:details}. With the two- and three-particle poles explicitly displayed, in Sec.~\ref{sec:voldep} we complete the decomposition of finite- and infinite-volume quantities by extending and applying various relations derived in Refs.~\cite{Luscher:1986pf, Luscher:1990ux, Kim:2005gf,Hansen:2014eka,Hansen:2015zga}.
Again, many technical details are collected in Appendix~\ref{app:FVTOPT}.
 Finally, in Sec.~\ref{sec:QC}, we identify the poles in $\mathcal M_L$ and thereby reach our quantization condition. 

To complete the derivation, in Sec.~\ref{sec:KtoM} we relate the generalized divergence-free $K$ matrix to the standard infinite-volume scattering amplitude. Our derivation here closely follows the approach of Ref.~\cite{Hansen:2015zga} but is complicated by the mixing of two- and three-body states. 
After deriving an expression for $\mathcal M_3$ in terms of the $K$ matrix in Sec.~\ref{subsec:KtoM},
we then invert the relation in Sec.~\ref{sec:MtoK}.
 Given a parametrization of the scattering amplitude, this allows one to determine the $K$ matrix and thus predict the finite-volume spectrum in terms of a given parameter set. Having given the general relation between finite-volume energies and coupled two- and three-particle scattering amplitudes, in Sec.~\ref{sec:approx} we study various limiting cases that simplify the general results. We conclude and give an outlook in Sec.~\ref{sec:conc}.

We include four appendixes. In addition to the two mentioned above,
Appendix~\ref{app:Hfunc} describes a specific example of the smooth cutoff functions, $H_2$ and $H_3$, 
that are used to simplify the results in various ways, in particular by removing disconnected two-to-three transitions,
while Appendix~\ref{app:time_rev} derives properties of the divergence-free $K$ matrix 
that follow from the parity and time-reversal invariance of the theory.

\section{Derivation of the Quantization Condition}

In this section we derive the main result of this work, a relation between the discrete finite-volume energy spectrum of a relativistic quantum field theory and that theory's physically observable, infinite-volume scattering amplitudes in the coupled two- and three-particle subspace. We restrict attention to theories with identical massive scalar particles, whose physical mass is denoted $m$. 
As we explain in more detail below, we must also assume that the two-particle $K$ matrices,
appearing due to two-particle subprocesses in the three-to-three scattering amplitude,
are only sampled at energies where they have no poles. 

The main result of this work, given in Eq.~(\ref{eq:QC}) below, is a quantization condition of the form
\begin{equation}
\label{eq:qcabb}
\Delta^{[\mathcal M]}(E, \vec P, L) = 0 \,.
\end{equation}
Here $\vec P$ is the total three-momentum of the system, 
and $L$ is the linear extent of the periodic, cubic spatial volume.
The superscript $\mathcal M$ indicates that the quantization condition 
depends on the infinite-volume scattering amplitudes of the theory.
For fixed values of $\vec P$ and $L$, solutions to Eq.~(\ref{eq:qcabb}) occur at
a discrete set of energies $E=E_1, E_2, E_3, \ldots$.
These give the finite-volume energy levels of the system,
up to exponentially suppressed corrections 
of the form $e^{- m L}$ that we neglect throughout.

We begin our derivation by introducing various kinematic variables. 
Since in general we work in a ``moving frame," with total energy-momentum $(E, \vec P)$, 
the energy in the center-of-mass (CM) frame is
\begin{equation}
E^* = \sqrt{E^2 - \vec P^2} \,.
\end{equation}
If the energy-momentum is shared between two particles, we denote the momentum
of one by $\vec p$, and that of the other by $\vec b_p = \vec P-\vec p$.
We add primes to these quantities if there are multiple two-particle states.
If the particles are on shell, we denote their energies as $\omega_p$
and $\omega_{Pp}$, respectively, with
\begin{equation}
\omega_{p} = \sqrt{\vec p^2 + m^2} \ \ {\rm and}\ \
\omega_{Pp} = \sqrt{(\vec P - \vec p)^2 + m^2 } = \sqrt{{\vec b_p}^{\;2} + m^2}
\,.
\end{equation}
If both particles are on shell, then when we boost to the CM frame, their energy-momentum four-vectors 
become $(\omega_p^*,   \vec p^*)$ and $(\omega_p^*, -\vec p^*)$, respectively, with
$\omega_p^*=E^*/2$ and $p^* \equiv \vert \vec p^* \vert = q^*$, where
\begin{equation}
q^* = \sqrt{E^{*2}/4 - m^2} \,.
\end{equation}
Thus the only remaining degree of freedom, with $(E, \vec P)$ fixed, is the direction of CM frame momentum $\hat p^*$. Throughout this work we use $\hat p^*$ to parametrize an on-shell two-particle state. 

A similar description applies when three particles share the total energy-momentum. 
The generic names we use for their momenta are
$\vec k$, $\vec a$ and $\vec b_{ka}=\vec P - \vec k -\vec a$.
If these particles are on shell, their energies are denoted $\omega_k$, $\omega_a$ and
$\omega_{Pka}$, respectively, with
\begin{equation}
\omega_{Pka} = \sqrt{(\vec P - \vec k - \vec a)^2 + m^2}  = \sqrt{\vec b_{ka}^{\;2} + m^2} \,.
\end{equation}
We will often consider the situation in which one of the particles, say that with momentum $\vec k$,
is on shell (and is referred to as the ``spectator"), 
while the other two may or may not be on shell (and are called the ``nonspectator pair").
In this situation, if we boost to the CM frame of the nonspectator pair,
the energy of this pair in this frame is denoted $E^*_{2,k}$ and is given by
\begin{equation}
E^*_{2,k} = \sqrt{(E - \omega_k)^2 - (\vec P - \vec k)^2} \,.
\end{equation}
If we further assume that all three particles are on shell, then the four-momenta
of the nonspectator pair boost to their CM frame as
 $(\omega_a, \vec a) \to (\omega_a^*, \vec a^*)$, 
$(\omega_{Pka}, \vec b_{ka}) \to (\omega_{a}^*,- \vec a^*)$,
where  $\omega_{a^*} = E^*_{2,k}/2$ and $a^* \equiv \vert \vec a^* \vert = q_k^*$,
with
\begin{equation}
q_k^* = \sqrt{E_{2,k}^{*2}/4 - m^2} \,.
\end{equation}
Thus the degrees of freedom for three on-shell particles 
with total energy-momentum $(E, \vec P)$ fixed can be parametrized
by the ordered pair $\vec k, \hat a^*$---i.e.~a spectator momentum and the direction
of the nonspectator pair in their CM frame. 

\bigskip

\begin{figure}
 \begin{center}
 \includegraphics[width = \textwidth]{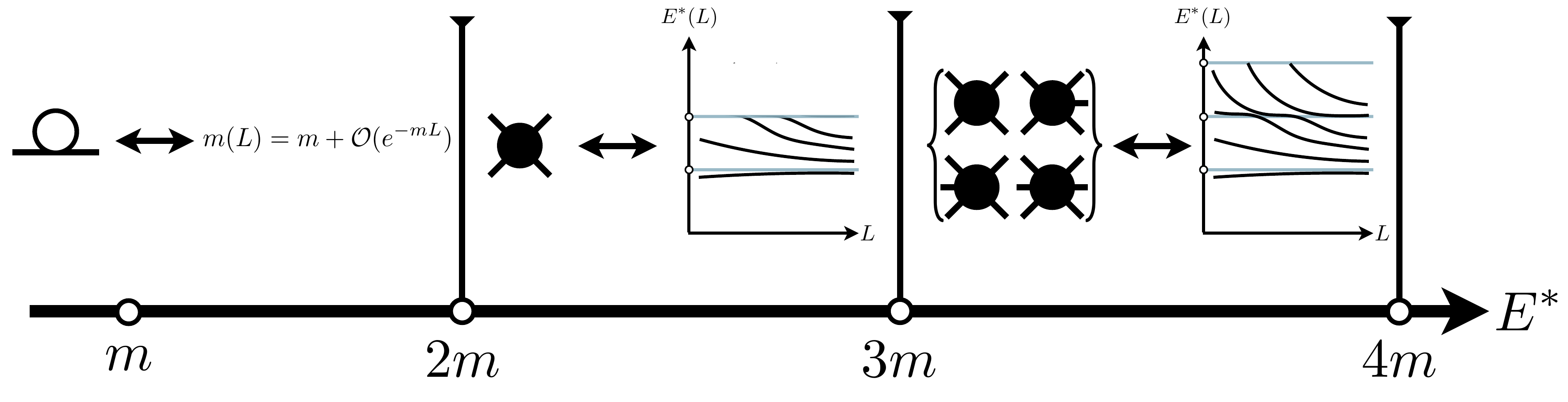}
 \caption{Summary of the range of center-of-mass frame energy, $E^*$, accommodated, and the scattering channels open in the various regions.  \label{fig:Erange}}
 \end{center}
 \end{figure}

The quantization condition derived in this work is valid for CM energies in the range\footnote{%
Strictly speaking, the quantization condition is valid also for $E^* < m$, 
but we do not expect this to be of practical interest as there are, in general, 
no finite-volume states in this region.
The quantization condition will have a solution for $E^*=m + \mathcal O(e^{-mL})$,
corresponding to a single-particle pole, but the exponentially suppressed finite-volume corrections
in the position of this pole will be incorrect. This is because we do not systematically control
such corrections. This is in contrast to finite-volume corrections to the mass of a two-particle
bound state, which are proportional to $e^{-\kappa L}$, with $\kappa$ the binding momentum.
These are correctly reproduced by the quantization condition.
}
\begin{equation}
\label{eq:kinwindow}
m < E^* < \mathrm{min}[4m, m + M_p]  \,.
\end{equation}
Here $M_p$ is the energy of the lowest lying pole in the two-particle $K$ matrix
(in the two-particle CM frame).
In practice we expect the region of practical utility to run from just below the 
two-particle threshold at $E^*=2m$, where there may be bound states, up to energies below the quoted upper limit. We caution that at energies below but near the upper limit, i.e.~at $E^* = \mathrm{min}[4m,m+M_p] - \kappa^2/m$ with $\kappa \ll m$, neglected corrections of the form $e^{-c \kappa L}$ [with $c$ a constant of $\mathcal O(1)$]
can become important. This indicates the transition into the new kinematic region where four-particle states (or $K$ matrix poles) must be included.

To explain the kinematic range quoted in Eq.~(\ref{eq:kinwindow}), we work
though the different regimes in $E^*$.
The following discussion is summarized schematically in Fig.~\ref{fig:Erange}.
In the range  $m < E^* < 3m$, the infinite-volume system 
is described solely by the two-to-two scattering amplitude,
and in finite volume this amplitude is sufficient to determine the spectral energies.
This is done with the quantization condition of L\"uscher~\cite{Luscher:1986pf,Luscher:1990ux},
and its generalizations. 

The major new result of the present work is to provide the quantization condition 
for $3m < E^*  < 4m$.  (For ease of discussion we assume first that
the two-particle $K$ matrix is smooth for the energies considered.) 
In this region, both two- and three-particle states can go on shell, 
and the dynamics of the infinite-volume system are governed by the coupled two- and three-particle scattering
amplitudes. Thus, one would expect that these same amplitudes determine the
finite-volume spectrum.
In this work we demonstrate that this is in fact the case and give the 
detailed form of the resulting quantization condition. 
Above $4m$, four-particle states become important. 
We do not include the effects of these and are thus limited by the four-particle production threshold. In fact, depending on the dynamics of the system, contributions from four-particle states might become important below threshold, as already discussed above. 

Finally, we note that within the three-to-three scattering amplitude, two-to-two scattering can occur as a subprocess with the third particle spectating. If the spectator is at rest in the three-particle CM frame, then the two-to-two amplitude is sampled at the highest possible two-particle CM frame energy, $E^* - m$. However, in our derivation of the quantization condition, we assume that the two-particle $K$ matrix is a smooth function of the two-particle energies sampled.
 Thus, if the $K$ matrix {\em does} have a pole at some two-particle CM energy $M_p$, 
 then our result holds only when $E^* - m < M_p  \Longrightarrow E^* < m + M_p$. 
 This explains the additional restriction in Eq.~(\ref{eq:kinwindow}).

\bigskip

\begin{figure}
 \begin{center}
 \includegraphics[width = \textwidth]{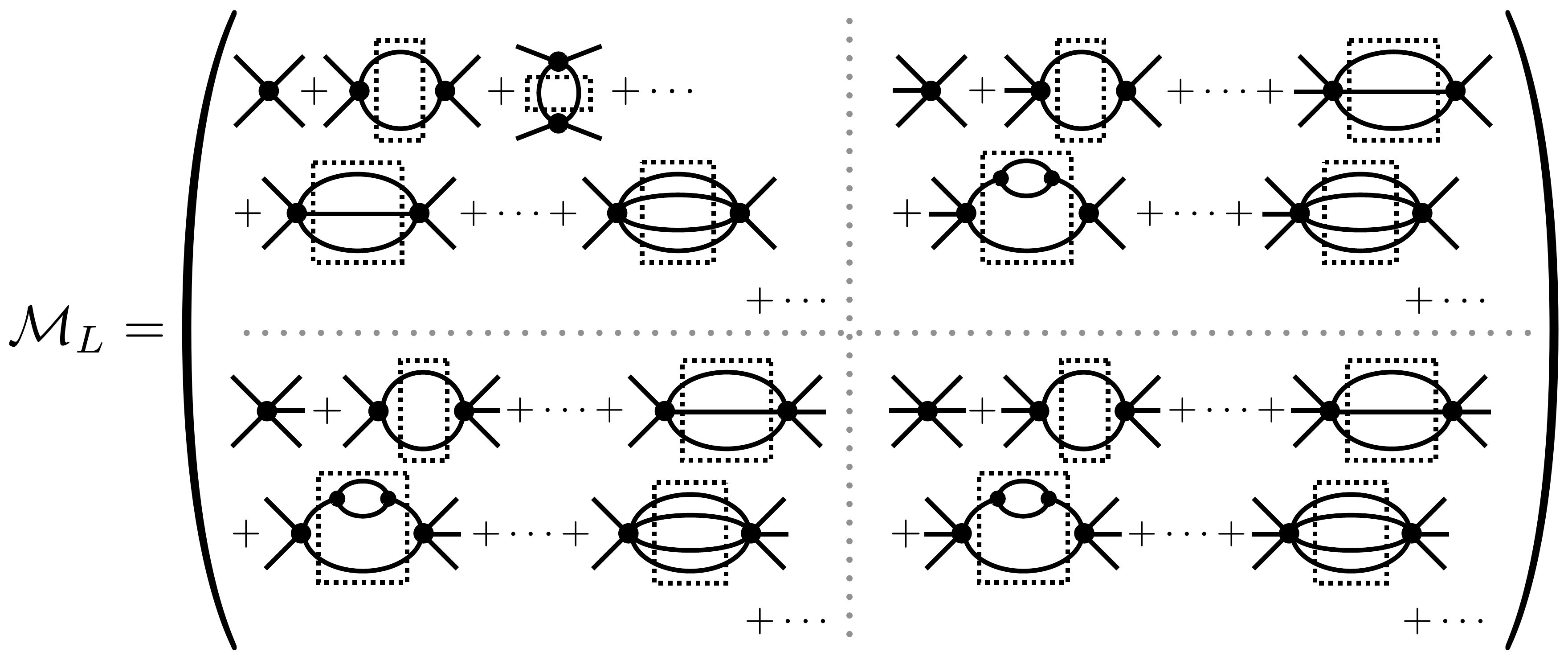}
 
 \caption{Examples of diagrams contributing to $\mathcal M_L$. 
External lines are amputated and evaluated on shell. Dashed boxes indicate that spatial loop
momenta are summed over the finite-volume set. 
\label{fig:MLdef}}
 \end{center}
 \end{figure}

We now introduce the key object used in our derivation of the quantization condition, 
a matrix of finite-volume correlators denoted $\mathcal M_L$,
\begin{equation}
\mathcal M_L \equiv \begin{pmatrix}  \mathcal M_{L,22}  & \mathcal  M_{L,23} \\ \mathcal M_{L,32} & \mathcal  M_{L,33}  \end{pmatrix} \,.
\label{eq:MLdef}
\end{equation}
$\mathcal M_{L,ij}$ is defined to be
the sum of all amputated, on-shell, connected diagrams with $j$ incoming and $i$ outgoing legs, evaluated in finite volume.
This is illustrated in Fig.~\ref{fig:MLdef}.
The restriction to finite volume implies that all spatial loop momenta are summed, rather
than integrated, with the sum running over $\vec q = 2\pi \vec n/L$, where $\vec n$ is a vector
of integers.\footnote{%
We sometimes refer to the set of all such momenta as the ``finite-volume set."}
The entries in $\MLL$ depend on the coordinates introduced above that
parametrize either two or three on-shell particles. In particular,
\begin{align}
 \mathcal M_{L,22} & \equiv  \mathcal M_{L,22}(\hat p'^{*}; \hat p^*) \,, \\
  \mathcal M_{L,23} & \equiv  \mathcal M_{L,23}(\hat p'^{*}; \vec k, \hat a^*) \,, \\
   \mathcal M_{L,32} & \equiv  \mathcal M_{L,32}(\vec k' , \hat a'^* ; \hat p^*) \,, \\
    \mathcal M_{L,33} & \equiv  \mathcal M_{L,33}(\vec k', \hat a'^*; \vec k, \hat a^*) \,.
\end{align}
These are extensions of the quantities $\mathcal M_{2,L}$ and $\mathcal M_{3,L}$ introduced
in Ref.~\cite{Hansen:2015zga}. Indeed, the latter correspond, respectively,
to $\ML{22}$ and $\ML{33}$ 
in a theory having a $\mathbb Z_2$ symmetry (in which case $\ML{23}=\ML{32}=0$).

It is clear from their definition that the $\ML{ij}$ 
are finite-volume versions of the infinite-volume scattering 
amplitudes. Indeed, as discussed in Sec.~\ref{sec:KtoM}, if the limit $L\to\infty$ is taken
in an appropriate way, $\MLL$ goes over to the infinite-volume scattering matrix.
Because of this, we loosely refer to the entries of $\MLL$ as ``finite-volume scattering amplitudes,"
recognizing that this is an imprecise description since there are no asymptotic states for finite $L$.

As defined, the external momenta  of $\MLL$ (including $\vec P$) must lie in the finite-volume set.
In this case $\MLL$ is a bona fide finite-volume correlation function whose poles occur at the
energies of the finite-volume spectrum, a property that is crucial for our derivation of the
quantization condition.
In order to relate $\MLL$ to its infinite-volume counterpart, however,
we will need to extend its definition so as to allow arbitrary external momenta.
As discussed in Ref.~\cite{Hansen:2015zga}, this extension is straightforward 
using the diagrammatic definition.
In every loop, the external momentum is routed such that only one loop momentum lies
outside the finite-volume set. A consistent choice of which momenta lie outside this set
can be made.

In many of the previous studies concerned with deriving such
quantization conditions (see for example
Refs.~\cite{Kim:2005gf,Briceno:2012yi,Hansen:2014eka}) it is standard
to first construct a skeleton expansion that expresses the
finite-volume correlator as a series of diagrams built from
Bethe-Salpeter kernels connected by
fully dressed propagators.  The utility of this approach is that it
explicitly displays the loops of particles that can go on shell, and
it turns out that only these long-distance loops lead to the power-law
finite-volume effects that we are after.  It also leads to a final
expression where all quantities can be defined in terms of
relativistically covariant amplitudes constructed from Feynman
diagrams.

\begin{figure}
 \begin{center}
 
\includegraphics[width = \textwidth]{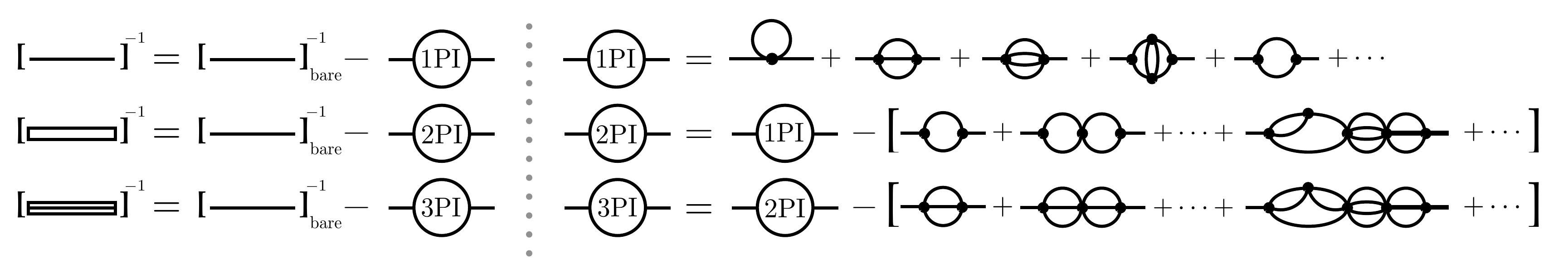}
 
 \caption{Summary of the three types of propagators used in our construction of Feynman diagrams:
fully dressed (or 1PI dressed), 2PI dressed and 3PI dressed. 
\label{fig:intro2PI}}
 \end{center}
 \end{figure}

In the present case, however, we find it simpler to follow a somewhat
different approach, based more extensively on TOPT.  This avoids the necessity of introducing
a large number of different Bethe-Salpeter kernels.  Instead of using
a skeleton expansion, we start from an all-orders diagrammatic
expansion for $\mathcal M_L$ in terms of an arbitrary collection of
contact interactions, including all possible derivative structures.
At this stage, the only place where we group diagrams together into composite
building blocks is in the propagators. Here we take all propagators to
be fully dressed with two classes of exceptions. The first applies
to propagators appearing in a two-particle loop carrying the total energy-momentum 
$(E, \vec P)$.  Then, instead of standard fully dressed propagators defined
via the one-particle irreducible (1PI) self energy diagrams,
we use a modified propagator defined via the {\em two}-particle
irreducible (2PI) self energy (see Fig.~\ref{fig:intro2PI}). This is necessary because if one of the particles in the two-particle loop
splits into two, then this leads to a three-particle state that
carries the total energy and momentum and can thus go on shell. 
We refer to such propagators as ``2PI dressed.'' The second exception occurs for diagrams in which a single propagator carries
the total energy-momentum. Such a propagator must be built from self-energies that are {\em three}-particle irreducible (3PI) (see Fig.~\ref{fig:intro2PI}). This is done so that all two- and
three-particle intermediate states are kept explicit, and we call the resulting propagator ``3PI dressed''.
The possibility of self-energy diagrams leading to on-shell three-particle
states is, in fact, one of the central complications of this work.


A second nonstandard aspect of our construction, 
closely related to the use of 2PI and 3PI propagators, 
is our use of a  ``diagram-by-diagram'' renormalization procedure. 
All diagrams are regulated in the ultraviolet (UV) 
using a regulator that we do not need to explicitly specify.
Counterterms  are then broken into an infinite series of terms designed to cancel the UV
divergences of each individual diagram, as well as certain finite pieces.
We then define each diagram to be implicitly accompanied by its counterterm so that the 
divergence is canceled immediately. 
In fact, this construction is only crucial for self-energy diagrams. 
Let $D^R_i$ denote the renormalized $i$th self-energy diagram in some labeling scheme 
$i = 1, 2, \ldots$. We then require that the counterterms are chosen such that
\begin{equation}
D^R_i(m^2) = 0 \,, \ \ \ \ \ \ \frac{d}{dp^2} D^R_i(p^2) \bigg \vert_{p^2 = m^2} = 0 \,,
\label{eq:renormscheme}
\end{equation}
implying that each self-energy diagram scales as $(p^2 - m^2)^2$ near the pole. This ensures 
that the 1PI, 2PI, 3PI and bare propagators all coincide at the one-particle pole. 
This choice is not strictly necessary, since our final result is renormalization scheme independent, but it greatly simplifies the analysis. 

\subsection{Identification of two- and three-particle poles: Na\"ive approach}
\label{sec:naive}
In this section we use TOPT to give an expression for $\MLL$ in which all the two- and
three-particle poles are explicit. However, the resulting expression turns out to be difficult
to use to determine the volume dependence, due to technical issues related
to self-energy insertions. This is why we call the approach taken here na\"ive. 
The technical issues are resolved in the following section, and its accompanying appendix,
but we think that it is useful pedagogically to separate the basic
structure of the derivation, along with the needed notation, from the technicalities.

\begin{figure}
 \begin{center}
 
\includegraphics[width = \textwidth]{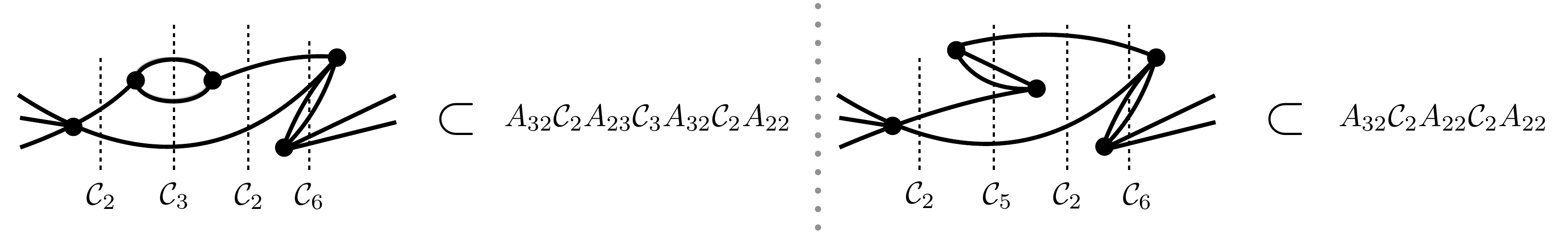}

 \caption{Examples of TOPT diagrams contributing to $\ML{32}$. The vertical dashed lines
 indicate intermediate states, which come with the $n$-cut factor $\cC_n$. 
 For the sake of clarity, we have not distinguished between the different types of propagators,
 an issue that is discussed at length in the text.
 We also do not show the diagrams containing counterterms that are associated with
 these diagrams.
 These two diagrams are both time orderings of the same underlying Feynman diagram,
 and yet contribute to different  parts of the result (\ref{eq:decom}), 
 as indicated by the expressions right of the figures.
   \label{fig:2to3TOPTex}}
 \end{center}
 \end{figure}

We give a brief recap of the essential features of TOPT in Appendix~\ref{app:TOPT}.
In essence, one evaluates all energy integrals in a Feynman diagram, arriving at a sum of terms, each of which is expressed as a set of integrals over only spatial momenta. This works equally well in finite volume, since we are taking the time direction to be infinite so that energy remains continuous. In the finite-volume case, the spatial momentum integrals are replaced by sums. Each term corresponds to a particular time ordering of vertices, 
between which are intermediate states,
each coming with an energy denominator. 
An example of such a time-ordered diagram is shown in Fig.~\ref{fig:2to3TOPTex}.
In an abuse of notation we refer to the intermediate
states as ``$n$-cuts" if they contain $n$ particles. 

 In an amputated diagram, the factor associated with an $n$-cut is proportional to
\begin{equation}
\cC_n \propto
\frac1{n!} \left( \prod_{i=1}^n \frac1{2\omega_i}\right)
 \frac1{E- \sum_{i=1}^n \omega_i}
\,,
\label{eq:ncutfactor}
\end{equation}
where $\omega_i$ is the on-shell energy of the $i$'th particle in the cut. 
The $1/n!$ is the symmetry factor for identical particles, 
and the factors of $1/(2\omega_i)$ result from on-shell propagators. 
The key point is that,
other than the factors appearing in Eq.~(\ref{eq:ncutfactor}) associated with the intermediate states,
all contributions to a TOPT diagram are smooth, nonsingular functions of the momenta. 
Thus, for the kinematic range we consider [given in Eq.~(\ref{eq:kinwindow})]
the only singularities in the diagrams arise from two- and three-cuts, and have the respective forms
\begin{equation}
\frac{1}{E - \omega_p - \omega_{Pp}} \ \ {\rm and}\ \
 \frac{1}{E - \omega_k - \omega_a - \omega_{Pka}}  \,.
 \label{eq:2and3poles}
\end{equation}
Our aim here is to obtain an expression for $\MLL$ in which all such factors are explicit.

If a summed momentum does not enter one of these two pole structures at least once, 
then we infer that for this coordinate the summand is a smooth function of characteristic width $m$.
For such a smooth function $s(\vec k)$, 
the difference between the sum and corresponding integral 
is exponentially suppressed,
\begin{equation}
\bigg [ \frac{1}{L^3} \sum_{\vec k} - \int_{\vec k}     \bigg ] s(\vec k)  = \mathcal O (e^{- m L}) \,,
\label{eq:expsuppr}
\end{equation}
Here the sum runs over the finite-volume set and $\int_{\vec k} = \int d^3 k/(2 \pi)^3$.
It follows that we may replace sums with integrals in all coordinates that do not 
enter two- and three-particle poles. 
This applies for loops with all $n$-cuts having $n\ge 4$, 
and so we are left with the finite-volume
dependence arising only from loops involving two- and three-cuts.
This procedure is illustrated in Fig.~\ref{fig:2to3TOPTex}.


Following this procedure and organizing all terms leads to the following result:
\begin{equation}
\label{eq:decom}
\mathcal M_L = A \sum_{j = 0}^\infty [\mathcal C A]^j - \tI 
= A (1-\cC A)^{-1} - \tI\,.
\end{equation} 
Each of the quantities on the right-hand side is a $2\times 2$ matrix, like $\MLL$.
The notation is highly compact, and is explained in detail below.
The basic content of the equation is, however, simple to state:
$\MLL$ can be written as a sum of terms built from alternating insertions of smooth functions, 
collected into the matrix $A$, and two- and three-particle poles, collected into the matrix $\mathcal C$.  
$A$ contains all time orderings lying between adjacent two- or three-cuts,
and includes $n$-cuts with $n\ge 4$. 
The same matrix $A$ always appears between any pair of factors of $\cC$ or external states,
because the same set of time orderings always appears.
The elements of $A$ are the analog of the Bethe-Salpeter kernels in the standard skeleton expansion approach.

The last term in Eq.~(\ref{eq:decom}) is the subtraction, $\tI$.
This arises because of the presence of disconnected terms in $A$.
That such terms are present is easily seen from Fig.~\ref{fig:2to3TOPTex}.
In the left-hand diagram, the contribution to $A_{23}$ is disconnected, 
since it involves a particle that runs between $\cC_2$ and $\cC_3$ without interacting. 
Similarly, the rightmost $A_{32}$ obtains a disconnected contribution.
The other two contributions (to the leftmost $A_{32}$ and to $A_{22}$) are connected.
In the right-hand figure the contribution to $A_{22}$ is disconnected.
Disconnected contributions are characterized by
containing one or two Kronecker deltas setting initial and final momenta equal,
each multiplied by factors of $2\omega L^3$. 
When such disconnected contributions are combined in $A + A \, \cC A + \cdots$, 
some of the resulting TOPT diagrams are themselves disconnected.
This is most obvious for the leading term, i.e.~$A$ itself.
Since $\MLL$ is, by definition, fully connected, such terms must be removed by hand,
and $\tI$ is simply defined to be the sum of all disconnected contributions in
$A [1 - \mathcal C A]^{-1}$.

It will turn out that we do not need a more detailed expression for $\tI$.
What will be important, however, is that $\tI$ only has diagonal entries, 
\begin{equation}
\tI \equiv \begin{pmatrix} \tI_{22} & 0 \\ 0 & \tI_{33}   \end{pmatrix} \,.
\end{equation}
This is because off-diagonal disconnected pieces in $\mathcal M_L$ necessarily
involve a $\textbf 1\to \textbf 2$ or $\textbf 2\to \textbf 1$ transition in which all external legs are on shell,
and this is not kinematically possible for stable particles.
We stress, however, that $A$ itself does contain off-diagonal disconnected 
contributions, because its external legs are in general not on shell.

An important property of $A$ is that all loops contained within it are integrated, rather than summed.
For the connected component of $A$, this implies that it is an infinite-volume object
(albeit not Lorentz invariant).
This holds also for the disconnected part, up to the volume dependence in the 
explicit factors of $L^3$ accompanying the Kronecker deltas mentioned above.

We now give precise definitions of the quantities entering Eq.~(\ref{eq:decom}), beginning with $\mathcal C$. 
Like all quantities in Eq.~(\ref{eq:decom}), 
$\mathcal C$ is a two-by-two matrix on the space of two- and three-particle scattering channels. 
In contrast to $\mathcal M_L$ and $A$, $\mathcal C$ (and also $\tI$, as we have explained above)
 is diagonal
\begin{equation}
\mathcal C \equiv \begin{pmatrix} \mathcal  C_{2; p' ; p } & 0 \\ 0 & \mathcal C_{3; k'a' ;k a}   \end{pmatrix} \,.
\end{equation}
The diagonal entries are matrices defined on the space of off-shell finite-volume momenta. 
For example, $\mathcal C_2$ has two indices of the form $\vec p \in (2 \pi/L) \mathbb Z^3$. 
We abbreviate this with the subscript $p';p$ as shown. The definition is
\begin{equation}
\mathcal C_{2;p';p} \equiv  - \delta_{p'p} \frac{1}{2} \frac{1}{L^3} 
\frac{1}{2 \omega_{Pp} 2 \omega_p (E - \omega_p - \omega_{Pp})} \,,
\end{equation}
which we recognize as containing the energy denominator of Eq.~(\ref{eq:ncutfactor}), 
as well as other factors. These additional factors are 
(i) $\delta_{p'p}$, which equals 1 for $\vec p{\,'} = \vec p$, and 0 otherwise, 
and is present because the cut does not change loop momenta; 
(ii) $1/L^3$, which is always associated with a loop sum; (iii) a symmetry factor of $1/2$ because the two intermediate particles are identical; and
(iv) the overall minus sign, which arises from keeping track of powers of $i$ in the
Feynman propagators and vertices before decomposing into TOPT diagrams.
Similarly, the three-cut factor is
\begin{equation}
\mathcal C_{3;k'a';k a} \equiv - \delta_{k' k} \delta_{a'a} \frac{1}{6} \frac{1}{L^6} \frac{1}{2 \omega_{a} 2 \omega_k 2 \omega_{Pka} (E - \omega_a - \omega_{k} - \omega_{Pka})} \,,
\end{equation}
where the indices include two finite-volume momenta,\footnote{%
Here we are choosing $\vec k$ and $\vec a$ to lie in the finite-volume set, so that, if the
external momenta do not lie in this set, the remaining momentum $\vec b_{ka}$ also
lies outside the set. The apparent asymmetry in this choice is removed by the fact that
the entries of $A$ are symmetric under particle exchange.}
with $k a$ standing for $\{\vec k,\vec a\}$.

The definition of the matrix $A$ depends on its location in the product. 
If it appears between two factors of $\mathcal C$, 
$A$ is defined as a matrix on the same space as $\mathcal C$,
\begin{equation}
A = \begin{pmatrix}   A_{22;p';p} & A_{23;p' ;ka} \\ A_{32;k'a';p} & A_{33;k'a';ka}
   \end{pmatrix} \,, \ \ \ \ \ \ (\mathrm{between\ two\ factors\ of\ }\mathcal C ) \,.
\label{eq:offshellA}
\end{equation}
If the $A$ lies at the left-hand end of a chain in Eq.~(\ref{eq:decom}),
so that it only abuts a $\mathcal C$ on the right, 
then it has finite-volume indices on the right but on-shell momenta on the left,
\begin{equation}
A = A(\hat p'^*;\vec k', \hat a'^*)  \equiv
 \begin{pmatrix}   A_{22; p}(\hat p'^*) &   A_{23; ka }(\hat p'^*)  \\  
     A_{32; p}(\vec k', \hat a'^*)    &    A_{33; ka}(\vec k', \hat a'^*)   \end{pmatrix}   ,
     \ \ \ \ \ \ (\mathcal C \mathrm{\ only\ on\ the\ right} ) \,.
\end{equation}
This is mirrored if the $A$ appears on the far right end of a chain,
\begin{equation}
A = A (\hat p^*, \vec k, \hat a^*) \equiv 
\begin{pmatrix}   A_{22; p'}(\hat p^*) &    A_{23; p'}(\vec k, \hat a^*)    \\  A_{32; k'a' }(\hat p^*)     &    A_{33; k'a'}(\vec k, \hat a^*)   \end{pmatrix}   ,
\ \ \ \ \ \ (\mathcal C \mathrm{\ only\ on\ the\ left} ) \,.
\end{equation}
Finally, the $j=0$ term  in Eq.~(\ref{eq:decom}) contains no factors of $\mathcal C$ and 
is evaluated only with on-shell momenta:
\begin{equation}
A =  \begin{pmatrix}   A_{22}(\hat p'^{*}; \hat p^*)   &   A_{23}(\hat p'^{*}; \vec k, \hat a^*) \\  A_{32}(\vec k', \hat a'^*;  \hat p^*) &   A_{33}(\vec k', \hat a'^*; \vec k, \hat a^*)  \end{pmatrix}  ,
 \ \ \ \ \ \ (\mathcal C\mathrm{-independent\ term} )\,.
\end{equation}

The various definitions of $A$ are all closely related and can all be 
determined from a ``master function,'' 
\begin{equation}
A(\vec p \, ', \vec k', \vec a'; \vec p, \vec k, \vec a) =   \begin{pmatrix}   A_{22}(\vec p \, '; \vec p)   &   A_{23}(\vec p \, '; \vec k, \vec a) \\  A_{32}(\vec k', \vec a'; \vec p) &   A_{33}(\vec k', \hat a'; \vec k, \vec a)  \end{pmatrix}  ,
\label{eq:masterA}
\end{equation}
by applying various coordinate-space restrictions. 
The master function depends on unrestricted momenta.
It is obtained from the fully off-shell matrix form of $A$, Eq.~(\ref{eq:offshellA}),
by continuing the momenta away from finite-volume values.
As discussed earlier, this continuation impacts the integrands inside $A$ in a well-defined
and smooth way.
For a two-particle state only one momentum, $\vec p$, is specified.
 We then define two restrictions of this coordinate. 
 To restrict to on-shell momenta we require that $\vec p$ is such that
  $E = \omega_p + \omega_{Pp}$. 
  This leaves only a directional degree of freedom, denoted $\hat p^*$. 
  Alternatively, to restrict to finite-volume momenta we require $\vec p \in (2 \pi/L) \mathbb Z^3$ 
  and represent the momentum as an index, $p$. 
  For a three-particle state we begin with two momenta $\vec k$, $\vec a$. 
  The restriction to on-shell states is effected by requiring 
  $E = \omega_k + \omega_a + \omega_{ka}$, leading to the degrees of freedom $\vec k, \hat a^*$. 
  The restriction to finite-volume momenta, $\vec k, \vec a \in (2 \pi/L) \mathbb Z^3$,
   is denoted with the index pair $ka$.

This notation allows one to easily construct various finite-volume sums. To give a concrete example we write out the term from Eq.~(\ref{eq:decom}) that is linear in $\mathcal C$,
\begin{align}
A(\hat p''^*, \vec k'', \hat a''^*) \mathcal C A(\hat p^*, \vec k, \hat a^*) & =  \sum_{ p'} 
 [A_{22; p'}(\hat p''^*) +  A_{32; p'}(\vec k'', \hat a''^*)  ] \mathcal C_{2;p';p'} 
 [    A_{22; p'}(\hat p^*) +    A_{23; p'}(\vec k, \hat a^*)   ] \\
& +  \sum_{k',  a'}  [A_{23; k'a' }(\hat p''^*)  +    A_{33; k'a'}(\vec k'', \hat a''^*)   ] 
\mathcal C_{3,k'a';k'a'} [     A_{32; k'a' }(\hat p^*)     +    A_{33; k'a'}(\vec k, \hat a^*)     ]  \,, 
\\
& = - \frac12 \frac{1}{L^3} \sum_{ \vec p \, '}   
\frac{ [A_{22}(\hat p''^*; \vec p\,') +  A_{32}(\vec k'', \hat a''^*; \vec p\,')  ]   [    A_{22}(\vec p\,';\hat p^*) +    A_{23}(\vec p\,'; \vec k, \hat a^*)   ]  }{2 \omega_{Pp'} 2 \omega_{p'} (E - \omega_{p'} - \omega_{Pp'})} \\
& - \frac16 \frac{1}{L^6}  \sum_{\vec k', \vec a'}
 \frac{ [A_{23 }(\hat p''^*; \vec k', \vec a')  +    A_{33 }(\vec k'', \hat a''^*; \vec k', \vec a')   ]     
 [     A_{32  }( \vec k', \vec a'; \hat p^*)     +    A_{33 }( \vec k', \vec a'; \vec k, \hat a^*)     ]    }
 {2 \omega_{Pk'a'} 2 \omega_{k'} 2 \omega_{a'} (E - \omega_{k'} - \omega_{a'} - \omega_{Pk'a'})} 
  \,. 
\end{align}
The simplest contribution is the product of two $A_{22}$ factors,
\begin{equation}
A \mathcal C A \supset - \frac{1}{2}\frac{1}{L^3} \sum_{ \vec p \, '}   \frac{ A_{22}(\hat p''^*; \vec p\,')       A_{22}(\vec p\,';\hat p^*)    }{2 \omega_{Pp'} 2 \omega_{p'} (E - \omega_{p'} - \omega_{Pp'})} \,. 
\end{equation}
The external momenta $\hat p''^*$ and $\hat p^*$ are fixed and the internal coordinate $\vec p\,'$ is summed over all finite-volume values.

\bigskip
Disconnected terms in $A$ complicate the determination of the volume dependence of
$\MLL$. Indeed, the analysis of Ref.~\cite{Hansen:2014eka} was largely concerned with
understanding the impact of such contributions.
Thus we would like to remove them to the extent possible.
This turns out to be possible for the off-diagonal disconnected parts of $A$, as we now explain.

We begin by recalling that finite-volume dependence arises when
one of the intermediate states goes on shell. As already noted in the discussion of $\widetilde I$, 
however,
it is not kinematically possible for both a two- and a three-particle state to be simultaneously
on shell if one of the particles has a common momentum. 
This implies that any disconnected
component in $A_{23}$ or $A_{32}$ cannot simultaneously lead to finite-volume effects
from both the adjacent cuts. 
This suggests including factors in the pole terms in $\cC$ such that this property is built
in from the beginning, rather than discovered at the end.

To formalize this idea, we introduce two functions $H_2(\vec p)$ and $H_3(\vec k, \vec a)$.
These depend, respectively,  on the momenta in a two- and three-particle off-shell intermediate state.
These functions have four key properties.
First, they are smooth functions of the momenta.
Second, they are symmetric under interchange of the particles in their respective intermediate
states, i.e. 
\begin{align}
H_2(\vec p)&=H_2(\vec b_p)\,,
\label{eq:H2sym}
\\
H_3(\vec k,\vec a)& =H_3(\vec a, \vec k) = H_3(\vec a, \vec b_{ka}) = 
H_3(\vec b_{ka},\vec a) = H_3(\vec k, \vec b_{ka}) = H_3(\vec b_{ka},\vec k)\,.
\label{eq:H3sym}
\end{align}
Third, they equal unity when all particles in a given intermediate state are on shell.
And, finally, they have no common support if one momentum is shared
between the two intermediate states.
As an equation, the ``nonoverlap" property is 
\begin{equation}
H_2(\vec p) H_3(\vec p, \vec a)=0\,.
\label{eq:H2H3}
\end{equation}
Further discussion of these properties and an explicit example of functions that
satisfy them are given in Appendix~\ref{app:Hfunc}. 
The reason that they can be defined is that there is a separation of ${\cal O}(m)$ between the individual momenta of the particles in an on-shell two-particle state and the corresponding momenta in an on-shell three-particle state.

We now rewrite Eq.~(\ref{eq:decom}) using these smooth cutoff functions. Specifically, we 
separate $\cC$ into a singular part, $\cC^H$, and a pole-free part, $\cC^\infty$,
\begin{equation}
\label{eq:Csep}
\mathcal C = \mathcal C^H + \mathcal C^{\infty} \,,
\end{equation}
where
\begin{align}
\mathcal C^H &\equiv \begin{pmatrix}  H_2(\vec p)  \mathcal C_{2; p'; p } & 0 
\\ 0 &  H_3(\vec k,  \vec a) \mathcal C_{3; k'a'; k a}   \end{pmatrix} \,,
\label{eq:CHdef}
\\
\mathcal C^\infty &\equiv \begin{pmatrix}  [1\!-\!H_2(\vec p)]  \mathcal C_{2; p'; p } & 0 
\\ 0 &  [1\!-\!H_3(\vec k,  \vec a)] \mathcal C_{3; k'a'; k a}   \end{pmatrix} \,.
\label{eq:Cinfdef}
\end{align}
$\cC^\infty$ is nonsingular because the factors of $1\!-\!H_i$ cancel their respective poles.
Substituting Eq.~(\ref{eq:Csep}) into Eq.~(\ref{eq:decom}), and collecting terms according
to the power of $\cC^H$, we arrive at
\begin{equation}
\label{eq:CHdecomtmp}
\MLL  = \widetilde A\sum_{n=0}^\infty [    \mathcal C^H \widetilde A ]^n - \tI \,,
\end{equation}
where $\widetilde A$ is given by
\begin{equation}
\widetilde A =   A \sum_{n=0}^\infty [    \mathcal C^\infty  A  ]^n \,.
\label{eq:Atildedef}
\end{equation}
This result (\ref{eq:CHdecomtmp}) is identical in form to Eq.~(\ref{eq:decom}), but with the
poles now ``regulated" by the $H$ functions, and with the kernels suitably modified.
The additional terms that have been added to obtain $\widetilde A$ from $A$
[i.e.~the $n>0$ terms in the sum in Eq.~(\ref{eq:Atildedef})]
all involve sums over intermediate momenta that have nonsingular summands,
so that these sums can be replaced by integrals ($1/L^3 \sum_k \longrightarrow \int_{\vec k}$).
Thus $\widetilde A$ remains an infinite-volume, smooth kernel, aside from
the above-mentioned Kronecker deltas accompanied by factors of  $L^3$.

The reason for this reorganization can now be understood.
$\widetilde A = A + A \cC^\infty A + \cdots $ contains disconnected parts, 
built up from the disconnected parts of $A$ discussed above.
However, it is easy to see that the off-diagonal disconnected parts of $\widetilde A$ 
do not contribute to $\MLL$.
This is because, 
if one of the $\widetilde A$'s in the expansion of Eq.~(\ref{eq:CHdecomtmp}) lies between two
factors of $\cC^H$, then its off-diagonal parts 
will be multiplied by $H_2 H_3$. But this factor vanishes for any disconnected parts,
by construction. The same is true if one or both sides of the $\widetilde A$ are at the
end of the chain, because then the external particles are on shell.\footnote{%
In more detail, the argument in this case goes as follows. We are free to multiply the
on-shell external states by a factor of $H_i$ (with $i$ the number of particles in the state),
since this factor is unity. Thus off-diagonal terms in $\widetilde A$ also come with a factor of
$H_2 H_3$ here.}
Thus, with no approximation, in Eq.~(\ref{eq:CHdecom}) we can drop the disconnected
parts of $\widetilde A_{23}$ and $\widetilde A_{32}$.

\bigskip
Having derived the formula (\ref{eq:CHdecomtmp}) we now explain why it is not
yet in a form that allows the determination of the volume dependence of $\MLL$ using the
methods of Refs.~\cite{Kim:2005gf,Hansen:2014eka,Hansen:2015zga}.
The problems are related to self-energy diagrams and the presence of disconnected
contributions.
We provide here only a brief sketch of the problems, without explaining all the
technical details, since in the end we avoid them by using an alternative approach
described in the following section.

The first issue arises in self-energy insertions on propagators present in two-particle $s$-channel loops.
An example is provided by the central loop of both diagrams 
in Fig.~\ref{fig:2to3TOPTex}.
The difference between these two diagrams is that the two vertices in
the self-energy loop have a different time ordering, leading to a different
sequence of cuts. Focusing on the central region between the two factors of $\cC_2$,
the left diagram contributes to
$A_{23} \cC_3 A_{32}$, while that on the right contributes directly to $A_{22}$.
When we change $A$ to $\widetilde A$ the two time orderings are recombined as
\begin{equation}
\widetilde A_{22} = A_{22} + A_{23} \; \cC_3 (1\!-\!H_3) A_{32} + \cdots\,.
\end{equation}
The sum over momenta that comes with $\cC_3$ can be
converted into an integral because it is multiplied by $1\!-\!H_3$.
Furthermore, since $\widetilde A_{22}$ lies between two factors of either $\cC_2 H_2$
or external on-shell states, we can set $H_3$ to zero.
Thus the two time orderings are recombined in $\widetilde A_{22}$ without
any regulator functions.
At this point we would like to say that adding these two orderings will lead to
the full, Lorentz invariant one-loop self-energy, which is proportional to $(p^2-m^2)^2$,
given our renormalization conditions.
If so, the double zero would cancel the poles in both factors of $\cC_2$, 
so that such diagrams would not in fact lead to finite-volume dependence from the two-particle loop.
In this way we would not have to worry about the self-energy insertion, except for its
contribution to three-cuts with a factor of $H_3$.

However, this argument is incorrect. To obtain the full one-loop self-energy, one needs
to include additional time orderings in which the vertices in the self-energy loop lie either before
or after the bracketing $\cC_2$ cuts. Without these, it turns out that the sum of the two diagrams that
are included only vanishes as $(p^2-m^2)$, and thus only cancels the poles in one of the $\cC_2$ factors. Thus the loop does contribute finite-volume effects.
Similarly, additional self-energy insertions on the propagators in the two-particle loop must
also be kept. This requires consideration of an infinite class of diagrams that does not arise
in the treatments of Refs.~\cite{Kim:2005gf,Hansen:2014eka,Hansen:2015zga}.

The second issue concerns Feynman diagrams contributing to $\MLL$ that are 1PI in the
$s$ channel, i.e.~have all the energy-momentum flowing through a single particle.
As noted above, the propagator of this particle must be 3PI. It turns out
that this leads to a new type of disconnected contribution to $A_{33}$ that is not a smooth function
of the external momenta. 
This is explained in Appendix~\ref{app:3PI}.
Such contributions cannot be handled using the methods
of Refs.~\cite{Kim:2005gf,Hansen:2014eka,Hansen:2015zga}, which rely on certain
smoothness properties of the kernels.
The issue with the 3PI propagators must be addressed at the level of Feynman diagrams,
before turning to TOPT.

\subsection{Identification of two- and three-particle poles: Improved approach}
\label{sec:correctder}

In this section we sketch the derivation of a replacement for
Eq.~(\ref{eq:CHdecomtmp}) that has an identical form but 
contains modified kernels $\tM$ (replacing $\widetilde A$),
and a modified subtraction $I$ (in place of $\tI$),
\begin{equation}
\mathcal M_{L} = \tM \frac{1}{1 - \mathcal C \tM} - I \,.
\label{eq:CHdecom}
\end{equation}
The issues described at the end of
the previous section do not apply to the new formulation,
and thus the methods
of Refs.~\cite{Kim:2005gf,Hansen:2014eka,Hansen:2015zga} can be applied to
analyze Eq.~(\ref{eq:CHdecom}).
The derivation is rather technical and lengthy and so is only sketched here.
It is explained in detail in Appendix~\ref{app:details}. 

We begin by following the same path as in the previous section,
constructing the diagrammatic expansion for $\mathcal M_L$ in terms of all possible contact interactions and the three types of dressed propagators. The latter can
be replaced by their infinite-volume counterparts, as they contain no on-shell intermediate states. 
This is described in more detail in Appendix~\ref{app:diagrammatic}, where we also explain why tadpole diagrams
can be absorbed into vertices to further simplify the set of allowed diagrams.

We then deviate from the na\"ive approach 
in the class of diagrams containing self-energy insertions on
propagators in two-particle $s$-channel loops [see Fig.~\ref{fig:twoloopsep} below]. 
As described in Appendix~\ref{app:partialreduction}, by inserting 
$1 = H_2(\vec p) + [1 \!-\! H_2(\vec p)]$ in such loops, we find that self-energies can
be ignored for the part with $H_2$, because they cancel poles and collapse the propagators
to local interactions. The $1\!-\!H_2$ terms remain, but they do not have any two-particle
cuts. This resolves the first complication described at the end of the previous section.

We next resolve the second complication from the
previous section involving 3PI-dressed propagators.
As described in Appendix~\ref{app:3PI}, these propagators can effectively be
shrunk to point vertices that cannot be cut.

After taking stock of the remaining classes of diagrams in Appendix~\ref{app:loopclassification},
we next switch to using TOPT. 
In Appendix~\ref{app:TOPT}, we explain how TOPT applies to our amputated on-shell correlators 
involving dressed propagators.
We thus reach a result corresponding to Eq. (18) in the na\"ive approach, but with kernels that are better behaved, and with a subtraction only needed for the 33 component.

Next, in Appendix~\ref{app:removedisc}, we
separate the cut functions $\cC$ as in Eq.~(\ref{eq:Csep}), and use the
identity in (\ref{eq:H2H3}) to reduce the number of resulting terms. 
In this and the following section of the appendix we show diagrammatically 
how the result Eq.~(\ref{eq:CHdecom}) arises.
The key properties of the kernel are that the $\tM_{22}$, $\tM_{23}$, and $\tM_{32}$ 
components contain
no disconnected parts, and are smooth, infinite-volume quantities,
while $\tM_{33}$ has disconnected parts corresponding to the two-to-two scattering
subprocess. The explicit form of the disconnected part is given in Eq.~(\ref{eq:tM33disc}).

\subsection{Volume dependence of $\MLL$}
\label{sec:voldep}
In this section we use the decomposition of the finite-volume scattering amplitude, given in Eq.~(\ref{eq:CHdecom}),
to determine the volume dependence of $\MLL$. 
Our aim is to piggyback on the methods and results of
Refs.~\cite{Kim:2005gf,Hansen:2014eka, Hansen:2015zga}, and it turns out that we can do so to a considerable extent.
However, since these works do not use TOPT to decompose finite-volume amplitudes,
some effort is needed to map their approach into the one used here.

We begin by reorganizing the series in (\ref{eq:CHdecom}) so as to separate the
contributions from the diagonal and off-diagonal elements of $B$.
Specifically, we introduce
\begin{equation}
B_D = \begin{pmatrix} B_{22} & 0 \\ 0 & B_{33}   \end{pmatrix}
\ \ {\rm and}\ \ 
B_{T} = \begin{pmatrix} 0 & B_{23} \\ B_{32} & 0   \end{pmatrix} \,, 
\label{eq:tMDTdef}
\end{equation}
such that $B=B_D+B_T$. We then rearrange Eq.~(\ref{eq:CHdecom}) into
\begin{equation}
\MLL = B_D + B_T + ( B_D +  B_T)  
\Xi  \sum_{n=0}^\infty     [  B_T  \Xi ]^n           ( B_D +  B_T) - I \,,
\label{eq:MLDT}
\end{equation}
where
\begin{equation}
\Xi \equiv  \CH \frac{1}{1 -  B_D \CH} \equiv \begin{pmatrix} \Xi_{22} & 0 \\ 0 & \Xi_{33} \end{pmatrix} \,.
\label{eq:Xidef}
\end{equation}
In this way all off-diagonal entries of $B$ are kept explicit, while the diagonal entries are
resummed into the diagonal matrix $\Xi$. 
The latter contains all the intermediate-state factors ${\cal C}^H$.

The key observation is that $\Xi$ has exactly the form that arises in the analyses of
Refs.~\cite{Kim:2005gf,Hansen:2014eka, Hansen:2015zga}. More specifically, $\Xi_{22}$ (which contains only two-cuts)
arises in Ref.~\cite{Kim:2005gf}, while $\Xi_{33}$ (containing only three-cuts) arises in Refs.~\cite{Hansen:2014eka, Hansen:2015zga}.
The only subtlety is that the result for $\Xi$ depends on the nature of the
$B$ factors on either side, i.e.~whether they are $B_D$ or $B_T$.
This dependence arises because $B_D$ (or, more precisely, $B_{33}$) contains
disconnected parts. Physically, these correspond to two-to-two subprocesses, and the
form of the result depends on whether such processes occur at the ``ends" or not.

To keep track of the different environments of the factors of $\Xi$, we introduce
superscripts indicating which type of $B$ is on either side. For example,
$\Xi^{(D,T)}$ implies that there is a $B_D$ on the left and a $B_T$ on the right.
We stress that this is only a notational device, allowing us to make substitutions that
depend on the environment (as will be explained below).
Using this notation, we further decompose $\MLL$ as
\begin{multline}
\MLL =  B_D +  B_D \Xi^{(D,D)} B_D  - I
+ B_D \Xi^{(D,T)}\sum_{n=0}^\infty\left[B_T \Xi^{(T,T)}\right]^n B_T  \Xi^{(T,D)} B_D 
\\
+ B_T
+ B_D \Xi^{(D,T)} \sum_{n=0}^\infty \left[  B_T  \Xi^{(T,T)}\right]^n   B_T 
+ B_T \sum_{n=0}^\infty \left[ \Xi^{(T,T)} B_T\right]^n  \Xi^{(T,D)} B_D
+ B_T \sum_{n=0}^\infty \left[ \Xi^{(T,T)} B_T \right]^n  \Xi^{(T,T)}  B_T\,.
\label{eq:MLdecomp}
\end{multline}
Our aim is to determine the appropriate substitutions for the four different types of
$\Xi$ factors appearing in this form.

We begin with the diagonal quantity that contains no factors of $B_T$,
\begin{equation}
X \equiv B_D + B_D \Xi^{(D,D)} B_D - I = B_D \sum_{n=0}^\infty \left[\CH B_D\right]^n - I
\equiv \begin{pmatrix} X_{22} & 0 \\ 0 & X_{33}   \end{pmatrix}\,.
\label{eq:Xdef}
\end{equation}
In terms of the components we have
\begin{align}
X_{22} &= B_{22} \sum_{n=0}^\infty \left[ \CH_2 B_{22}\right]^n \,,
\label{eq:X22}
\\
X_{33} &= B_{33} \sum_{n=0}^\infty \left[ \CH_3 B_{33}\right]^n - I_{33} \,.
\label{eq:X33}
\end{align}
These two quantities are chosen to have very similar forms to the finite-volume amplitudes
analyzed previously in Refs.~\cite{Kim:2005gf} and Refs.~\cite{Hansen:2014eka, Hansen:2015zga}, respectively,
so that we can make use of the results of these publications.

We focus first on $X_{22}$. This is the part of $\MLL$ with two-particle external
states in which, by hand, we allow only two-cuts. 
$X_{22}$ is not a physical quantity, 
since three-cuts that are present in $\mathcal M_{L,22}$ have been removed in its definition.
We note that $X_{22}$ is not only unphysical above the three-particle threshold (where we have removed physical three-particle intermediate states) but also below (where virtual three-particle contributions to $\mathcal M_{L,22}$ have been dropped). In this regard, we see that, in deriving a formalism that works both above and below the three-particle threshold, we are left with subthreshold expressions that are more complicated than the standard results describing that region. In particular, below $E^*=3m$ one can study the amplitude taking into account only the two-cuts, 
and this is indeed the approach used in Ref.~\cite{Kim:2005gf}.

Despite the unphysical nature of $X_{22}$, it has nevertheless been
constructed to have the same form as 
the physical subthreshold finite-volume two-to-two amplitude. 
In particular, $X_{22}$ is built of alternating smooth quantities ($B_{22}$) and two-cuts ($\CH_2$).
This allows us to apply the methods of Ref.~\cite{Kim:2005gf}, as explained in Appendix~\ref{app:X22}.
We show there that
\begin{equation}
X_{22}(E, \vec P) = \cK_{22,D}(E,\vec P) \frac1{1 + F_2(E,\vec P) \cK_{22,D}(E,\vec P)} 
\,,
\label{eq:X22res}
\end{equation}
where $\cK_{22,D}$ is an unphysical $K$ matrix discussed below,
and $F_2$ is the moving-frame L\"uscher zeta function\footnote{%
In Ref.~\cite{Hansen:2014eka} what we call $F_2$ here is called simply $F$. Here we reserve
$F$ for the slightly different quantity defined in Eq.~(\ref{eq:Fdef}).}
\begin{equation}
F_{2; \ell' m'; \ell m} (E,\vec P) \equiv \frac{1}{2} \bigg [ \frac{1}{L^3} \sum_{\vec p} - {\mathrm{PV}} \int \frac{d^3 p}{(2 \pi)^3} \bigg ] \frac{4 \pi Y_{\ell' m'}(\hat p^*)    
Y^*_{\ell,m}(\hat p^*)}{2 \omega_p 2 \omega_{Pp} (E -  \omega_p -  \omega_{Pp})} 
\left( \frac{p^*}{q^*}\right )^{\ell + \ell'}
h(\vec p) \,.
\label{eq:F2def}
\end{equation}
$h(\vec p)$ is a UV cutoff function, the details of which do not matter, except that
it must equal unity when $E=\omega_p+\omega_{Pp}$. 
Different choices for the cutoff function are given in Ref.~\cite{Kim:2005gf} and Refs.~\cite{Hansen:2014eka, Hansen:2015zga}.
``PV" indicates the use of the principal-value  prescription for the integral
over the pole. For $E^*>2m$ this is standard (given, for example,
by the real part of the $i\epsilon$ prescription), while for $E^*<2m$ 
we define PV such that the
result is obtained by analytic continuation from the above threshold.
This corresponds, for example, to the definition given in Refs.~\cite{Luscher:1986pf, Luscher:1990ux}.

The derivation in  Appendix~\ref{app:X22} leads
to an explicit expression for $\cK_{22,D}$, Eq.~(\ref{eq:K2Dres}).
We stress that the appearance of an unphysical $K$ matrix here
is analogous to the appearance of the unphysical quantity, $\Kdf$, in the three-particle quantization condition of Ref.~\cite{Hansen:2014eka}. 
This is not a concern, because in the end (Sec.~\ref{sec:KtoM})
we will be able to relate the unphysical quantities to physical scattering amplitudes.

\bigskip
We now turn to the quantity $X_{33}$, defined in Eq.~(\ref{eq:X33}).
This is the part of $\MLL$ with three-particle external states that contains only three-cuts.
It is unphysical at all energies since the physical amplitude always has two-cuts.
Nevertheless, it has the same structure as the finite-volume amplitude
considered in Ref.~\cite{Hansen:2015zga}, denoted $\cM_{3,L}$. This quantity is defined for theories with a 
$\mathbb Z_2$ symmetry forbidding even-odd transitions (and thus forbidding two-cuts).
Thus we can hope to reuse results from that work.
As for $X_{22}$, however, we cannot do so directly, because the analysis leading to
these results uses Feynman diagrams, whereas here we are using TOPT. Since
we are dropping cuts by hand, we cannot in any simple way recast the TOPT result 
(\ref{eq:X33}) into one using Feynman diagrams. Instead, in order to use the
results from Ref.~\cite{Hansen:2015zga}, we have to redo the analysis of Refs.~\cite{Hansen:2014eka, Hansen:2015zga}
using TOPT.

In a theory with a $\mathbb Z_2$ symmetry we have $B_{23}=B_{32}=0$, 
so $X_{33}$ is simply equal to $\cM_{3,L}$ and is thus physical.
The TOPT derivation given above still applies (and indeed is simplified by the absence
of $\textbf 2\leftrightarrow \textbf 3$ mixing) so the result Eq.~(\ref{eq:X33}) for $X_{33}$ still holds.
Although $B_{33}$ will differ in detail from that in our $\mathbb Z_2$-less theory,
its essential properties are the same.
In particular, it can be separated into connected and disconnected parts
\begin{equation}
B_{33} = B_{33}^{\rm conn} + B_{33}^{\rm disc}\,,
\label{eq:tM33decomp}
\end{equation}
with the latter containing all contributions in which two particles interact
while the other particle remains disconnected.
Determining the finite-volume dependence arising from 
these disconnected contributions was the major challenge in
the analysis of Refs.~\cite{Hansen:2014eka, Hansen:2015zga}.

Thus we must start with Eq.~(\ref{eq:X33}) rather than the Feynman diagram skeleton expansion.
This turns out to be a rather minor change. Both approaches have the same sequences of
cuts alternating with either connected or disconnected kernels. Working through
the derivation of Refs.~\cite{Hansen:2014eka, Hansen:2015zga} we find that all steps still go through,
the only change being in the precise definition of the kernels.
This is a tedious but straightforward exercise that we do not reproduce in detail, although
we collect some technical comments on the differences caused by using TOPT
in Appendix~\ref{app:X33}.
The outcome is that the final result, Eq.~(68) of Ref.~\cite{Hansen:2015zga}, still holds,
but with some of the quantities having different definitions.
Applying this result to $X_{33}$ in the $\mathbb Z_2$-less theory, we find\footnote{%
Note that we use an italic $L$ to denote finite volume, while calligraphic $\cL$ and
$\cR$ denote left and right, respectively.}
\begin{align}
X_{33}&= \cD_{L,3} +  \cS_{\cL,3}\left\{
\cL^{(u,u)}_{L,3} \mathcal K_{\rm df,33,D} \frac1{1+F_3 \mathcal K_{\rm df,33,D}} \cR^{(u,u)}_{L,3} \right\}\cS_{\cR,3}\,,
\label{eq:X33res}
\\
\cD_{L,3} &= \cS_{\cL,3}\left\{ \cD_{L,3}^{(u,u)}\right\}\cS_{\cR,3}\,,
\label{eq:DL3res}
\\
\cD_{L,3}^{(u,u)} &= -
\frac1{1+\cM_{2,L} G^H} \cM_{2,L}  G^H \cM_{2,L} [2\omega L^3] \,,
\label{eq:DL3uures}
\\
\cL^{(u,u)}_{L,3} &= \frac13 - \frac1{1+\cM_{2,L} G^H} \cM_{2,L} F
\label{eq:LL3res}
\,,\\
\cR^{(u,u)}_{L,3} &= \frac13 - 
\frac{F}{2\omega L^3}\frac1{1+\cM_{2,L} G^H} \cM_{2,L} [2\omega L^3]
\label{eq:RL3res}
\,,\\
F_3 &= \frac{F}{2\omega L^3} \cL^{(u,u)}_{L,3} = \cR^{(u,u)}_{L,3} \frac{F}{2\omega L^3}
\label{eq:F3res}
\,.
\end{align}
Here $\cS_{L,3}$ and $\cS_{R,3}$ are symmetrization operators acting respectively on the
arguments at the left and right ends of expressions within curly braces.
They are defined in Eqs.~(36) and (37) of Ref.~\cite{Hansen:2015zga}.\footnote{%
In Ref.~\cite{Hansen:2015zga} $\cS_{L,3}$ and $\cS_{R,3}$ were combined into a single
symmetrization operator $\cS$. Here it is convenient to separate the two operations.
}
The superscripts involving $u$ are explained in Ref.~\cite{Hansen:2014eka}.
$\mathcal K_{\rm df,33,D}$ is an unphysical, three-particle $K$ matrix that is a smooth function of its arguments,
and is given by Eq.~(\ref{eq:Kdf3Ddef}).
It takes the place of the quantity $\Kdf$ that appears in the theory with a $\mathbb Z_2$ symmetry,
in an analogous way to the replacement of $\cK_{2}$ with $\cK_{22,D}$ in $X_{22}$
described above.
$F$, which is defined in Ref.~\cite{Hansen:2014eka}, is similar to $F_2$, but includes an extra index to account for the third particle,\footnote{%
This form of $F$ differs from that defined in Ref.~\cite{Hansen:2014eka} by the 
choice of UV regulator in the sum-integral difference. Here we use $h(\vec p)$ [see Eq.~(\ref{eq:F2def})],
whereas in Ref.~\cite{Hansen:2014eka} a product of two $H$ functions is used. Since both regulators equal unity
at the on-shell point, the change in regulator only leads to differences of $\mathcal O(e^{-mL})$.} 
 \begin{equation}
F_{k'\ell' m';k \ell m} = \delta_{k' k} H(\vec k) F_{2;\ell' m';\ell m}(E-\omega_k,\vec P-\vec k)
\,,
\label{eq:Fdef}
\end{equation}
where the additional factor of $H$ arises from the definition of $H_3$.

The two remaining quantities that need to be defined are $\cM_{2,L}$ and $G^H$.
The former is the finite-volume two-particle scattering amplitude below the three-particle threshold,
except with an extra index for the third particle
\begin{equation}
\cM_{2,L;k'\ell' m';k \ell m} = \delta_{k' k} \left[\cK_{2}(E-\omega_k,\vec P-\vec k) 
\frac1{1 + F_2(E-\omega_k,\vec P-\vec k) \cK_{2}(E-\omega_k,\vec P-\vec k)} \right]_{\ell' m';\ell m}
\label{eq:M2Ldef}.
\end{equation}
It is important to distinguish this quantity from the two-particle finite-volume scattering amplitude, which we denote as $\cM_{L,22}$. A key feature of this result is that
it is the physical $K$ matrix, $\cK_{2}$, that appears in this expression
(rather than the unphysical $\cK_{22,D}$, for example)
as long as $E^*< 4m$.
This nontrivial result is explained in Appendix~\ref{app:X33}.
It implies that $\cD_{L,3}$, $\cL_{L,3}^{(u,u)}$, $\cR_{L,3}^{(u,u)}$, and $F_3$
are the same as those appearing in Refs.~\cite{Hansen:2014eka, Hansen:2015zga}.
The only unphysical quantity in $X_{33}$ is thus $\mathcal K_{\rm df,33,D}$.
We do not have an explicit expression for this rather complicated quantity, but this
does not matter as it will be related to the physical scattering amplitudes 
in Sec.~\ref{sec:KtoM} below.

Finally, we define $G^H$. This is almost identical to the matrix $G$ defined in Refs.~\cite{Hansen:2014eka, Hansen:2015zga}
[see, for example, Eq.~(A2) of Ref.~\cite{Hansen:2014eka}], except that it contains an additional
cutoff function. The necessity of this change is discussed in Appendix~\ref{app:X33},
and the explicit form is given in Eq.~(\ref{eq:GHdef}).
This is a minor technical change that has no impact on the general formalism.

\bigskip
The results for $X_{22}$ and $X_{33}$ can be conveniently combined by introducing
the matrices
\begin{gather}
\cD_L =  \begin{pmatrix} 0 & 0 \\ 0 & \cD_{L,3}  \end{pmatrix} \,,\ \ \
\cS_{\cL} =  \begin{pmatrix} 1 & 0 \\ 0 & \cS_{\cL,3}  \end{pmatrix} \,,\ \ \
\cS_{\cR} =  \begin{pmatrix} 1 & 0 \\ 0 & \cS_{\cR,3}  \end{pmatrix} \,,\ \ \
\cF = \begin{pmatrix} F_2 & 0 \\ 0 & F_3  \end{pmatrix} \,, \ \ \ 
\label{eq:DLSLSRFdef}
\\
\KD  = \begin{pmatrix} \cK_{22,D} & 0 \\ 0 & \cK_{\mathrm{df},33,D}  \end{pmatrix}  \,, \ \ \ 
\cL^{(u)}_L = \begin{pmatrix}  1 & 0 \\ 0 & \cL_{L,3}^{(u,u)}  \end{pmatrix} \ \ {\rm and}\ \
\cR^{(u)}_L = \begin{pmatrix}  1 & 0 \\ 0 & \cR_{L,3}^{(u,u)}  \end{pmatrix} \,.
\label{eq:KdfDLLRLdef}
\end{gather}
Then we have
\begin{equation}
X = \cD_L + \cS_{\cL}\left\{\cL^{(u)}_L \KD \frac1{1 + \cF \KD}\cR^{(u)}_L \right\} \cS_{\cR}
\,.
\label{eq:Xres}
\end{equation}

\bigskip
Our  next step is to determine the result for $\Xi^{(T,T)}$. This
lies between factors of $B_T$, so the two contributions we need to calculate are
\begin{align}
Y_{22} \equiv B_{32} \Xi_{22} B_{23} \,,
\label{eq:Y22def}
\\
Y_{33} \equiv B_{23} \Xi_{33} B_{32} \,.
\label{eq:Y33def}
\end{align}
$Y_{22}$ differs only slightly from $X_{22}$ and is calculated in
Appendix~\ref{app:Y22}, with the result
\begin{equation}
Y_{22} = B_{32}
\left[ \cD_{C,2} 
- \cD_{A',2} F_2 \frac1{1+ \cK_{22,D} F_2} \cD_{A,2}\right]B_{23}\,.
\label{eq:Y22res}
\end{equation}
The volume  dependence enters through the factors of $F_2$.
$\cD_{C,2}$, $\cD_{A',2}$ and $\cD_{A,2}$ are 
infinite-volume integral operators, whose explicit forms are given in 
Eqs.~(\ref{eq:DC2def})-(\ref{eq:DA2def}).
$\cD_{A',2}$ acts to the left, $\cD_{A,2}$ to the right,
while $\cD_{C,2}$ acts in both directions.
We refer to them collectively as decoration operators.

Turning now to $Y_{33}$, we note that this is similar to $X_{33}$, as can be seen by
comparing Eq.~(\ref{eq:X33}) to the following:
\begin{equation}
Y_{33} = B_{23}  \sum_{n=0}^\infty \left[\CH_3 B_{33}\right]^n \CH_3 B_{32}
\,.
\end{equation}
The major difference is that $Y_{33}$ has factors of $B_{23}$ or $B_{32}$ on the
ends, while $X_{33}$ has factors of $B_{33}$. This is an important difference
because $B_{23}$ and $B_{32}$ do not have disconnected parts, while $B_{33}$ does.
This means that $Y_{33}$ is analogous to the correlation function studied in Ref.~\cite{Hansen:2014eka},
in which there are three-particle connected operators at the ends (called $\sigma$ and
$\sigma^\dagger$ in that work). We thus need to repeat the analysis of Ref.~\cite{Hansen:2014eka} using
the TOPT decomposition of the correlation function. This is a subset of the work already
done for $X_{33}$ (where the presence of a disconnected component in the kernels 
on the ends leads to additional complications, as studied in Ref.~\cite{Hansen:2015zga}).
The result is that we can simply read off the answer from Eq.~(250) of Ref.~\cite{Hansen:2014eka},
\begin{equation}
Y_{33} = B_{23} \left[ \cD_{C,3} - \cD_{A',3} F_3 \frac1{1 + \mathcal K_{\rm df,33,D} F_3} \cD_{A,3}\right] B_{32}
\,.
\label{eq:Y33res}
\end{equation}
Here $\cD_{C,3}$, $\cD_{A',3}$ and $\cD_{A,3}$ are decoration operators, whose 
definition can be reconstructed from Ref.~\cite{Hansen:2014eka} taking into account the difference
between the Feynman-diagram analysis used there and the TOPT used here. 
We will, in fact, not need the definitions and so do not reproduce them here.

We observe that the form of the result is very similar to that for $Y_{22}$, Eq.~(\ref{eq:Y22res}).
The two can be combined into a matrix equation
\begin{equation}
\Xi^{(T,T)} = \cD_C - \cD_{A'}  \cF \frac{1}{1 +  \mathcal K_{\rm df,33,D} \cF } \cD_A \,,
\label{eq:TTsub}
\end{equation}
if we use the definitions
\begin{equation}
\mathcal D_C \equiv \begin{pmatrix} \mathcal D_{C,2} & 0 \\ 0 & \mathcal D_{C,3} \end{pmatrix} \,, \ \ \
\mathcal D_{A'} \equiv \begin{pmatrix} \mathcal D_{A',2} & 0 \\ 0 & \mathcal D_{A',3} \end{pmatrix} \,, \ \ \
\mathcal D_A \equiv \begin{pmatrix} \mathcal D_{A,2} & 0 \\ 0 & \mathcal D_{A,3} \end{pmatrix} \,.
\end{equation}

\bigskip
The final quantities we need to determine are $\Xi^{(D,T)}$ and its ``reflection" $\Xi^{(T,D)}$.
This requires that we calculate
\begin{align}
Z_{23} &\equiv B_{22} \Xi_{22} B_{23}\,,
\label{eq:Z23def}
\\
Z_{32} &\equiv B_{33} \Xi_{33} B_{32}\,,
\label{eq:Z32def}
\end{align}
and their reflections.
The former is obtained in Appendix~\ref{app:Y22} by a simple extension of the
analysis for $X_{22}$ and $Y_{22}$. The result is
\begin{equation}
Z_{23} + B_{23} = \left[ \frac1{1+\cK_{22,D} F_2} \cD_{A,2} \right] B_{23}\,.
\label{eq:Z23res}
\end{equation}
The calculation of $Z_{32}$ requires a more nontrivial extension of the analysis for
$X_{33}$ and $Y_{33}$. This is because $\Xi_{33}$ connects a kernel with a disconnected
component  ($B_{33}$) to one without ($B_{32}$),
and such correlators were not explicitly considered in Refs.~\cite{Hansen:2014eka, Hansen:2015zga}.
We work out the extension in Appendix~\ref{app:Z32}, finding
\begin{equation}
Z_{32} + B_{32} = \cS_{\cL,3}\left\{\cL^{(u,u)}_{L,3} \frac1{1 + \mathcal K_{\rm df,33,D} F_3} \cD_{A,3} 
B_{32} \right\} 
\,.
\label{eq:Z32res}
\end{equation}
Combining Eqs.~(\ref{eq:Z23res}) and (\ref{eq:Z32res}) into matrix form yields
\begin{equation}
  B_D \Xi^{(D,T)}    =
   \cS_{\cL} \!  \left [  \cL^{(u)}_L \frac{1}{1 + \KD\cF}  \cD_A \right ]  - 1    \,.
 \label{eq:DTsub}
\end{equation}
A similar analysis leads to the following result for the reflected quantity:
\begin{align}
 \Xi^{(T,D)}     B_D    &=
 \left [     \cD_{A'}    \frac{1}{1 + \cF \mathcal K_{\rm df,D} }   \cR^{(u)}_L \right ]  \!  \cS_{R}  - 1  \,, 
\label{eq:TDsub}
\end{align}

\bigskip
We have now determined the volume dependence of all factors of $\Xi$ appearing
in the expression (\ref{eq:MLdecomp}) for $\MLL$. Substituting Eqs.~(\ref{eq:Xres}),
(\ref{eq:TTsub}), (\ref{eq:DTsub}) and (\ref{eq:TDsub}) into this expression, 
expanding, and rearranging, we find the final result of this subsection,
\begin{equation}
\MLL =  \cD_L +  \cS_{\cL} \! \left [ \cL^{(u)}_L 
\cK_{\mathrm{df}} \frac{1}{1 + \cF \cK_{\mathrm{df}}} \cR^{(u)}_L \right ]
 \! \cS_{\cR} \,.
\label{eq:MLfinal}
\end{equation}
Here the modified matrix of  $K$ matrices is given by
\begin{equation}
\mathcal K_{\mathrm{df}}  =  \mathcal K_{\mathrm{df},D}  
+ \mathcal D_A      \sum_{n=0}^\infty [  B_T  \mathcal D_C  ]^n  
 B_T  \mathcal D_{A'}\,.
\label{eq:finalKdfdef}
\end{equation}
We stress that the second term in $\cK_{\rm df}$, which is induced by the presence of
$\textbf 2\to \textbf 3$ and $\textbf 3\to \textbf 2$ transitions, contains both diagonal and off-diagonal parts
(the former having an even number of factors of $B_T$ and the latter an odd number).

It is worth noting that, given the notation we use, the form of $\MLL$ in Eq.~(\ref{eq:MLfinal}) qualitatively resembles that of the three-particle sector in the presence of the $\mathbb Z_2$ symmetry. 

\subsection{Quantization condition}
\label{sec:QC}

The result (\ref{eq:MLfinal}) allows us to determine the energy levels of the theory
in a finite volume. This is because $\MLL$ is simply a (conveniently chosen) matrix of correlation
functions through which four-momentum $(E,\vec P)$ flows. 
It will thus diverge whenever $E$ equals the energy of a finite-volume state.\footnote{%
In general, this means that all elements of the matrix will diverge, unless there are
symmetry constraints.}
In general, such a divergence cannot come from $\cD_L$, because this quantity depends only
on the two-particle $K$ matrix, while the spectrum should depend on both two- and three-particle
channels. Since symmetrization will not produce a divergence, it must be that
the quantity in square brackets in Eq.~(\ref{eq:MLfinal}) diverges.
For the same reason as for $\cD_L$, 
divergences in $\cL_L^{(u)}$ and $\cR_L^{(u)}$ cannot correspond to 
finite-volume energies. A divergence in the matrix $\cK_{\rm df}$ will not lead to a divergent $\MLL$, 
since the former appears in both numerator and denominator.
Thus a divergence in $\MLL$ can come, in general, only from the factor $(1+ \cF \cK_{\rm df})^{-1}$.
Since this is a matrix, it will diverge whenever $\det(1+ \cF \cK_{\rm df})$ vanishes.
Thus we find the quantization condition
\begin{equation}
\mathrm{det} \left[\begin{pmatrix}  1 & 0 \\ 0 & 1\end{pmatrix} + \begin{pmatrix}  F_2 & 0 \\ 0 & F_3 \end{pmatrix} \begin{pmatrix} \mathcal K_{ 22} &  \mathcal K_{ 23} \\  \mathcal K_{ 32} &   \mathcal K_{\mathrm{df},33} \end{pmatrix} \right ] = 0 \,,
\label{eq:QC}
\end{equation}
where $\mathcal K_{22}$, $\mathcal K_{23}$, $\mathcal K_{32}$ and $\mathcal K_{\mathrm{df},33}$
 are entries in the matrix $\mathcal K_{\mathrm{df}}$ defined in Eq.~(\ref{eq:finalKdfdef}). 

We stress that each of the entries in Eq.~(\ref{eq:QC}) is itself a matrix, containing angular-momentum indices and (for the three-particle cases) also a spectator-momentum index.
The angular momentum indices run over an infinite number of values,
so the quantization condition involves an infinite-dimensional matrix. To use it in practice
one must truncate the angular-momentum space. This will be discussed further in 
Sec.~\ref{sec:approx}. We also emphasize that Eq.~(\ref{eq:QC}) separates finite-volume dependence,
contained in $F_2$ and $F_3$, from infinite-volume quantities, contained in $\cK_{\rm df}$.

The generalized quantization condition has a form that is a relatively simple generalization of
those that hold separately for two  and three particles in the case that there is a $\mathbb 
Z_2$
symmetry. Indeed, this case can be recovered simply by setting $\cK_{23}=\cK_{32}=0$.
However, we recall that, in the absence of the $\mathbb Z_2$ symmetry, the elements of $\cK_{\rm df}$
are complicated quantities, as can be seen from Eq.~(\ref{eq:finalKdfdef}).
They are also unphysical, as they depend on the cutoff functions.
In particular, $\cK_{22}$ is {\em not} equal to the physical two-particle $K$ matrix.
In fact, all we know about the elements of $\cK_{\rm df}$ is that they are smooth functions
of their arguments. In a practical application they would need to be parametrized in
some way.

By contrast, we do know $F_2$---it is given in Eq.~(\ref{eq:F2def})---and $F_3$ can
be determined from the spectrum of two-particle states below the three-particle threshold,
$E^* < 3m$.
Thus it can be determined first, before applying the full quantization condition in the regime
$3m<E^* < 4m$.
This means that by determining enough energy levels, both in the two- and three-particle regimes, one can in principle use the quantization condition to determine the parameters in any smooth ansatz for $\cK_{\rm df}$.
How to go from these parameters to a result for the physical two- and three-particle scattering
amplitudes is the topic of the next section.

\section{Relating $\mathcal{K}_{\rm df}$ to the scattering amplitude \label{sec:KtoM}}

In this section we derive the relation between $\mathcal K_{\rm df}$
and the physically observable scattering amplitude in the coupled two-
and three-particle sectors. The quantization condition derived in the
previous section depends on $\mathcal K_{\rm df}$ and also on the
finite-volume quantities $F_2$ and $F_3$. The two-particle
finite-volume factor, $F_2$, is a known kinematic function, whereas its
three-particle counterpart, $F_3$, depends on kinematic factors as
well as the two-to-two scattering amplitude at two-particle energies
below the three-particle threshold. Thus, if one uses the standard
L\"uscher approach to determine the two-to-two scattering amplitude in
the elastic region, then both $F_2$ and $F_3$ are known functions and
each finite-volume energy above the three-particle threshold gives a
constraint on $\mathcal K_{\rm df}$.

It follows that one can, in principle, use LQCD,
or other finite-volume numerical techniques,
to determine the divergence-free $K$ matrix via Eq.~(\ref{eq:QC}). 
As we have already stressed, this infinite-volume quantity is unphysical
in several ways. First, the $i\epsilon$ pole prescription is replaced
by the modified principal value prescription. Second, the $K$ matrix depends
on the cutoff functions $H_2$ and $H_3$. And, finally, the physical
singularities that occur at all above-threshold energies
in the three-to-three scattering amplitude are subtracted to define
a divergence-free quantity.

To relate $\mathcal K_{\rm df}$ to physical scattering amplitudes,
we take a carefully defined infinite-volume limit of the result for
$\MLL$ given in Eq.~(\ref{eq:MLfinal}), such that $\MLL$ goes over to
a matrix of infinite-volume scattering amplitudes.
This is the approach taken in Ref.~\cite{Hansen:2014eka}
to derive a relation between $\mathcal K_{{\rm df},3}$ and the three-particle
scattering amplitude in theories
with a $\mathbb Z_2$ symmetry preventing two-to-three transitions.
The extension here is that we must consider
a coupled set of equations with
both two- and three-particle channels.

As a warm-up, we briefly review the procedure for determining 
the two-particle scattering amplitude, $\mathcal{M}_{22}$, 
below the three-particle threshold, 
from its finite-volume analogue, $\mathcal{M}_{L,22}$. 
 The latter has the same functional form as $X_{22}$ appearing in Eq.~(\ref{eq:X22res}), with the unphysical $\cK_{22,D}$ replaced by $\cK_2$, the physical two-body $K$ matrix below the three-body threshold,
 \begin{align}
\mathcal M_{L,22}(E, \vec P) = \cK_{2}(E,\vec P) \frac1{1 + F_2(E,\vec P) \cK_{2}(E,\vec P)}  \,,
\qquad\qquad (E^* < 3 m)
\,.
\end{align}
To obtain $\mathcal{M}_{22}$, we first make the replacement
$E \to E + i \epsilon$ in the poles that appear in the finite-volume sum
contained in $F_2$, Eq.~(\ref{eq:F2def}).
Then we send $L\to\infty$ with $\epsilon$ held fixed and positive, 
and finally send $\epsilon\to0$. This converts the finite-volume Feynman diagrams
into infinite-volume diagrams with the $i\epsilon$ prescription,  which are
exactly those diagrams building up $\mathcal M_{22}$.
The result is 
\begin{align}
\mathcal{M}_{22}(E, \vec P)=
\lim_{L\to\infty}\bigg|_{i\epsilon}\mathcal{M}_{L,22}(E, \vec P)
=\mathcal{K}_{2}(E, \vec P)  \frac{1}{1+\rho_2(E, \vec P)\mathcal{K}_{2}(E, \vec P) }
\,,
\qquad\qquad (E^* <3 m), 
\label{eq:M2_K2_PVtilde}
\end{align}
where we have used~\cite{Hansen:2015zga}
\begin{align}
\lim_{L\to\infty}\bigg|_{i\epsilon}F_2(E,\vec P) &= \rho_2(E,\vec P)\,,
\label{eq:F2torho2}
\\
\label{eq:rhodef}
\rho_{2;\ell',m';\ell,m}(E, \vec P)& \equiv 
\delta_{\ell',\ell} \delta_{m',m}   \tilde\rho(E^*)\,,
\\
\label{eq:rhodef2}
\tilde\rho(E^*) &\equiv \frac{1}{16 \pi  E^*} \times
\begin{cases} 
-  i \sqrt{E^{*2}/4-m^2} & (2m)^2< E^{*2} \,, 
\\ 
\vert \sqrt{E^{*2}/4-m^2} \vert &   0<E^{*2} \leq (2m)^2 \,.
\end{cases}
\end{align}
Equation~(\ref{eq:M2_K2_PVtilde}) is just the standard relation between the two-particle $K$ matrix
and scattering amplitude.

\subsection{Expressing $\mathcal M$ in terms of $\mathcal K_{\rm df}$}
\label{subsec:KtoM}

To relate the generalized divergence-free $K$ matrix to the scattering amplitudes
we take the infinite-volume limit of Eq.~(\ref{eq:MLfinal}) using the same prescription 
as that given in Eq.~(\ref{eq:M2_K2_PVtilde}),
\begin{equation}
\label{eq:compactKtoM}
\begin{pmatrix} \mathcal M_{22} & \mathcal M_{23} \\ \mathcal M_{32} & \mathcal M_{33} \end{pmatrix} = \lim_{\epsilon \to 0} \lim_{L \to \infty} 
\left\{ \mathcal D_L +  \mathcal S_{\mathcal L} \! \left [ \mathcal L^{(u)}_L \mathcal K_{\mathrm{df}} \frac{1}{1 + \cF \mathcal K_{\mathrm{df}}} \mathcal R^{(u)}_L \right ] \! \mathcal S_{\mathcal R}
 \right \} \,.
\end{equation}
We stress that one must replace $E \to E + i \epsilon$ 
in all two- and three-particle poles appearing in finite-volume sums.
In principle this expression gives the desired relation but in very compact notation. 
The remainder of this section is dedicated to explicitly displaying 
the integral equations encoded in this result. In doing so, we take over several results from Ref.~\cite{Hansen:2015zga}.

We begin by studying the infinite-volume limit of $\mathcal D_L$, 
which is given in Eq.~(\ref{eq:DLSLSRFdef}),
and whose only nonzero element is ${\cal D}_{3,L}$.
The latter, defined in Eq.~(\ref{eq:DL3res}), is the symmetrized form of
${\cal D}^{(u,u)}_{L,3}$, given in Eq.~(\ref{eq:DL3uures}). 
The infinite-volume limit of the latter quantity,
\begin{equation}
\lim_{L \rightarrow \infty} \bigg \vert_{i \epsilon}
{\cal D}^{(u,u)}_{L,3;p \ell' m';k \ell m}
\equiv
{\cal D}^{(u,u)}_{3;\ell' m; \ell m}(\vec p, \vec k)\,,
\end{equation}
satisfies the integral equation~\cite{Hansen:2015zga}
\begin{equation}
\label{eq:Duuinteq}
 \mathcal D^{(u,u)}_3(\vec p, \vec k) =
- \mathcal{M}_{22}(\vec p)  G^\infty(\vec p,\vec k) \mathcal{M}_{22}(\vec k) 
-
\int_{\vec r\,'} \frac1{2\omega_{r'}}  \mathcal{M}_{22}(\vec p)  G^\infty(\vec p,\vec r\,' \,) 
{\cal D}^{(u,u)}_3(\vec r\,',\vec k) \,,
\end{equation}
where 
\begin{equation}
\label{eq:Ginfdef}
G_{\ell' m' ; \ell m}^\infty(\vec p, \vec k)
 \equiv
\left(\frac{k^*}{q_p^*}\right)^{\ell'} 
\frac{4 \pi Y_{\ell' m'}(\hat  k^*) 
H_3(\vec p,  \vec k)
Y_{\ell m}^*(\hat p^*)} 
{2 \omega_{Pkp} (E - \omega_k - \omega_p - \omega_{Pkp}+i\epsilon)}
\left(\frac{p^*}{q_k^*}\right)^\ell \,.
\end{equation}

Note that in Eq.~(\ref{eq:Duuinteq}) we are following the compact notation of 
Ref.~\cite{Hansen:2015zga}, in which the dependence on the spectator momenta
 is made explicit but the angular-momentum indices are suppressed. 
 Each element appearing in Eq.~(\ref{eq:Duuinteq}) is a matrix in 
 angular momentum space with two sets of $\ell m$ indices, 
 contracted in the standard way. For example, the first term is explicitly given by 
\begin{equation}
\mathcal D^{(u,u)}_{3;\ell' m' ; \ell m}(\vec p, \vec k) \,\supset\,
-\mathcal{M}_{22;\ell' m' ; \ell_1 m_1}(\vec p) 
\,
G^\infty_{\ell_1 m_1 ; \ell_2 m_2}(\vec p,\vec k) 
\,
\mathcal{M}_{22;\ell_2 m_2 ; \ell m}(\vec k) \,.
\end{equation} 

We next evaluate the infinite-volume limits of the three-particle end cap functions
 $\mathcal{L}^{(u,u)}_{3,L}$ and $\mathcal{R}^{(u,u)}_{3,L}$, 
 defined, respectively,  in Eqs.~(\ref{eq:LL3res}) and (\ref{eq:RL3res}).
 These are the only nontrivial elements of the matrices $\cL^{(u)}_L$
 and $\cR^{(u)}_L$ [see Eq.~(\ref{eq:KdfDLLRLdef})].
 Defining
\begin{align}
\lim_{L \rightarrow \infty} \bigg \vert_{i \epsilon}  
\mathcal L^{(u,u)}_{3,L; p \ell' m'; k \ell m}    & \equiv \mathcal L^{(u,u)}_{3; \ell' m'; \ell m}(\vec p, \vec k) \,, \\
\lim_{L \rightarrow \infty} \bigg \vert_{i \epsilon}   \mathcal R^{(u,u)}_{3,L; p \ell' m'; k \ell m}   & \equiv \mathcal R^{(u,u)}_{3; \ell' m'; \ell m}(\vec p, \vec k) \,, 
\end{align}
we find~\cite{Hansen:2015zga}
\begin{align}
 \mathcal L^{(u,u)}_3(\vec p, \vec k) &=  
\left(\frac13 - \mathcal{M}_{22}(\vec p\,) \rho_3(\vec p\,) \right)
(2 \pi)^3 \delta^3(\vec p - \vec k) 
- \mathcal D^{(u,u)}_3(\vec p, \vec k) \frac{\rho_3(\vec k)}{2 \omega_k} \,,
\label{eq:Luures}
\\
 \mathcal R^{(u,u)}_3(\vec p, \vec k) &=  
\left(\frac13 - \rho_3(\vec p\,) \mathcal{M}_{22}(\vec p\,) \right)
(2 \pi)^3  \delta^3(\vec p - \vec k) 
-   \frac{\rho_3(\vec p \,)}{2 \omega_p}  
\mathcal D^{(u,u)}_3(\vec p, \vec k\,) \,.
\label{eq:Ruures}
\end{align} 
Here we have used\footnote{%
What we call $\rho_3$ here is denoted simply $\rho$ in Ref.~\cite{Hansen:2015zga}.}
\begin{align}
\lim_{L \rightarrow \infty} \bigg \vert_{i \epsilon}  F &= \rho_3\,,
\\\
\rho_{3;\ell' m';\ell m}(\vec k) &\equiv 
\delta_{\ell' \ell}\, \delta_{m' m}\,  H(\vec k) \tilde\rho(E_{2,k}^*)\,.
\label{eq:rho3def}
\end{align}
We also reiterate that, in Eqs.~(\ref{eq:Luures}) and (\ref{eq:Ruures}), $\cM_{22}$ is
needed only below the three-particle threshold, so that, according to our assumptions,
it is a known quantity.

These end caps must be combined with the infinite-volume limit of the middle factor
in Eq.~(\ref{eq:compactKtoM}),
\begin{align}
\T& \equiv \lim_{L \rightarrow \infty} \bigg \vert_{i \epsilon}\T_L \,,
\\
\T_L &= 
\mathcal K_{\df} \frac{1}{1+ \mathcal F \mathcal K_{\df}}\,.
\end{align} Here both $\T_L$ and its infinite-volume counterpart, $\T$, are matrices in the space of two- and three-particle channels
\begin{align}
\T_L & \equiv
\begin{pmatrix}
\T_{22, L ; \ell_2' m_2'; \ell_2 m_2}   &  \T_{23, L ; \ell_2' m_2'; k \ell_3 m_3} 
 \\
 \T_{32, L ; k' \ell_3' m_3'; \ell_2 m_2}  &  \T_{33, L ; k' \ell_3' m_3'; k \ell_3 m_3} 
\end{pmatrix} \,, \\[5pt]
\T & \equiv
\begin{pmatrix}
\T_{22; \ell_2' m_2'; \ell_2 m_2}   &  \T_{23; \ell_2' m_2'; \ell_3 m_3}(\vec k)
 \\
 \T_{32; \ell_3' m_3'; \ell_2 m_2}(\vec k' ) &  \T_{33; \ell_3' m_3'; \ell_3 m_3}(\vec k'; \vec k)
\end{pmatrix} \,.
\end{align}
We have given different labels for the angular-momentum indices on the two- and three-particle states to stress that these are independent quantities. To take the infinite-volume limit of $\T_L$, it is more convenient to use one
of the following two matrix equations:
\begin{align}
\T_L &= \mathcal K_{\df} -\mathcal K_{\df}  \mathcal F   \T_L \,,
\label{eq:TL_R}\\
&=
\mathcal K_{\df} 
-
\T_L  \mathcal F \mathcal K_{\df} \,.
\label{eq:TL_L}
\end{align}
These go over to integral equations for $\T$ in the infinite-volume limit.

The nonzero components of the matrix $\cF$ are $F_2$ and $F_3$
[see Eq.~(\ref{eq:DLSLSRFdef})]. 
The infinite-volume limit of $F_2$ is given in Eq.~(\ref{eq:F2torho2}),
while to obtain that for $F_3$ it is convenient to rewrite it as~\cite{Hansen:2015zga}
\begin{align}
F_3=\frac{F}{2\omega L^3}\left[\frac{1}{3}-\mathcal{M}_{L,22}F-\mathcal{D}_L^{(u,u)}\frac{F}{2\omega L^3}\right] \,,
\label{eq:F3newform}
\end{align}
which allows the limit to be constructed from those for $F$, $\cM_{L,22}$ and $\cD_L^{(u,u)}$
given above.

We now have all the components to proceed. 
Taking the infinite-volume limits of Eqs.~(\ref{eq:TL_R}), (\ref{eq:TL_L}) and
(\ref{eq:F3newform}), expanding out the $2\times2$ matrices, and performing
some simple algebraic manipulations, we find
\begin{align}
 \T_{22}
 &=
\left[1+ \mathcal{K}_{22}\,{ \rho_2}\, \right]^{-1}
 \left[\mathcal{K}_{22}
  -
\int_{\vec r\,'}
\int_{\vec r}
\,
 \mathcal{K}_{23}(\vec r\,'\,)\, \frac{\rho_3(\vec r\,'\,)}{2 \omega_{r'}} \,
 {\cal L}_{3}^{(u,u)}(\vec r\,',\vec r\,)\,  \T_{32}(\vec r\,) \right]\,,
\label{eq:T22}
\\
 \T_{23}(\vec k \,)
 &=
\left[1+ \mathcal{K}_{22}\,{ \rho_2}\right]^{-1}
\left[ \mathcal{K}_{23}(\vec k \,)
  -
\int_{\vec r\,'}
\int_{\vec r}
 \mathcal{K}_{23}(\vec r\,' \,)\, 
 \frac{ \rho_3(\vec r\,' \,)}{2 \omega_{r'}}\, 
 {\cal L}_{3}^{(u,u)} (\vec r\,' ,\vec r\,) \T_{33} (\vec r ,\vec k\,)  \right]\,,
\label{eq:T23}
\\
 \T_{32}(\vec k' \,)
&=
\left[ \mathcal{K}_{32}(\vec k' \,)
-
\int_{\vec r\,'}
\int_{\vec r}
\T_{33}(\vec k', \vec r\,'\,)\,
{\cal R}_{3}^{(u,u)} (\vec r\,' ,\vec r\,)\,
\frac{ \rho_3(\vec r \,)}{2 \omega_{r}}\, 
 \mathcal{K}_{32} (\vec r \,)  \right]
 \left[1+{ \rho_2}\, \mathcal{K}_{22}\right]^{-1}\,,
\label{eq:T32_v2}
\\
 \T_{33}(\vec k' ,\vec k\,)
&=
 \mathcal{K}_{{\rm df}, 33}(\vec k' ,\vec k\,)
-
 \mathcal{K}_{32}(\vec k' \,)\,{ \rho_2 }\,  \T_{23}(\vec k\,)
 -
 \int_{\vec r\,'}
\int_{\vec r}
 \mathcal{K}_{\rm df,3}(\vec k' ,\vec r\,')\, 
 \frac{  \rho_3(\vec r\,')}{2 \omega_{r'}}\,
 {\cal L}_{3}^{(u,u)}(\vec r\,' ,\vec r\,)  \T_{33}(\vec r,\vec k\,) \,.
\label{eq:T33}
\end{align}
Substituting Eq.~(\ref{eq:T23}) in Eq.~(\ref{eq:T33}), 
and performing some further manipulations, we arrive at an integral equation for $\T_{33}$ alone
\begin{align}
 \T_{33}(\vec k' ,\vec k\,)
&=
V_{33}(\vec k' ,\vec k\,)
 -
 \int_{\vec r\,'}
\int_{\vec r}
V_{33}(\vec k' ,\vec r\,'\,)\, 
 \frac{  \rho_3(\vec r\,'\,)}{2 \omega_{r'}}\,
 {\cal L}_{3}^{(u,u)}(\vec r\,' ,\vec r\,)  \T_{33}(\vec r,\vec k\,) \,,
\label{eq:T33_v2}
\end{align}
where
\begin{align}
V_{33}(\vec k' ,\vec k\,)
= \mathcal{K}_{{\rm df}, 33}(\vec k' ,\vec k\,)
-
 \mathcal{K}_{32}(\vec k' \,)\,{ \rho_2 }\,  
 \left[1+ \mathcal{K}_{22}\,{ \rho_2}\right]^{-1}
 \mathcal{K}_{23}(\vec k \,). 
\end{align}
Given $\T_{33}$ we can then perform the integrals in Eqs.~(\ref{eq:T23}) and (\ref{eq:T32_v2})
to obtain $\T_{23}$ and $\T_{32}$, respectively, and finally perform the integral in
Eq.~(\ref{eq:T22}) to obtain $\T_{22}$.
We emphasize that all these equations involve on-shell quantities evaluated at fixed total energy and momentum, $(E,\vec P)$.

Finally, we can combine the results for $\T$, the end caps ($\mathcal L_L^{(u)}$ and $\mathcal R_L^{(u)}$), and $\cD_3$, to read off
the results for the four components of the scattering amplitude from Eq.~(\ref{eq:compactKtoM}),
\begin{align}
\mathcal{M}_{22}(\hat p'^{*}; \hat p^*)
&=
\T_{22}(\hat p'^{*}; \hat p^*)\,,
\label{eq:M2_final}
\\
\mathcal M_{23}(\hat p'^{*};  \vec{k},\hat{a}^*)
 &=
\left\{
\int_{\vec r}
\,
\T_{23}( \vec r\,)
\, 
 {\cal R}_{3}^{(u,u)} (\vec r,  \vec k\,)
 \right\}\mathcal S_{\mathcal R}
\,,
\label{eq:M23_final}
\\
\mathcal M_{32}(\vec k' , \hat a'^* ; \hat p^*)&=\mathcal S_\mathcal{L} 
\left\{\int_{\vec r\,'} {\cal L}_3^{(u,u)}(\vec{k}\,',\vec{r}\,' ) \T_{32}(\vec{r}\,' )\,   \, \right\}\,,\label{eq:M32_final}\\
\mathcal M_{33}(\vec k', \hat a'^*; \vec k, \hat a^*)
&=
{\cal D}_3(\vec k', \hat a'^*; \vec k, \hat a^*)
+
\mathcal S_{\mathcal L}
\left\{\int_{\vec r}
\int_{\vec r\,'}
 \ {\cal L}_{3}^{(u,u)}(\vec k\,',  \vec r\,)
 \T_{33}(\vec r,  \vec r\,'\,)\, 
 {\cal R}_{3}^{(u,u)}(\vec r\,',  \vec k\,)
\right\}
\mathcal S_{\mathcal R}
  \,
 .
\label{eq:M33_final}
\end{align}
In these expressions we have contracted the external harmonic indices with spherical harmonics to reach functions of momenta with no implicit indices, and symmetrized $\cD_3^{(u,u)}$ to obtain $\cD_3$.

To summarize, given $\cK_{\rm df}$ at a given value of $(E,\vec P)$,
together with knowledge of $\cM_{22}$ below the three-particle threshold,
we can obtain $\mathcal M$ at this same total four-momentum by
solving the integral equations (\ref{eq:Duuinteq}) for $\cD_{3}^{(u,u)}$
and (\ref{eq:T33_v2}) for $\T_{33}$, and then doing integrals, matrix multiplications
and symmetrizations. All the integrals are of finite range due to the presence of
the UV cutoff $H(\vec k)$ in $\rho_3$. The angular-momentum matrices have infinite size, and thus for
practical applications one must truncate them, as will be discussed in Sec.~\ref{sec:approx}.

We see from Eqs.~(\ref{eq:T22}) and (\ref{eq:M2_final}) that the two-body scattering amplitude no longer satisfies Eq.~(\ref{eq:M2_K2_PVtilde}) above the three-particle threshold.\footnote{%
If we use the full formalism below the three-particle threshold, then it is not
obvious from our results how one regains the two-particle form of
Eq.~(\ref{eq:M2_K2_PVtilde}). We return to this issue in the conclusions.
}
It is reassuring to apply the $\mathcal K_{23}\to0$ limit to Eq.~(\ref{eq:T22}) 
\begin{align}
\lim_{\mathcal K_{23}\to0} 
\mathcal{M}_{22}
&=
\left[1+ \mathcal{K}_{22}\,{ \rho_2}\right]^{-1}
\mathcal{K}_{22}
\,,
\end{align}
in which we recover the elastic two-particle unitarity form, Eq.~(\ref{eq:M2_K2_PVtilde}).

In Appendix~\ref{app:time_rev} we explore the consequences of time-reversal and parity invariance for these quantities. We conclude that, for theories with these symmetries, the two off-diagonal components of both $\mathcal K_{\rm df}$ and the scattering amplitude are simply related, so that only one of the two need be explicitly calculated.

\subsection{Expressing $\mathcal{K}_{\rm df}$ in terms of $\mathcal{M}$ \label{sec:MtoK}}

In this subsection we give a method for determining $\mathcal K_{\rm df}$ from the scattering amplitude, $\mathcal M$. 
In other words, we invert the expressions derived in the previous subsection.
The motivation for doing so is that we can imagine having a parametrization of $\cM$,
containing a finite number of parameters, from which we want to predict the finite-volume
spectrum. To do so, we need first to be able to convert from $\cM$ to $\cK_{\rm df}$,
so as to be able, in a second step, to use
the quantization condition, Eq.~(\ref{eq:QC}), to calculate the energy levels.

In the two-particle sector, applying the quantization condition in this manner
has allowed lattice practitioners to disentangle partial waves that mix due 
to the reduction of rotational symmetry~\cite{Dudek:2012xn, Dudek:2012gj},
as well as the different components in coupled-channel scattering~\cite{Dudek:2016cru, Moir:2016srx, Wilson:2015dqa, Wilson:2014cna, Dudek:2014qha}. 
This is done by parametrizing the scattering amplitudes, 
deducing how the finite-volume energy levels depend on a given parametrization 
and then performing global fits of the energy levels extracted from various volumes, boosts, and irreducible representations of the various little groups associated with the different total momenta. This technique was proposed and tested in Ref.~\cite{Guo:2012hv} for the study of coupled-channel two-particle systems. Given the parallels between coupled-channel systems with only two-particle states and the coupled two-to-three system considered here, this approach is likely to be required in an implementation of the present formalism as well. 

We again follow closely the derivation of Ref.~\cite{Hansen:2015zga} and use results
from that work.
We begin by defining the divergence-free three-to-three scattering amplitude
\begin{equation}
\mathcal M_{\rm df, 33}(\vec k', \hat a'^*; \vec k, \hat a^*) \equiv
 \mathcal M_{33}(\vec k', \hat a'^*; \vec k, \hat a^*) - {\cal D}_3(\vec k', \hat a'^*; \vec k, \hat a^*) \,,
\end{equation}
and expressing this in terms of building blocks introduced in the previous subsection
\begin{align}
\label{eq:M33dfwithLR}
\mathcal M_{\rm df, 33}(\vec k', \hat a'^*; \vec k, \hat a^*)
&=
\mathcal S_{\mathcal L}
\left\{\int_{\vec r}
\int_{\vec r\,'}
 \ {\cal L}_{3}^{(u,u)}(\vec k\,',  \vec r\,'\,)
 \T_{33}(\vec r\,',  \vec r\,)\, 
 {\cal R}_{3}^{(u,u)}(\vec r,  \vec k\,)
\right\}
\mathcal S_{\mathcal R},
  \,
\\
&=
\int_{\vec r}
\int_{\hat{b}^*}
\int_{\vec r\,'}
\int_{\hat{b}'^*}
  \left\{
(2\pi)^3 \delta^3(\vec k\,'- \vec r\,'\,)4\pi\delta^2(\hat a'^*-\hat b'^*)
+ \Delta_{\cal L}(\vec k\,',\hat a'^*;\vec r\,',\hat b'^*) \right\}
\nn\\
&\hspace{2cm}\times
 \T_{33}(\vec r\,' ,   \hat b'^*  ;  \vec r, \hat b^* )\, 
  \left\{
(2\pi)^3 \delta^3(\vec k- \vec r\,)4\pi\delta^2(\hat a^*-\hat b^*)
+ \Delta_{\cal R}(\vec r,\hat b^*;\vec k,\hat a^*) \right\}\,.
\label{eq:M33df_final}
\end{align}  
In the second form of the result we have written $\T_{33}$ in terms of on-shell momenta
rather than the spherical harmonic indices used in the first form.
The kernels $\Delta_\mathcal{R}$ and $\Delta_\mathcal{L}$ are 
taken from Ref.~\cite{Hansen:2015zga} and their definition can be inferred by comparing
Eqs.~(\ref{eq:M33dfwithLR}) and (\ref{eq:M33df_final}).
Here and below, all angular integrals are normalized to unity, 
i.e.~$\int_{\hat a^*} = \int d \Omega_{\hat a^*}/(4 \pi)$. 

Similar relations hold for $\mathcal M_{23}$ and $\mathcal M_{32}$
\begin{align}
\mathcal M_{23}(\hat p'^{*};  \vec{k},\hat{a}^*)
&=
\int_{\vec r}
\int_{\hat{b}^*}
 \T_{23}(  \vec r\,)\, 
  \left\{
(2\pi)^3 \delta^3(\vec k- \vec r\,)4\pi\delta^2(\hat a^*-\hat b^*)
+ \Delta_{\cal R}(\vec r,\hat b^*;\vec k,\hat a^*) \right\},
\\
\mathcal M_{32}(\vec{k}\,',\hat{a}'^*;\hat p^{*} )
&=
\int_{\vec r'}
\int_{\hat{b}'^*}
  \left\{
(2\pi)^3 \delta^3(\vec k\,'- \vec r\,'\,)4\pi\delta^2(\hat a'^*-\hat b'^*)
+ \Delta_{\cal L}(\vec k\,',\hat a'^*;\vec r\,',\hat b'^*) \right\}
 \T_{32}(  \vec r\,')\, .
\label{eq:M32_inverse}
\end{align}

Now, using the kernels $I_{\cal L}$ and
$I_{\cal R}$ defined in Ref.~\cite{Hansen:2015zga} via the integral equations,
\begin{align}
\label{eq:ILdef}
 I_{\cal L}(\vec k\,',\hat a'^*;\vec k,\hat a^*) &= (2\pi)^3 \delta(\vec k\,'-\vec k\,)4\pi\delta^2(\hat a'^*-\hat a^*)
-  \int_{r'} \int_{\hat b^*}
I_{\cal L}(\vec k\,',\hat a'^*;\vec r\,',\hat b^*) \Delta_{\cal L}(\vec r\,',\hat b^*;\vec k,\hat a^*) \,,
\\
\label{eq:IRdef}
 I_{\cal R}(\vec k\,',\hat a'^*;\vec k,\hat a^*) &= (2\pi)^3 \delta(\vec k\,'-\vec k\,)4\pi\delta^2(\hat a'^*-\hat a^*)
-  \int_{r'} \int_{\hat b^*}
\Delta_{\cal R}(\vec k\,',\hat a'^*;\vec r\,',\hat b^*) I_{\cal R}(\vec r,\hat b^*;\vec k,\hat a^*) 
\,,
\end{align}
we derive the following expressions for $\T_{23}$, $\T_{32}$, and $\T_{33}$ in terms of ${\cal M}_{23}$, ${\cal M}_{32}$,  and ${\cal M}_{{\rm df},33}$ respectively:
\begin{align}
\label{eq:T23_M23}
4\pi Y^*_{\ell' m'}(\hat p'^*){\T_{23;\ell' m';\ell m}}(\vec k\,) Y_{\ell m}(\hat a^*)
&= 
\int_r\int_{\hat b^*}
 {\cal M}_{23}(\hat p'^*;\vec r,\hat b^*)
I_{\cal R}(\vec r,\hat b^*;\vec k,\hat a^*)
\,,
\\
\label{eq:T32_M32}
4\pi Y^*_{\ell' m'}(\hat a'^*){\T_{32;\ell' m';\ell m}}(\vec k\,') Y_{\ell m}(\hat p^*)
&= 
\int_r\int_{\hat b^*}
I_{\cal L}(\vec k\,',\hat a'^*;\vec r,\hat b^*)
 {\cal M}_{32}(\vec r,\hat b^*;\hat p^*)
\,,
\\
4\pi Y^*_{\ell' m'}(\hat a'^*){\T_{33;\ell' m';\ell m}}(\vec k\,';\vec k\,) Y_{\ell m}(\hat a^*)
&= \int_{r'} \int_{\hat b'^*} \int_r\int_{\hat b^*}
I_{\cal L}(\vec k\,',\hat a'^*;\vec r\,',\hat b'^*) {\cal M}_{{\rm df},33}(\vec r\,',\hat b'^*;\vec r,\hat b^*)
I_{\cal R}(\vec r,\hat b^*;\vec k,\hat a^*)
\,,
\label{eq:T33_M33}
\end{align}
while $\T_{22}=\mathcal{M}_{22}$ from Eq.~(\ref{eq:M2_final}).

These expressions allow one to obtain the various components of $\T$ from the scattering amplitude. The final task is to invert Eqs.~(\ref{eq:T22}), (\ref{eq:T23}) and (\ref{eq:T33}), to determine $\mathcal K_{\rm df}$ given $\T$. One simple way to do this is to start with the inverted finite-volume relation and again take the infinite-volume limit, as in Eqs.~(\ref{eq:TL_R}) and (\ref{eq:TL_L}). This gives
\begin{align}
 \mathcal{K}_{22}
 &=
\left[1- \T_{22}\,{ \rho_2}\, \right]^{-1}
 \left[\T_{22}
  +
\int_{r'}
\int_r
\,
 \T_{23}(\vec r\,'\,)\, \frac{\rho_3(\vec r\,'\,)}{2 \omega_{r'}} \,
 {\cal L}_{3}^{(u,u)}(\vec r\,',\vec r\,)\,  \mathcal{K}_{32}(\vec r\,) \right]\,,
\label{eq:K22_T22}
\\
 \mathcal{K}_{23}(\vec k\,)
 &=
\left[1- \T_{22}\,{ \rho_2}\, \right]^{-1}
 \left[\T_{23}
  +
\int_{r'}
\int_r
\T_{23}(\vec r\,' \,)\, 
 \frac{ \rho_3(\vec r\,' \,)}{2 \omega_{r'}}\, 
 {\cal L}_{3}^{(u,u)} (\vec r\,' ,\vec r\,) \,
  \mathcal{K}_{\rm df,33} (\vec r ,\vec k\,)
\right]\,,
\label{eq:K23_T23}
\\
 \mathcal{K}_{32}(\vec k'\,)
 &=
 \left[\T_{32}
  +
\int_{r'}
\int_r
 \mathcal{K}_{\rm df,33}(\vec k\,';\vec r\,' \,)\, 
 {\cal R}_{3}^{(u,u)} (\vec r\,' ,\vec r\,) \,
  \frac{ \rho_3(\vec r \,)}{2 \omega_{r}}\, 
 \T_{32} (\vec r \,)
\right]
\left[1- { \rho_2}\,\T_{22}\, \right]^{-1}
\,,
\label{eq:K32_T32}
\\
\mathcal{K}_{\rm df,33}(\vec k' ,\vec k\,)
&=
W_{33}(\vec k' ,\vec k\,)
 +
 \int_{r'}
\int_r
W_{33}(\vec k' ,\vec r\,'\,)\, 
 \frac{  \rho_3(\vec r\,'\,)}{2 \omega_{r'}}\,
 {\cal L}_{3}^{(u,u)}(\vec r\,' ,\vec r\,)  \,
 \mathcal{K}_{\rm df,33}(\vec r,\vec k\,) \,,
\label{eq:K33_T33}
\end{align}
where
\begin{align}
W_{33}(\vec k' ,\vec k\,)
= \T_{33}(\vec k' ,\vec k\,)
+
\T_{32}(\vec k' \,)\,{ \rho_2 }\,  
 \left[1- \T_{22}\,{ \rho_2}\right]^{-1}
\T_{23}(\vec k \,). 
\end{align}
This completes the expression for $\mathcal K_{\rm df}$ in terms of $\mathcal M$.

\bigskip

In summary, given $\mathcal{M}$, one can determine the finite-volume energies as follows:
\begin{itemize}
\item Using $\mathcal M_{22}$ below the three-particle threshold, solve the integral equation (\ref{eq:Duuinteq}) to determine $\mathcal D^{(u,u)}_3(\vec p, \vec k)$.
\item Substitute this into Eqs.~(\ref{eq:Luures}) and (\ref{eq:Ruures}) to determine $\mathcal L^{(u,u)}_3(\vec p, \vec k)$ and $\mathcal R^{(u,u)}_3(\vec p, \vec k)$ and from these infer $\Delta_{\mathcal L}$ and $\Delta_{\mathcal R}$ via Eqs.~(\ref{eq:M33dfwithLR}) and (\ref{eq:M33df_final}).
\item Using $\Delta_{\mathcal L}$ and $\Delta_{\mathcal R}$ as inputs, solve the integral equations (\ref{eq:ILdef}) and (\ref{eq:IRdef}), and thereby determine $I_{\mathcal L}$ and $I_{\mathcal R}$.
\item Use these, in turn, in Eqs.~(\ref{eq:T23_M23})-(\ref{eq:T33_M33}) to deduce the two-by-two matrix $\mathcal T$ from the scattering amplitude.
\item Inserting $\mathcal T$, $\mathcal L^{(u,u)}_3$ and $\mathcal R^{(u,u)}_3$ into Eqs.~(\ref{eq:K22_T22})-(\ref{eq:K33_T33}), calculate the generalized divergence-free $K$ matrix, $\mathcal K_{\rm df}$, corresponding to the input scattering amplitude.
\item Substitute $\mathcal K_{\rm df}$ into Eq.~(\ref{eq:QC}) and solve for all roots in $E$ at fixed values of $\vec P$ and $L$.
\end{itemize}
Up to neglected terms that scale as $e^{- m L}$, these solutions correspond to the unique finite-volume energies associated with the input scattering amplitudes. Performing this procedure for a particular parametrization of $\mathcal M$, one may fit the parameter set to a large number of finite-volume energies and thereby determine the coupled two- and three-particle scattering amplitudes from Euclidean finite-volume calculations.

\section{Approximations}
\label{sec:approx}

 In order to use Eq.~(\ref{eq:QC}) in practice, it is necessary to
 truncate the matrices appearing inside the determinant.

To systematically understand the various truncations that one might apply it is useful to ``subduce'' the quantization, i.e.~to block diagonalize $1 + \mathcal K_{\rm df} \mathcal F$ and identify the quantization conditions associated with each sector. The divergence-free $K$ matrix is an infinite-volume quantity and is diagonal in the total angular momentum of the system. By contrast the finite-volume quantities $F_2$ and $F_3$ couple different angular-momentum states, a manifestation of the reduced rotational symmetry of the box. 
At the same time, the residual symmetry of the finite volume still provides important restrictions on the form of $F_2$ and $F_3$. For a given boost, these can be block diagonalized, with each block corresponding to an irreducible representation of the symmetry group. One can then truncate each block by
 assuming that all partial waves above some $\ell_{\rm max}$ do not
 contribute. This subduction procedure is well understood for the
 two-particle system~\cite{Dudek:2012gj}, and is expected to
  carry through to three-particle systems.
 
In this work we do not further discuss the subduction of the quantization condition but instead consider two simple approximations applied directly to the main result. These approximations were also discussed in Refs.~\cite{Hansen:2014eka, Hansen:2015zga}. First, we
consider the case of $\ell_{2, \rm max} = \ell_{3, \rm max} =0$, in which all two-particle angular momentum components beyond the $s$ wave are assumed to
vanish. In the two-particle sector, this
implies that all quantities that were previously matrices in angular
momentum are replaced with single numbers. The three-particle states, by contrast, still carry dependence on the spectator momentum so that the index space is reduced from $k, \ell, m$ to $k$. 
We refer to this as the $s$ wave approximation.

Using the  same arguments as in Ref.~\cite{Hansen:2014eka}, one can show that the
presence of the cutoff function $H_3$ in $F$ and $G^H$ implies that only a finite
number of spectator momenta contribute to the quantization condition.
Labeling the set of allowed momenta $\{k_1,k_2,\ldots,k_N \}$,
we can write the condition out explicitly in the $s$ wave approximation,
\begin{align}
 \hspace{-2cm} \det\left(
\begin{array}{llllll}
1+F^s_{2} \mathcal{K}_2^s &[F^s_{2} \mathcal{K}_{23}^s]_{k_1}
&[F^s_{2} \mathcal{K}_{23}^s]_{k_2} & &\left[F^s_{2}
  \mathcal{K}_{23}^s\right]_{k_N} \\ \left[F^s_{3}
  \mathcal{K}_{32}^s\right]_{k_1} & 1+[F^s_{3} \mathcal{K}_{\rm
    df,33}^s]_{k_1;k_1} & [F^s_{3} \mathcal{K}_{\rm
    df,33}^s]_{k_1;k_2} &\cdots & [F^s_{3} \mathcal{K}_{\rm
    df,33}^s]_{k_1;k_N} \\ \left[F^s_{3}
  \mathcal{K}_{32}^s\right]_{k_2} & [F^s_{3} \mathcal{K}_{\rm
    df,33}^s]_{k_2;k_1} & 1+[F^s_{3} \mathcal{K}_{\rm
    df,33}^s]_{k_2;k_2} & & [F^s_{3} \mathcal{K}_{\rm
    df,33}^s]_{k_2;k_N} \\ & \vdots&&\ddots \\ \left[F^s_{3}
  \mathcal{K}_{32}^s\right]_{k_N} & [F^s_{3} \mathcal{K}_{\rm
    df,33}^s]_{k_N;k_1} & [F^s_{3} \mathcal{K}_{\rm
    df,33}^s]_{k_N;k_2} & & 1+[F^s_{3} \mathcal{K}_{\rm
    df,33}^s]_{k_N;k_N}
\end{array}
\right)=0\,.
\label{eq:QC_s}
  \end{align}

The ``$s$" superscripts indicate that $\ell=0$ for the two-particle states and also for one of the particle pairs within the three-particle states. The explicit definitions for the components of $\mathcal K_{\rm df}$ are
\begin{gather}
{\mathcal K}_{22}^{s}\equiv{\mathcal K}_{22;00;00}\,,\ \ \ \  {\mathcal
  K}_{23;k}^{s}\equiv{\mathcal K}_{23;00;k00}\,,\\ {\mathcal
  K}_{32;k'}^{s}\equiv{\mathcal K}_{32;k'00;00}\,,\ \ \ \  {\mathcal K}_{{\rm
    df},33;k';k}^{s}\equiv{\mathcal K}_{{\rm df},33;k'00;k00}\,.
\end{gather}
The various finite-volume quantities are then given by
\begin{align}
F_{2}^s & \equiv F^s_2(E, \vec P) \equiv \frac{1}{2} \bigg [ \frac{1}{L^3} \sum_{\vec a} -
  {\mathrm{PV}} \int \frac{d^3 a}{(2 \pi)^3} \bigg ]
\frac{h(\vec a)}{2 \omega_a 2 \omega_{Pa} (E - \omega_a -
  \omega_{Pa})} \,,
\label{eq:F2def_s}
\\
\label{eq:F3_s} 
F^s_{3;k';k} & \equiv  \left [  \frac{F^s}{6 \omega L^3} - 
\frac{F^s}{2\omega L^3}\frac1{1+\cM^s_{2,L} G^s} \cM^s_{2,L} F^s  \right ]_{k';k} \,,
\\
F_{k';k}^s &  \equiv \delta_{k' k} H(\vec k) F_2^s(E - \omega_k, \vec P - \vec k) \,,
\\
\label{eq:Gdef_s}
G_{k';k} ^s  & \equiv \frac{ H_3(\vec k', \vec k\,) }
{2 \omega_{P kk'} (E - \omega_{k} - \omega_{k'}-
  \omega_{P {k}{k'}})} \frac{1}{2 \omega_{k} L^3} \,, \\
  \cM^s_{2,L;k' ;k } & \equiv \delta_{k' k}  \, \cK^s_{2}(E-\omega_k,\vec P-\vec k) 
\frac1{1 + F^s_2(E-\omega_k,\vec P-\vec k) \cK^s_{2}(E-\omega_k,\vec P-\vec k)}  \,.
\end{align}
Thus in this approximation, there are $(N+1)^2$ unknown elements of $\cK_{\rm df}$,
a complete determination of which would require determining the same number of
energy levels.%
\footnote{The number of independent components is reduced if the theory is symmetric under time reversal and/or parity transformations.  For example, if the theory has both symmetries, the relations (\ref{eq:TPsymmK23}) and (\ref{eq:TPsymmK33}) imply that the
number of independent components is $(N+1)(N+2)/2$.} %
Assuming this has been achieved, the relations of Sec.~\ref{subsec:KtoM} that give $\cM$
in terms of $\cK_{\rm df}$ still hold, except that now all the previously implicit
spherical-harmonic indices are set to zero. 

Second, we consider the simplest possible case, 
referred to in Refs.~\cite{Hansen:2014eka, Hansen:2015zga} as
the isotropic approximation. In this approximation all components of $\mathcal K_{\rm df}$ are constant functions of the momenta of the incoming and outgoing particles. Compared to the $s$-wave-only limit discussed above, here we make the additional assumption that $\mathcal K_{23}$, $\mathcal K_{32}$ and $\mathcal K_{\mathrm{df},33}$ have the same values for all choices of the spectator momentum, i.e.~are constant functions of these coordinates,
\begin{align}
{\mathcal K}_{23}^{\rm iso}&={\mathcal K}_{23;00;k00}\,,\\ {\mathcal
  K}_{32}^{\rm iso}&={\mathcal K}_{32;k'00;00}\,,\\ {\mathcal K}_{{\rm
    df},33}^{\rm iso}&={\mathcal K}_{{\rm df},33;k'00;k00}\,,
\end{align}
for all spectator momenta. Within this approximation,
Eq.~(\ref{eq:QC_s}) simplifies further to
\begin{align}
\label{eq:QC_iso}
(1+F^s_{2} \mathcal{K}_2^s) (1+F^{\rm iso}_{3} \mathcal{K}_{\rm
  df,33}^{\rm iso}) =F^s_{2}  F^{\rm iso}_{3}  \mathcal{K}_{32}^{\rm
  iso} \mathcal{K}_{23}^{\rm iso}\,,
  \end{align}
where
\begin{equation}
\label{eq:F3_iso}
F^{\rm iso}_{3} \equiv \sum_{k',k} F^s_{3;k';k }  \,. 
\end{equation}
Additional simplifications to the relation between $\cK_{\rm df}$ and $\cM$ also occur,
but we do not give these explicitly as they are simple generalizations of those
derived in Ref.~\cite{Hansen:2015zga}.

It is worth noting that Eq.~(\ref{eq:QC_iso}) resembles the expression for two coupled two-particle channels each projected to a single partial wave~\cite{He:2005ey, Hansen:2012tf, Briceno:2012yi}. In the limit that the $\textbf 2\leftrightarrow \textbf 3$ coupling vanishes, one recovers the spectrum for $s$ wave two-particle states together with that obtained in Ref.~\cite{Hansen:2014eka} for three-particle states in the isotropic approximation. Turning on the two-to-three coupling then shifts the levels and also splits any degeneracies between two- and three-particle states, as is
shown schematically in the rightmost panel of Fig.~\ref{fig:Erange}.

\section{Conclusions and outlook}
\label{sec:conc}

In this paper we have obtained the finite-volume quantization condition 
for a  general theory of identical scalar particles,
in the regime where both two- and three-particle states contribute
($3m < E^* < 4m$).
In other words, we have generalized the quantization conditions of Refs.~\cite{Luscher:1986pf, Luscher:1990ux, Rummukainen:1995vs, Kim:2005gf, Hansen:2014eka, Hansen:2015zga} to systems with general $\textbf 2 \leftrightarrow \textbf 3$ interactions. This opens the door for the first studies of particle production, a central aspect of relativistic quantum field theory, from finite-volume numerical calculations. The result also represents important progress toward our ultimate goal of relating the finite-volume spectrum and the $S$ matrix for all possible two- and three-particle systems.

Significant work is still required in order to make this formalism a practical tool for numerical lattice QCD. 
At this stage, the most important remaining restriction is that the quantization condition is valid for a given $E^*$, only if the two-particle $K$ matrix, $\cK_2$, 
is a smooth function for two-particle energies below $E^*-m$.
This is a crucial limitation as there are many examples of interesting
three-particle systems in particle and nuclear physics where $\cK_2$
does have such poles, due to the presence of narrow resonances.

In addition to the inclusion of singularities in $\cK_2$,
the quantization condition must be generalized to describe nonidentical particles and particles with intrinsic spin, and to accommodate multiple two- and three-particle channels. The importance of these extensions is exemplified by the case of the Roper resonance, which can decay into multiple two- ($N \pi$, $N \eta$) and three-particle ($N \pi \pi$) channels and for which poles in $\cK_2$ should arise in the three-particle channel due to $N \pi \pi \to \Delta \pi \to N \pi \pi$.
We expect that the generalizations in particle content will be relatively straightforward,
based on the experience with two particles. Work in this direction is underway.

The methodology adopted here differs from that used in
previous field-theoretic derivations of quantization conditions
(e.g.~that of Ref.~\cite{Hansen:2014eka})
because it relies on time-ordered perturbation theory in an essential way.
This approach has the advantage that it appears to naturally generalize to four or more particles. While such a generalization seems quite ambitious
at present, it is our ultimate goal as it will allow us to completely establish the relation between finite-volume energies and scattering observables. This in turn will allow us to study a large variety of hadronic resonances that decay into many-particle final states.

One result that we find surprising concerns the transformation, under
time reversal, of the auxiliary amplitude $\cK_{\rm df}$.
As shown in Appendix~\ref{app:time_rev}, $\cK_{\rm df}$ has exactly the same
transformation properties as $\cM$.
The complicated construction of $\cK_{{\rm df},3}$, described in Ref.~\cite{Hansen:2014eka}
for the case of no mixing with two-particle channels, and carried over here to the
case where two-to-three mixing does occur, 
includes a choice of ordering of loop integrals that seems to violate time reversal.
Nevertheless, any such violation must be canceled by the ``decorations" that are
applied to obtain the final form.
Thus $\cK_{\rm df}$ has properties that are closer to those of $\cM$ than previously expected.

One property that $\cK_{\rm df}$ does not share with $\cM$ is Lorentz invariance.
Our derivation violates manifest Lorentz invariance since it uses time-ordered 
perturbation theory. Nevertheless, as in the case of time-reversal symmetry, it
could have been the case that, at the end of the analysis, $\cK_{\rm df}$ turned out to
be Lorentz invariant. In fact, it nearly does. Looking at the relations
in Sec.~\ref{sec:KtoM}, one finds that the only violation of Lorentz invariance
comes from the denominator in $G^\infty$ [see Eq.~(\ref{eq:Ginfdef})].
The factor of $\omega_{Pkp}(E-\omega_k-\omega_p-\omega_{Pkp}+i\epsilon)$ is
manifestly noninvariant.\footnote{%
The remaining factors in $G^\infty$ are invariant as they always refer to the
CM frame of the nonspectator pair.
Were it not for the form of the denominator,
$\mathcal L_3^{(u,u)}(\vec p, \vec k) 2 \omega_k$
and $2\omega_p\mathcal R_3^{(u,u)}(\vec p, \vec k)$ would be Lorentz
invariant, as would $\cD_3^{(u,u)}$, and this would carry over to $\cK_{\rm df}$,
because all integrals would then be over Lorentz invariant phase space.}
We are investigating an alternative, Lorentz-invariant definition of $\mathcal K_{{\rm df},3}$, but
save the details for a future publication.

Finally, we highlight another feature of our formalism that deserves to be better
understood. This concerns what happens when $E^*$ passes through the three-particle
threshold at $E^*=3m$. When we are sufficiently far below this threshold, the two-particle
analysis should be valid leading to the quantization condition
$\det(1+F_2 \cK_2)=0$. However, as stressed earlier, we can also use our more
general approach in this regime, and it should lead to the same answer.
This equality is not, however, manifest. The issue is that $\mathcal K_{22}$ does not coincide with the standard two-particle $K$ matrix, even below the three-particle threshold. To study the subthreshold behavior of $\mathcal K_{22}$ one must use its relation to
the standard two-particle scattering amplitude given by Eqs.~(\ref{eq:T22}) and (\ref{eq:M2_final}). It should then be possible to express the quantization condition as the vanishing of $\det(1+F_2 \cK_2)$, up to corrections that are exponentially suppressed in $L$, but become enhanced near the three-particle threshold.

\section*{Acknowledgments}

RAB acknowledges support from U.S. Department
of Energy contract DE-AC05-06OR23177, under which Jefferson Science Associates,
LLC, manages and operates Jefferson Lab. SRS was supported in part by the United States Department
of Energy grant DE-SC0011637.

\appendix

\newpage
\section{Details of the smooth cutoff functions}

\label{app:Hfunc}
 
  \begin{figure}
 \begin{center}
 \includegraphics[scale=0.5]{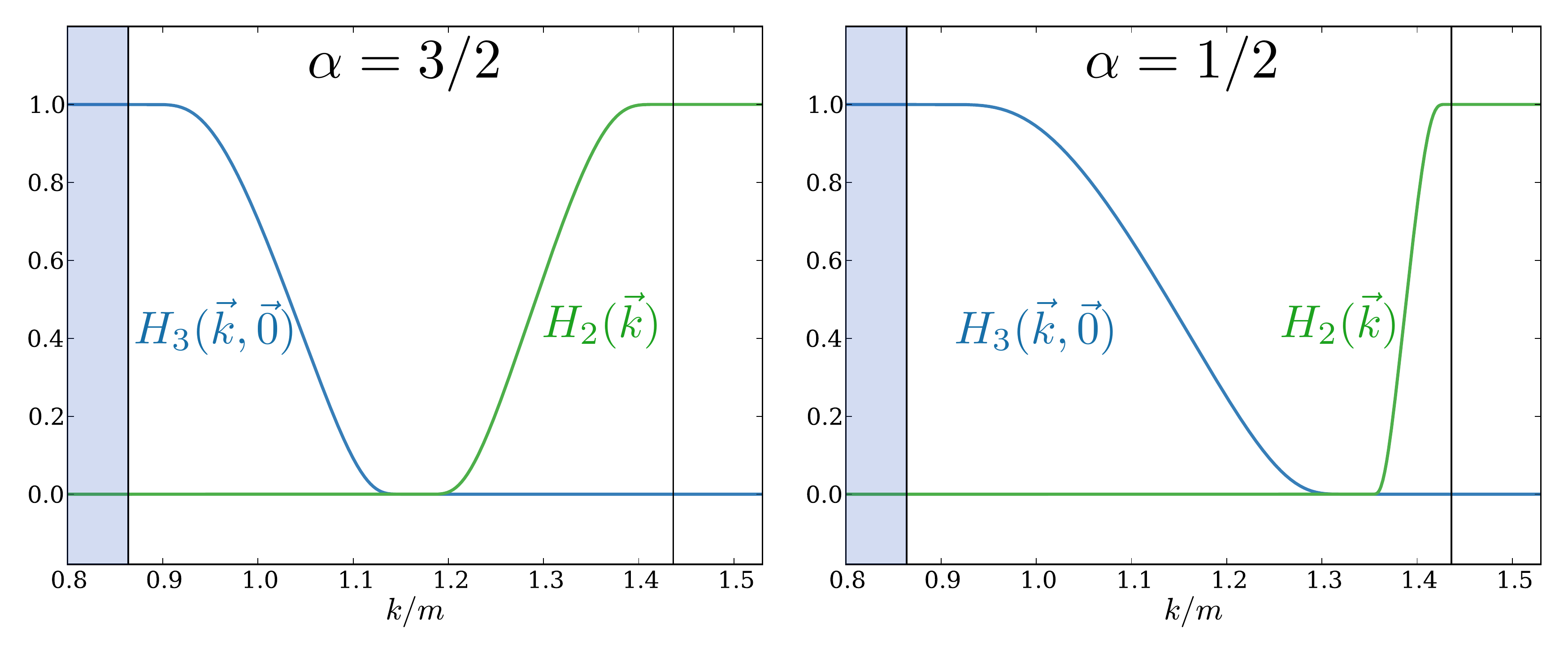}
 \caption{$H_3(\vec k, \vec 0)$ (falling, blue curve) and $H_2(\vec k)$ (rising,
   green curve) as a function of $k = \vert \vec k \vert$ for $\vec P = 0$ and $E=E^* = 3.5m$. 
   (For $\vec P=0$, $H_3(\vec k, \vec 0)$ and $H_2(\vec k)$ depend only on the magnitude of $\vec k$.)
  The shaded region on the left and the vertical line on the right indicate where
   on-shell states can occur. In the shaded region, $k$ is
   small enough that the nonspectator pair in the three-particle state
   can be on shell [$E_{2,k}^* \ge 2m \Longrightarrow H_3(\vec k, \vec 0) = 1$]. 
   Similarly, the right line indicates the $k$ value for which $E = 2\omega_k$, 
   i.e.~the value where the two-particle state goes
   on-shell [$2 \omega_k \ge E^* \Longrightarrow H_2(\vec k)=1$]. In the
   left plot we see that $\alpha=3/2$ gives the same characteristic
   width to both cutoff functions (and thus similar finite-volume effects). 
   In the right plot $\alpha=1/2$ broadens $H_3(\vec k, \vec 0)$, but
   at the expense of narrowing $H_2(\vec k)$, leading to enhanced
   finite-volume effects from the latter.}
 \label{fig:Hfunctions}
 \end{center}
 \end{figure}

In this appendix we give an explicit example of the smooth cutoff
functions used in the main text. 
These must satisfy the symmetry properties of Eqs.~(\ref{eq:H2sym}) and (\ref{eq:H3sym}),
as well as the ``nonoverlap" property of Eq.~(\ref{eq:H2H3}),
and must equal unity when the particles are on shell.

Our example uses the interpolating function $J(x)$ introduced in Ref.~\cite{Hansen:2015zta}.
This vanishes for $x\le0$, equals unity for $x\ge1$, and interpolates smoothly in between.
A specific example of such a function is
\begin{equation}
J(x) \equiv
\begin{cases}
0 \,, & x \le 0 \,; \\ \exp \left( - \frac{1}{x} \exp \left
[-\frac{1}{1-x} \right] \right ) \,, & 0<x < 1 \,; \\ 1 \,, & 1\le x
\,,
\end{cases}
\label{eq:Jdef}
\end{equation}
but our formalism works for any $J$ that satisfies the 
key property of being smooth for all $x$. 

Our example for the three-particle cutoff function is then given by
\begin{equation}
H_3(\vec k, \vec a) = H(\vec k) H(\vec a) H(\vec b_{ka})
\,,
\label{eq:H3def}
\end{equation}
where $\vec b_{ka} = \vec P - \vec k - \vec a$, and
\begin{equation}
H(\vec k) = J(z_3)\,,\qquad
z_3=  \frac{E_{2,k}^{*2} - (1 + \alpha) m^2}{(3 - \alpha ) m^2} \,.
\label{eq:Hdef}
\end{equation}
Here $\alpha$ is a parameter satisfying $-1< \alpha < 3$ that
we discuss in more detail below. The value $\alpha=-1$ corresponds to the cutoff used in
Refs.~\cite{Hansen:2015zta,Hansen:2016fzj}, but here we need a more general form.

To understand Eqs.~(\ref{eq:H3def}) and (\ref{eq:Hdef}),
recall that $E_{2,k}^{*2}=(E-\omega_k)^2-(\vec P-\vec k)^2$ is the energy of the
nonspectator pair in their CM frame, assuming that the spectator is on shell.
If all three particles are on shell, it follows that $E_{2,k}^{*2}\ge 4 m^2$.
In this case, $z_3 \ge 1$ 
(with $z_3=1$ at threshold for the nonspectator pair, $E_{2,k}^{*2}=4m^2$)
and so $H(\vec k)=1$. Similarly, the other two $H$ functions equal unity.
Thus $H_3=1$ if all three particles are on shell.\footnote{%
We note that the converse does not hold: $H_3=1$ does not imply that all
three particles are on shell, as can be seen from the simple example of
$\vec P=\vec k=\vec a=0$ with $E> 3m$.}
Now consider changing $\vec k$ (with $E$ and $\vec P$ fixed)
such that $E_{2,k}^{*2}$ drops below $4m^2$. Then $z_3$ drops below unity,
and $H(\vec k)$ falls smoothly,
vanishing when $E_{2,k}^{*2}$ reaches $(1+\alpha)\, m^2$,
and staying zero thenceforth.
Because of the symmetric product in Eq.~(\ref{eq:H3def}) it follows that
$H_3$ vanishes when any nonspectator pair has a CM squared energy that
lies $(3-\alpha)\, m^2$ below threshold.
We stress that this vanishing of $H_3$ always occurs when, with fixed $E$ and $\vec P$, any of the three momenta becomes sufficiently large. Thus $H_3$ acts as a UV cutoff.

We next describe our example for the two-particle cutoff function, $H_2(\vec p)$. This depends
only on a single momentum, since the momentum of the second particle is
fixed to $\vec b_{p}\equiv \vec P-\vec p$. The aim of $H_2$ is to ensure that,
if either $\vec p$ or $\vec b_p$ is equal to one of the three-particle momenta
$\vec k$, $\vec a$ or $\vec b_{ka}$, then $H_2(\vec p) H_3(\vec k,\vec a) = 0$.
The motivation for this condition is discussed in the main text.
We also need $H_2(\vec p)$ to equal unity if both particles are on shell.

A solution to these conditions is
\begin{equation}
H_2(\vec p) = J(z_p) J(z_b) \,, \qquad
z_p = \frac{E_{2,p}^{*2} - (1+\alpha) m^2}{(-\alpha m^2)} \,, \qquad
z_b = \frac{E_{2,b_p}^{*2} - (1+\alpha) m^2}{(-\alpha m^2)} 
\,.
\label{eq:H2def}
\end{equation}
Here $\alpha$ is the same parameter as above, except now satisfying $0 < \alpha < 3$.
In the two-particle case, $E_{2,p}^{*2}$ (given by the same expression as $E_{2,k}^{*2}$
except with $k$ replaced with $p$)
is the invariant mass squared of the particle with momentum $\vec b_{p}$,
assuming that with momentum $\vec p$ is on shell.
Similarly, $E_{2,b_p}^{*2}$ is the invariant mass squared of the particle with momentum
$\vec p$ if that with momentum $\vec b_p$ is on shell.
In general these two invariant masses are different.
One case when they are the same is if both particles are on shell, in which case
$E_{2,p}^{*2}=E_{2,b_p}^{*2}=m^2$. Then $z_p=z_b=1$, so that $H_2=1$, as required.

Now we consider what happens to $H_2$ as we vary $\vec p$ away from a value
leading to two on-shell particles. If $E_{2,p}^{*2}$ decreases below $m^2$, 
then $J(z_p)$ remains equal to unity.
If, instead, $E_{2,p}^{*2}$ increases above $m^2$, then $J(z_p)$ decreases,
vanishing for $E_{2,p}^{*2} \ge (1+\alpha) m^2$.
Thus $H_2$ vanishes when either $E_{2,p}^{*2}$ or
$E_{2,b_p}^{*2}$  reaches $(1+\alpha) m^2$, 
i.e. when one of these quantities lies
$\alpha m^2$ or more {\em above} threshold.

We can now see why $H_2 H_3=0$ if one of the two-particle momenta equals one of
the three-particle momenta. 
Consider first $\vec k=\vec p$, so that $E_{2,k}^{*2}=E_{2,p}^{*2}$.
If $E_{2,k}^{*2} \le (1+\alpha) m^2$, we have $H_2(\vec p) > 0$ and $H(\vec k)=0$,
while if
$E_{2,k}^{*2} \ge (1+\alpha) m^2$ we have $H_2(\vec p) =0$ and $H(\vec k)>0$.
$H_2 H_3 \propto H_2(\vec p) H(\vec k)$ vanishes in either case.
The symmetries of $H_2$ and $H_3$ ensure that this holds also if any other
pair of two- and three-particle momenta are equal.

Finally, we argue that $\alpha=3/2$ is a reasonable choice in order to
minimize exponentially suppressed finite-volume effects.
Such effects are generated by the difference between a sum and an integral over
the loop momenta with the integrand given by the cutoff functions multiplied by other smooth functions.
Generically, from the Poisson summation formula, we know that the suppression falls
as $\exp(-\Delta L)$,
where $\Delta$ characterizes the size of the region over which the summand/integrand varies.
Thus we want the cutoff functions to change from $0$ to $1$ over as large a region as possible.
Here this leads to two conflicting conditions. From $H_3$, we want $(3-\alpha) m^2$
[the range of $E_{2,k}^{*2}$ over which the variation in $H(\vec k)$ occurs] 
to be as large as possible, while from $H_2$ we want $\alpha m^2$ to be maximized.
The choice $\alpha=3/2$ sets these two distances from threshold equal.
We illustrate this optimization in Fig.~\ref{fig:Hfunctions}.

 We close this appendix by stressing that the forms we have given for $H_2$ and $H_3$ are
 not unique. We think that these are reasonable, somewhat optimized choices, but
 in a practical application it would be worthwhile investigating other options.

\section{Detailed derivation of Eq.~(\ref{eq:CHdecom})}

\label{app:details}

In this appendix we give the details of the derivation of the result Eq.~(\ref{eq:CHdecom})
for the finite-volume correlator, $\MLL$. 
This replaces the na\"ive analysis of Sec.~\ref{sec:naive}.
The outline of the new derivation has been sketched in Sec.~\ref{sec:correctder}.
We break the derivation into seven steps. 

\subsection{Diagramatic expansion}
\label{app:diagrammatic}

The first step is the same as in the na\"ive approach, namely to write out
a perturbative expansion in Feynman diagrams for $\MLL$.
This has been described in some detail in Sec.~\ref{sec:naive}, and here we
add a few further details.

We work with a general effective field theory (EFT) for our scalar field, 
with Lagrange density
\begin{multline}
\label{eq:lag}
\mathcal L(x) = \frac{1}{2} \phi(x) (\partial^2 + m^2 ) \phi(x) 
+ \sum_{n=3}^\infty \frac{\lambda_n}{n!} \phi(x)^n 
+ \sum_{n=3}^\infty \frac{g_n}{(n-1)!} [\partial^2 \phi(x)] \phi(x)^{n-1} + \cdots 
\\
+ \frac{1}{2} (\delta Z_\phi) \phi(x) \partial^2 \phi(x) 
+ \frac{1}{2} (\delta Z_m m^2) \phi(x)^2  
+ \frac{ \lambda_3}{3!}  (\delta Z_{\lambda_3}) \phi(x)^3
+ \cdots \,.
\end{multline}
The first ellipsis indicates additional interactions containing more derivatives, and the second indicates the counterterms corresponding to all included vertices. 
We imagine regulating Feynman diagrams using, for example, dimensional regularization,
and choose the counterterms so that,
in the limit that the UV regulator is removed,
all correlation functions are finite functions
of the mass, $m$, and the coupling constants, $\lambda_n, g_n,\ldots$.
We define $\delta Z_\phi$ and $\delta Z_m$ so that 
$m$ is the physical pole mass of the particle interpolated by
$\phi$ and the pole has unit residue
\begin{equation}
\frac{1}{i} \lim_{p^2 \to m^2} (p^2 - m^2) \int d^4 x \, e^{- i p x} 
\langle 0 \vert \phi(x) \phi (0) \vert 0 \rangle = 1 \,.
\end{equation}
We do not need to specify the precise definitions of the remaining counterterms---any 
scheme may be used, e.g.~the $\overline{\rm MS}$ scheme.

$\ML{ij}$ is formally defined as the sum of all connected finite-volume
Feynman diagrams with $j$ incoming and $i$ outgoing legs, amputated and
put on shell.
As described in the main text, we use a diagram-by-diagram renormalization scheme
in which the appropriate counterterm is combined with each divergent diagram.
This implies, in particular, that the combination of each self-energy Feynman diagram with
its counterterm satisfies the renormalization conditions of Eq.~(\ref{eq:renormscheme}).
How this generalizes when using TOPT will be discussed later.

\begin{figure}
\begin{center}
\includegraphics[width = \textwidth]{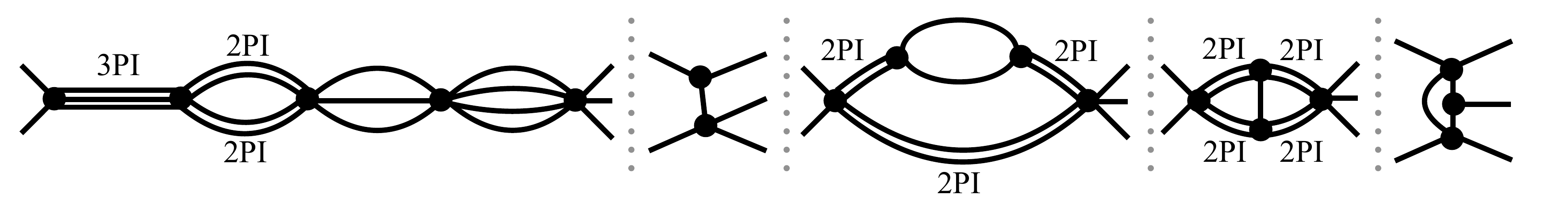}
\caption{Examples of diagrams contributing to $\ML{23}$, showing
the three different types of dressed propagators,
and the notation we use for them in subsequent diagrams.
The details of the vertices are
not specified---they are drawn from the interactions in Eq.~(\ref{eq:lag}) having the
appropriate number of fields.
External propagators are amputated, and unlabeled propagators are fully dressed.
\label{fig:MLLwithPIprop}}
\end{center}
\end{figure}

As noted in the main text, we sum self-energy insertions into dressed propagators of three
different types, shown in Fig.~\ref{fig:intro2PI}.
Here we describe in more detail where we use each type of dressed propagator.
The underlying rule is simple: All cuts in which two or three particles can go on shell must
be kept explicit. Here a cut must separate the diagram into two parts in the $s$ channel
and pass through at least one propagator that is not external. 
If a particular propagator appears only in cuts with three or more particles, it can be
fully dressed, i.e.~composed of 1PI self-energies.
This is because any cut through the self-energy loops would contain at least four particles.
Similarly, if the propagator can appear in cuts with two particles, 
then it must be composed of 2PI self-energies (and thus be 2PI dressed). 
Finally, if the propagator can appear in cuts with a single particle, then it must be
composed of 3PI self-energies (and thus be 3PI dressed).\footnote{Note that it is not possible for a given propagator to appear in both two- and one-particle $s$ channel cuts, so that our classification here is unambiguous.}
These three cases are illustrated in Fig.~\ref{fig:MLLwithPIprop}. 
Further examples appear in Figs.~\ref{fig:twoloopsep} and \ref{fig:twoloopnoselfenergy_v2} below.

An important observation is that all three types of dressed propagators have only
exponentially suppressed volume dependence and thus can be replaced by their
infinite-volume counterparts.
This is because the loops appearing (implicitly) in these propagators lead to four- or higher cuts
of the overall diagram, and thus do not have singularities in the kinematic range of
interest. Thus the summands are smooth and
the sum-integral difference is exponentially suppressed [see Eq.~(\ref{eq:expsuppr})].

\begin{figure}
\begin{center}
\includegraphics[width = \textwidth]{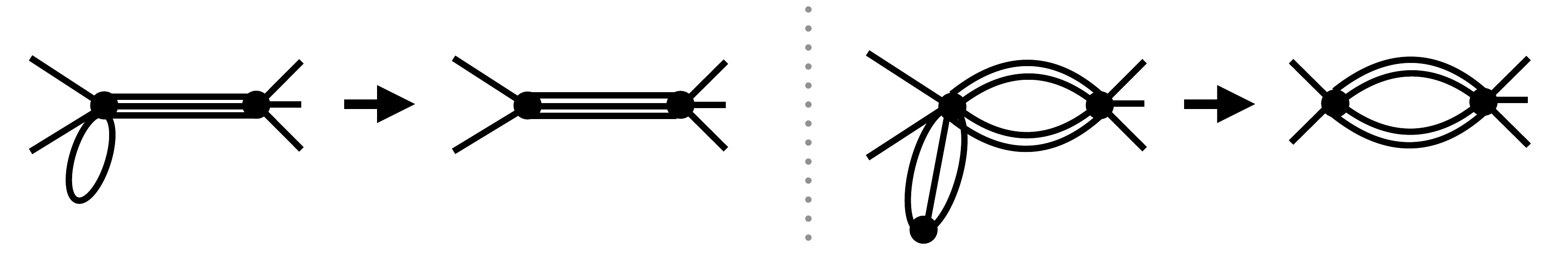}
\caption{Examples of tadpole diagrams and their absorption into the
adjoining vertices, as described in the text. Notation for propagators
is as in Fig.~\ref{fig:MLLwithPIprop}.
\label{fig:tadpoles}}
\end{center}
\end{figure}

A final comment concerns ``tadpole loops," i.e.~loops through
which no external four-momentum flows. Examples are shown in Fig.~\ref{fig:tadpoles}.
Such loops do not lead to on-shell intermediate states precisely because
no external momentum flows through the subdiagrams. They are thus uncuttable according to our rules. This is equivalent to the observation that the summands are nonsingular, so that the momentum sums can be replaced with integrals. In fact, from the point of view of determining finite-volume effects, we can simply absorb
these loops [along with their (implicitly) associated counterterms]
into the adjoining vertices. This reduction is illustrated in the figure.

\subsection{Partial reduction of two-particle self-energy bubbles}
\label{app:partialreduction}

\begin{figure}
\begin{center}
\subfigure[]{\includegraphics[width = \textwidth]{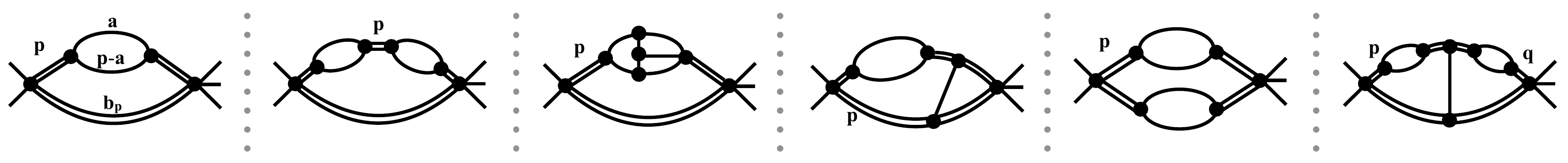}\label{fig:twoloopsep}}

\subfigure[]{\includegraphics[width = .8\textwidth]{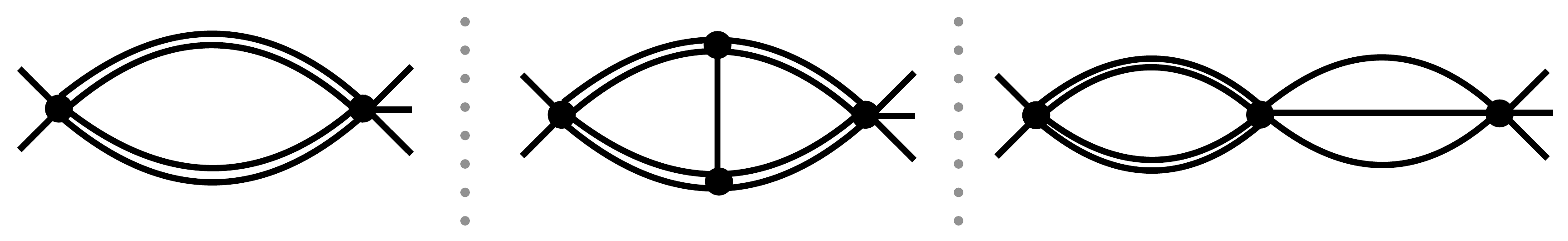}\label{fig:twoloopnoselfenergy_v2}}

\caption{ (a) Examples of diagrams contained in the class 2PI+.
2PI-dressed propagators are shown by double lines, and fully dressed propagators by single lines.
For each two-particle loop containing at least one self-energy insertion on a 2PI-dressed
propagator, we multiply the loop by $H_2 + [1\!-\!H_2]$, as described in the text.
(b) Examples of diagrams not included in the set 2PI+. }
\end{center}
\end{figure}

We now depart from the approach used in Sec.~\ref{sec:naive}. 
Rather than use TOPT immediately, we first sum up a class of Feynman diagrams.
These are the diagrams that contain at least one 2PI-dressed propagator
on which there is a self-energy insertion that is two-particle reducible.
Examples are shown in Fig.~\ref{fig:twoloopsep},
and we refer to them collectively as diagrams of class 2PI+.
The challenge here is that all such diagrams have three-particle cuts that lead
to finite-volume effects.
We stress that diagrams containing 2PI-dressed propagators without additional
self-energy insertions, such as those in Fig.~\ref{fig:twoloopnoselfenergy_v2},
are not included in the 2PI+ class of diagrams. However diagrams containing at least one two-particle loop with a self-energy insertion, as well as some number of two-particle loops without insertions, are included in 2PI+.
 
We next use the function $H_2(\vec p)$ (defined in Appendix~\ref{app:Hfunc}).
For each diagram in class 2PI+, we multiply each two-particle loop containing
at least one explicit two-particle self-energy insertion by 
\begin{equation}
\label{eq:H2iden}
1 = H_2(\vec p \hspace{1pt}) + [1\!-\!H_2(\vec p)] \,,
\end{equation}
and consider separately the $H_2$ and $1\!-\!H_2$ parts.
Here $\vec p\hsp1$ is the momentum of one of the propagators---we
can use either of the two momenta  in the loop as $H_2$ is symmetric.
It is important that only one such factor is inserted in a given loop, irrespective of how
many self-energy insertions are present. To illustrate these rules, we note that all of the diagrams of
Fig.~\ref{fig:twoloopsep} except the last are multiplied by 
$H_2(\vec p) + [1\!-\! H_2(\vec p)]$,
while the last diagram is multiplied by
$(H_2(\vec p) + [1\!-\!H_2(\vec p)])(H_2(\vec q) + [1\!-\!H_2(\vec q)])$.
We stress that, in the latter case, the momenta $\vec p$ and $\vec q$
are independent.

For the remainder of this subsection we consider two-particle loops that have been
multiplied by the $H_2$ part of Eq.~(\ref{eq:H2iden}).
The presence of $H_2$ leads to a key simplification: 
The sums inside all of the self-energies on the 2PI-dressed propagators
can be replaced with integrals.
This result holds because the function $H_2(\vec p)$ only has
support when the momenta in the three-particle state are far from going on-shell. 
To explain this, we consider the first diagram in Fig.~\ref{fig:twoloopsep}.
The three particles under consideration are those with momenta labeled 
$\vec a$, $\vec p-\vec a,$ and $\vec b_p=\vec P-\vec p$.
We recall that the function $H_3(\vec b_p, \vec a)$ has support
in a region around the on-shell manifold 
(those values of $\vec b_p$ and $\vec a$ for which all three particles can go on shell) 
of characteristic width $m$. 
But, by construction, $H_2(\vec p) H_3(\vec b_p, \vec a)= H_2(\vec b_p) H_3(\vec b_p, \vec a)=0$, 
implying that $H_2(\vec p)$ vanishes everywhere in this near-on-shell region. 
Thus $H_2(\vec p)$ forces the momentum in 
the self-energy loops to be well away from their on-shell values,
and thus well away from the $\cC_3$ pole associated with a three-particle
intermediate state.\footnote{%
We stress that this is not a direct constraint on the momentum in the
self-energy loop, i.e.~on $\vec a$ in our example. This momentum is freely summed/integrated.
The point is that, in the presence of $H_2(\vec p)$, the summand does not come close to
the three-particle singularity.}
The difference between momentum sums and integrals for such loops is 
therefore exponentially suppressed.

The self-energy insertions on the 2PI-dressed propagators
can also contain loops with more than two particles. 
An example is the third diagram in Fig.~\ref{fig:twoloopsep}.
Since particles in such loops cannot go on shell
(requiring an intermediate state containing four or more particles for the complete diagram), 
the momentum sums in these loops
can be also be replaced with integrals.
Thus we find the result claimed above: The entire self-energy can be evaluated in infinite volume.

\begin{figure}
\begin{center}
\includegraphics[width = \textwidth]{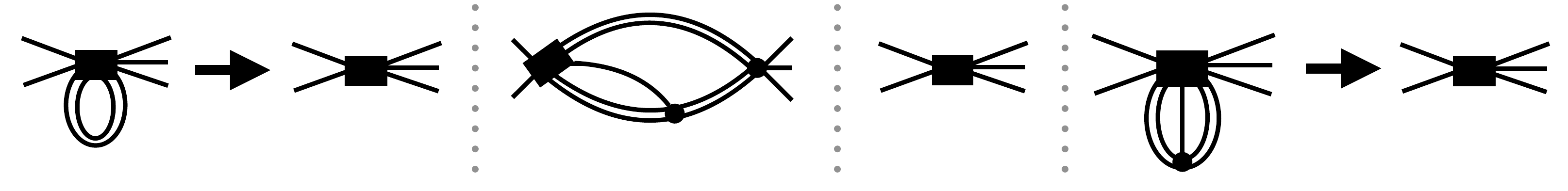}

\caption{Simplification of the class 2PI+ diagrams shown in Fig.~\ref{fig:twoloopsep} 
when the loops containing
two 2PI-dressed propagators with self-energy insertions are multiplied by $H_2$. 
Propagators containing self-energy insertions are shrunk to new vertices, 
shown by the filled rectangles.
The detailed form of the vertex represented by the rectangle depends on the diagram.
The first three diagrams in Fig.~\ref{fig:twoloopsep} are all simplified to the same form,
 and thus only one
diagram is shown. The remaining three are simplified in different ways.
In a second step, indicated by the arrows, tadpole diagrams are absorbed into the vertices.
 \label{fig:shrinking}}
\end{center}
\end{figure}

The resulting integrated self-energies are just particular examples of
the quantities $D^R_i(p^2)$ discussed in the main text. 
In particular, since the diagrams are accompanied by counterterms that enforce the conditions of
Eq.~(\ref{eq:renormscheme}), we know that they vanish quadratically as one goes on shell,
\begin{equation}
D^R_i(p^2) \underset{p^2 \to m^2}{\longrightarrow} c (p^2 - m^2)^2 \,.
\label{eq:quadratic}
\end{equation}
Thus each self-energy cancels the poles from the 2PI-dressed propagators on either side.
If there is a chain of self-energies then the poles are ``overcanceled''
leading to factors of $(p^2-m^2)$ in the numerator.
As a result, each 2PI-dressed propagator with self-energy insertions, in a cut that is accompanied by a factor of $H_2$,
gives only short-distance contributions. 
We can implement this diagrammatically by shrinking the propagator
to a new effective vertex, as shown in Fig.~\ref{fig:shrinking}.
This vertex is complicated---possibly involving nonanalytic functions of momenta and
containing $H_2(\vec p)$---but it satisfies the key property that it is
``uncuttable.'' In other words, it is a smooth function of real three-momenta and thus cannot lead to important finite-volume effects, which is also true for vertices in general.

As shown in Fig.~\ref{fig:shrinking}, shrinking propagators often lead to
tadpole loops. These loops can then be absorbed into vertices,
as discussed in the previous subsection.

The conclusion of this analysis is that we can effectively ignore self-energy insertions
on 2PI propagators when the factor $H_2$ is present. They give rise to additional vertices, 
which are special in that they occur only in certain topologies of diagrams and contain
factors of $H_2$. But since we are at no stage actually calculating the Feynman diagrams,
the presence of new vertices does not lead to any change in the diagrams 
to be considered.\footnote{%
The only exception to the statement that no new diagrams need to be considered
is that, after applying the shrinking procedure,
there are diagrams in which some of the propagators are 2PI dressed, whereas, if one applied
the rules discussed in Appendix~\ref{app:diagrammatic}, they would be fully dressed. 
An example is shown by the second diagram in Fig.~\protect\ref{fig:shrinking},
where the bottom propagator in the leftmost loop would be fully dressed according to
the general rules, but is in fact 2PI dressed.
This exception has, however, has no impact on determining finite-volume effects,
as both types of propagator have the same pole and residue.}

In summary, the analysis of this subsection allows us to avoid one of the problems with
the na\"ive result (\ref{eq:CHdecomtmp}), namely the fact that the quantity
$\widetilde A$ does not contain all time orderings needed to build up the full self-energy,
and so the result behaves as $(p^2-m^2)$ rather than the quadratic dependence of
Eq.~(\ref{eq:quadratic}). 
By working at this stage with Feynman diagrams we are, in effect, summing
all the time orderings, rather than the restricted set contained in $\widetilde A$.

\subsection{Shrinking 3PI-dressed propagators}
\label{app:3PI}

The second problem mentioned at the end of Sec.~\ref{sec:naive} in the main text concerned
contributions to $\MLL$ that involve 3PI-dressed propagators.
In this section we describe the problem in more detail and then explain how it can be avoided
by shrinking all 3PI-dressed propagators down to local vertices.

\begin{figure}
\begin{center}
\includegraphics[width = \textwidth]{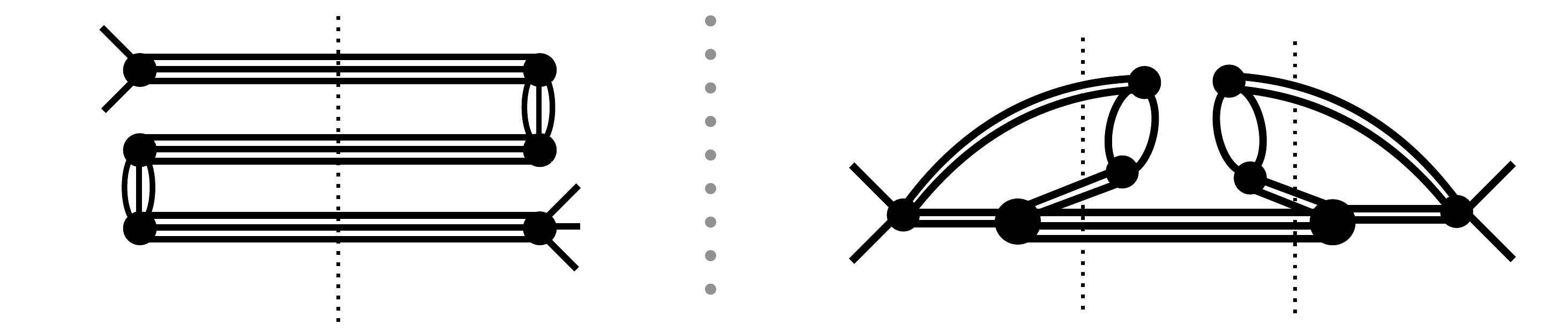}

\caption{Examples of spurious three-particle intermediate states 
arising when applying TOPT to diagrams involving 3PI-dressed propagators.
\label{fig:ugly3PI}}
\end{center}
\end{figure}

The problem arises once we switch from working with Feynman diagrams to using TOPT
(a change that is discussed more extensively in Appendix~\ref{app:TOPT} below).
We then discover that certain time orderings of diagrams containing 3PI-dressed propagators
have spurious three-particle intermediate states. 
Two examples are shown in Fig.~\ref{fig:ugly3PI}.
These are contributions to TOPT that have poles of the form $(E - \omega_a - \omega_k - \omega_{Pka})^{-1}$
and thus, in general, contribute to the kernels $A$ introduced in Sec.~\ref{sec:naive}.
These poles are spurious, however, because they cancel in the full Feynman diagrams.
This is clear in the examples shown because one can factorize the corresponding Feynman diagrams into a product of loops and propagators and the singularities arise only from these
individual factors, and not from overlapping cuts such as those shown.

In  principle one could continue with the TOPT analysis, keeping track of these spurious
contributions until they cancel in the end. This is difficult, however, as they contain
disconnected contributions involving Kronecker deltas. 
A better solution is to avoid these contributions from the beginning. This is possible
due to the fact that there are no on-shell intermediate states that involve the
3PI-dressed propagators in our kinematic range.
This is apparent from the initial Feynman diagram in which each 3PI-dressed propagator
appears factorized from the remainder of the diagram, and has singularities only
at $E^*=m$ and $E^*\ge 4m$.
Thus the 3PI-dressed propagators are uncuttable. 
They are also functions only of the fixed external four-momentum, $(E,\vec P)$,
and are thus themselves fixed.
It follows that, from the point of view of determining finite-volume dependence, we
can shrink them into the adjoining vertices.
With this done, none of the spurious cuts remain.
In the following we assume that such a procedure has been employed.

\subsection{Classification of remaining loops}
\label{app:loopclassification}

\begin{figure}
\begin{center}
\includegraphics[width = \textwidth]{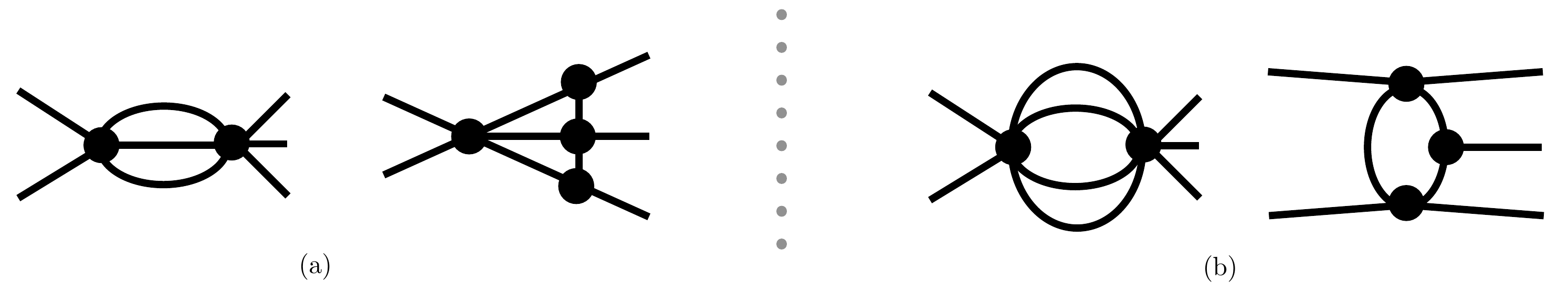}
\caption{Examples of  the (a) third and (b) fourth classes of loops 
that arise after shrinking propagators and tadpole loops. 
See the numbered list in the text for details. \label{fig:classes}}
\end{center}
\end{figure}

At this stage it is useful to take stock of the types of Feynman diagrams that remain 
after propagators and tadpole diagrams are shrunk as described above.
The remaining diagrams contain only fully dressed and 2PI-dressed propagators,
and are built from overlapping loops that fall into the four classes:
\begin{enumerate}
\item 
Loops containing a pair of 2PI-dressed propagators, on which there
are no self-energy insertions. 
Examples are shown in Fig.~\ref{fig:twoloopnoselfenergy_v2}. 
These loops are, at this stage, not multiplied by factors containing $H_2$.
\item 
Loops containing a pair of 2PI-dressed propagators in which at least one of these propagators
has a self-energy insertion. 
Such loops are contained in diagrams of class 2PI+ [see Fig.~\ref{fig:twoloopsep}]. 
All such loops are multiplied by $[1\!-\!H_2(\vec p)]$. 
The presence of this factor implies that these loops 
cannot give rise to two on-shell particles, 
but do give rise to three particles that all go on shell.
\item 
Loops that include sets of three particles that carry the total energy and momentum 
$(E, \vec P)$ (and can thus simultaneously go on shell)
 but are not included in the previous class.
Examples are shown in Fig.~\ref{fig:classes}(a).
\item 
Loops that give rise to no on-shell intermediate states, 
either because four or more particles carry the total energy and momentum or 
because the loops are in a $t$-channel-like structure and 
thus do not carry the total energy-momentum that flows through the diagram.
Examples are shown in Fig.~\ref{fig:classes}(b).
\end{enumerate}
The overall result is that we have removed all appearances of self-energy diagrams except where
they are needed because a physical on-shell cut can run through them, i.e.~in loops of class (2).

\begin{figure}
\begin{center}
\includegraphics[scale=0.5]{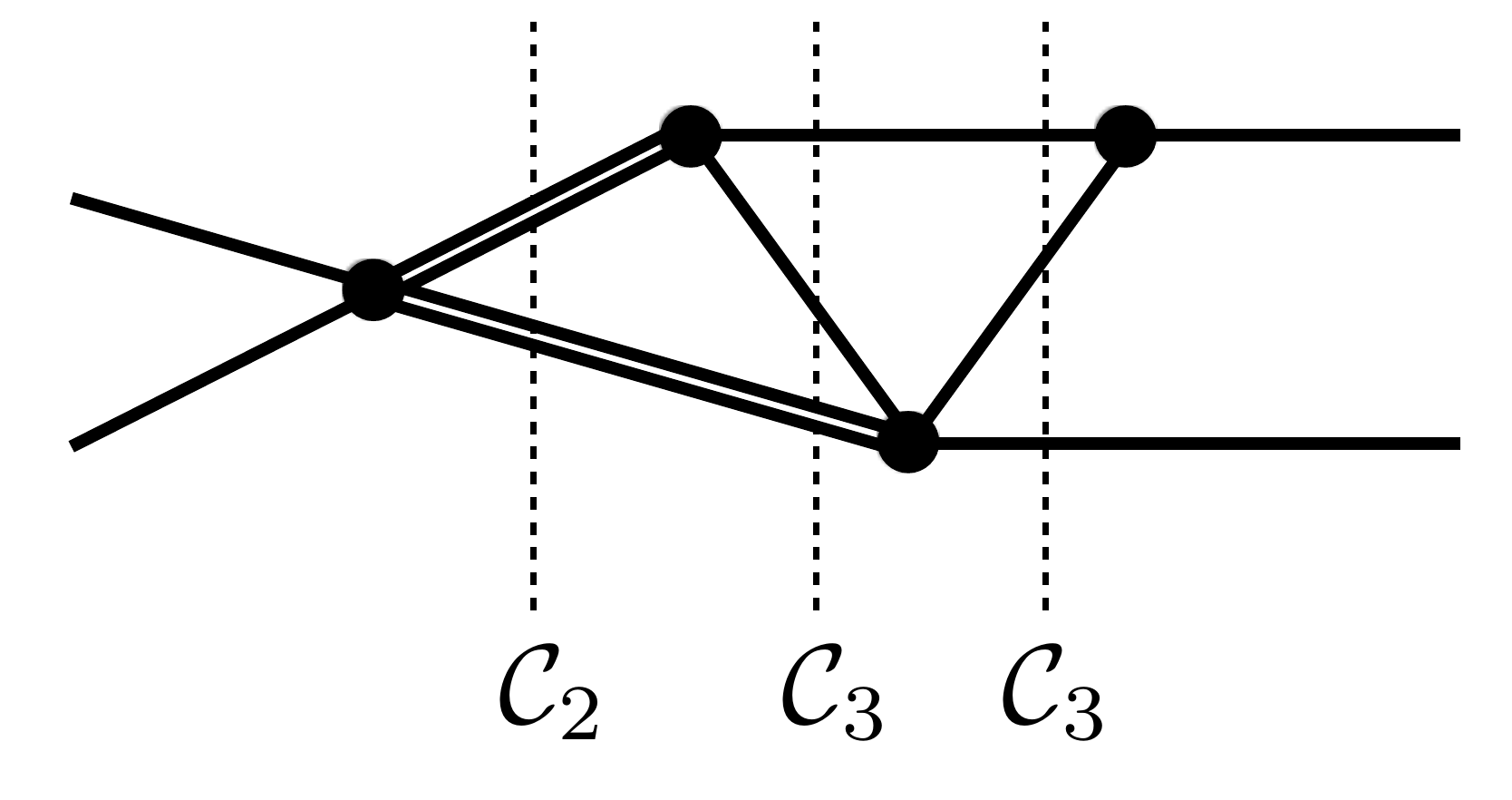} 
\caption{Example of a diagram with overlapping loops. The possible two- and three-particle cuts
are shown. The central cut is not associated uniquely with a single loop.
\label{fig:fish}}
\end{center}
\end{figure}

Finally, we observe that, because loops overlap, there is not a
one-to-one correspondence between loops and cuts. 
This is illustrated in Fig.~\ref{fig:fish}. As a result, we cannot study individual loops, or even finite sets of loops, and determine the important finite-volume effects. 
Indeed, in general, the singularity structure of a given diagram is quite complicated.
Since finite-volume dependence arises from two- and three-particle cuts,
what we need is a tool for breaking diagrams into multiple terms that
individually contain a specific sequence of cuts. 
This can be done straightforwardly using TOPT, to which we now turn.

\subsection{Applying time-ordered perturbation theory}
\label{app:TOPT}

At this stage we break up the Feynman diagrams into their component
time orderings. This can be achieved by evaluating all
energy integrals, and then partial fractioning the resulting products of poles.
A more direct method 
is to evaluate the Feynman diagrams using a mixed time-momentum representation
for the propagators, and then do the time integrals.\footnote{%
For a lucid explanation of this method, see Ref.~\cite{Stermantext}.}
The result---the TOPT expression---is a sum of terms each of which depend only on spatial momenta.
Since we work in finite volume, these momenta are summed over the finite-volume discrete set.

Our application of TOPT is slightly complicated by our use of dressed propagators.
We first describe the approach ignoring this complication, i.e.~using bare propagators,
and then return to the complications introduced by dressing. Consider a Feynman diagram with some number of on-shell, amputated external legs
and with total energy-momentum $(E, \vec P)$ flowing
from the initial to the final state. 
One then enumerates all ordered sequences of vertices in the
diagram between the initial and final states.\footnote{%
The requirement that all vertices must lie {\em between} the initial and final states
is a consequence of having on-shell, amputated external propagators.
One can think of this as occurring because the initial particles are
created at $t=-\infty$ and the final particles destroyed at $t=\infty$.}
Each individual ordering represents a mathematical expression determined as follows.
(1) Route a vertical line (i.e. a ``cut,'' $c$) 
between each pair of consecutive vertices in the ordering. 
(2) Define the factor $\sum_{i \in \{c\}} \omega_i$, given
by summing all of the on-shell energies of the propagators intersecting the cut.
(3) Calculate the product
\begin{equation}
\mathcal P_o = \prod_{c \in \{o\}}\left( \frac{1}{E - \sum_{i \in \{c\}}  \omega_i}\right) \,,
\end{equation}
where $o$ denotes a particular ordering, $\{o\}$ denotes the set of cuts
within the ordering, and $c$ denotes a particular cut.
(4) Multiply $\mathcal P_o$ by a factor of $1/(2\omega_j)$ for
each internal propagator, and by the expressions arising from each vertex,
as well as possible $1\!-\!H_2$ and symmetry factors.
This leads to the expression for the $n$-cut factor $\cC_n$
 given in Eq.~(\ref{eq:ncutfactor}).
Summing over all orderings then gives the value of the Feynman diagram.
Examples of time orderings are shown in Fig.~\ref{fig:2to3TOPTex}.

As noted in the main text,
when we apply TOPT in the kinematic range given in Eq.~(\ref{eq:kinwindow}),
the only singularities that can appear are the poles due
to two- and three-particle intermediate states, given in Eq.~(\ref{eq:2and3poles}).
Finite-volume effects arise only from momentum sums
that run over one or both of these poles.
All other sums can be converted to integrals.

The above discussion assumes a propagator of the form $i/(p^2-m^2+i\epsilon)$, and thus does not directly hold for the dressed propagators. 
Given the renormalization conditions of Eq.~(\ref{eq:renormscheme}),
however, both types of dressed propagator do have exactly this pole
structure, including the residue, for $p^0 \to \omega_p$.
The effect of dressing appears only in the constant and in terms of $\mathcal O(p^2-m^2)$, but such terms can be absorbed into the vertices as long as they remain smooth within
our kinematic range. Since the vertices are general, this leads to no additional complications.
Then it is legitimate to use TOPT ignoring the fact that
the propagators are dressed. This means that the distinction between fully and 2PI-dressed
propagators is no longer relevant. 

The remaining issue is thus whether there are additional singularities in the dressed
propagators within our kinematic range ($E^* < 4m$). 
The fully dressed propagator has a two-particle cut,
while the 2PI-dressed propagator has a three-particle cut. However, by construction, these 
both correspond to cuts with four or more particles in the full diagram. Thus these singularities
do not appear within our kinematic range.

A final technical complication concerns counterterms in TOPT. When we break up
a UV divergent loop into its various time orderings we also need to break up the
counterterms accordingly. An example is given by the self-energy loop in the center
of the diagrams of Fig.~\ref{fig:2to3TOPTex}: Its two vertices have different time orderings
in the two diagrams, and these are separately UV divergent. In fact, in general, since we
have broken Lorentz symmetry in TOPT, the individual counterterms needed for the
different time orderings will not be Lorentz invariant. Lorentz invariance is regained
only at the end when all time orderings are recombined.
In practice, one can always define the counterterms
operationally for each time ordering by using dimensional regularization and
removing the pole with a prescription such as $\overline{\rm MS}$
(up to finite corrections needed to satisfy renormalization conditions discussed previously).

In summary at this stage we have reduced every Feynman diagram to a
sum of terms each given by products of smooth functions and two-
and three-particle poles. Thus $\MLL$ can be written in the form given in Eq.~(\ref{eq:decom}) of
the main text, except that  the kernels between two- and three-cuts are now different.\footnote{%
Strictly speaking, we need to show that kernels that appear are independent of their
position in the chain of terms in Eq.~(\ref{eq:decom}). We return to this issue below.}
These differences  are  due to the presence of
factors of $[1\!-\!H_2]$ in diagrams with self-energy insertions, 
to the absence of 3PI-dressed propagators, 
and to the alterations in vertices arising from the shrinking procedure and from the tadpole loops and other smooth terms that have been absorbed. 

In what follows we denote the coordinates
that appear in the two- and three-particle poles as ``explicit''
whereas all coordinates that are integrated at this stage are buried
inside various smooth functions and are thus referred to as
``implicit''. Note that all $H_2(\vec p\,)$ functions at this point
are implicit with the exception of the $[1\!-\!H_2(\vec p)]$ factors
accompanying the two- and three-particle poles in class (2) loops.

 \subsection{Introduction of regulator functions on cuts}
\label{app:removedisc}

The next step is, as in Sec.~\ref{sec:naive}, to multiply each two- and three-cut
by unity written, respectively, as Eq.~(\ref{eq:H2iden}) and
\begin{align}
1 & = H_3(\vec k, \vec a) + [1\!-\!H_3(\vec k, \vec a)] \,.
\label{eq:H3iden}
\end{align}
The momenta here are the explicit summed coordinates appearing in the cut factors.
The only difference compared to the main text is that here 
we do not make this substitution in the two-cuts in class 2 loops,
since these loops already come with a factor of $[1 \!-\! H_2(\vec p)]$.

Having made these substitutions we then consider the parts containing $H_i$
and $1\!-\!H_i$ separately, so that the cuts that arise are $\cC_2^H$, $\cC_2^\infty$,
$\cC_3^H$, $\cC_3^\infty$ and higher-order cuts.
[See Eq.~(\ref{eq:CHdef}) for the definitions of these cuts.]
At this stage singularities arise only from factors of $\cC_2^H$ or $\cC_3^H$.
All other possibilities do not have poles within our kinematic regime. 
This implies that any loop momentum that does not appear
in either a $\cC_2^H$ or $\cC_3^H$ can be integrated rather than summed.

We can now make use of the important result that,
whenever a two-cut and a three-cut share a common propagator, 
then $H_2 H_3 = 0$ (as described in Appendix~\ref{app:Hfunc}).
In Sec.~\ref{sec:naive}, we used this result to drop disconnected parts from
$\widetilde A_{23}$ and $\widetilde A_{32}$.
Here we apply it at a slightly earlier stage.
The aim is to come up with a version of Eq.~(\ref{eq:CHdecomtmp}) that does not
suffer from the problems described in the main text.

\begin{figure}
\begin{center}

\includegraphics[width = \textwidth]{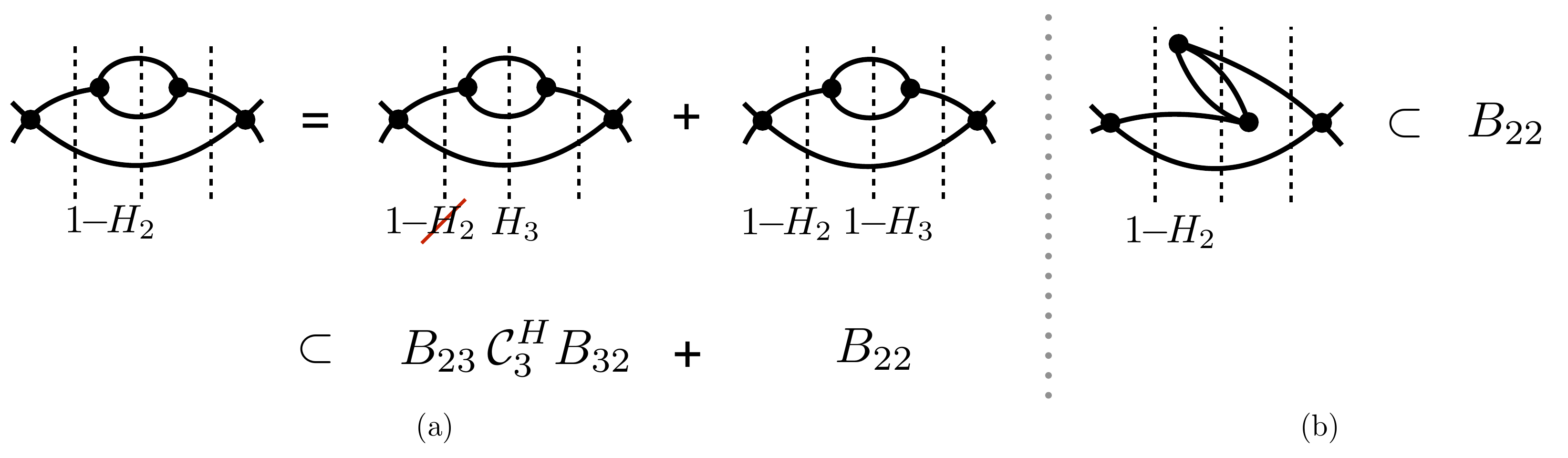}
\caption{(a) The reduction procedure for one time-ordering
of a diagram with a class 2 loop. Vertical dashed lines indicate
$n$-cuts. Fully dressed and 2PI-dressed propagators are both shown by single lines,
because the singularities arise only from the pole parts of these propagators, which
are identical.
The factor of $1\!-\!H_2$ is associated with the entire loop,
and not with a particular cut. (b) The reduction procedure for a different time ordering. See text for detailed discussion.\label{fig:loopreduce_all}} 
\end{center}
\end{figure}

\begin{figure}
\begin{center}
\includegraphics[width = \textwidth]{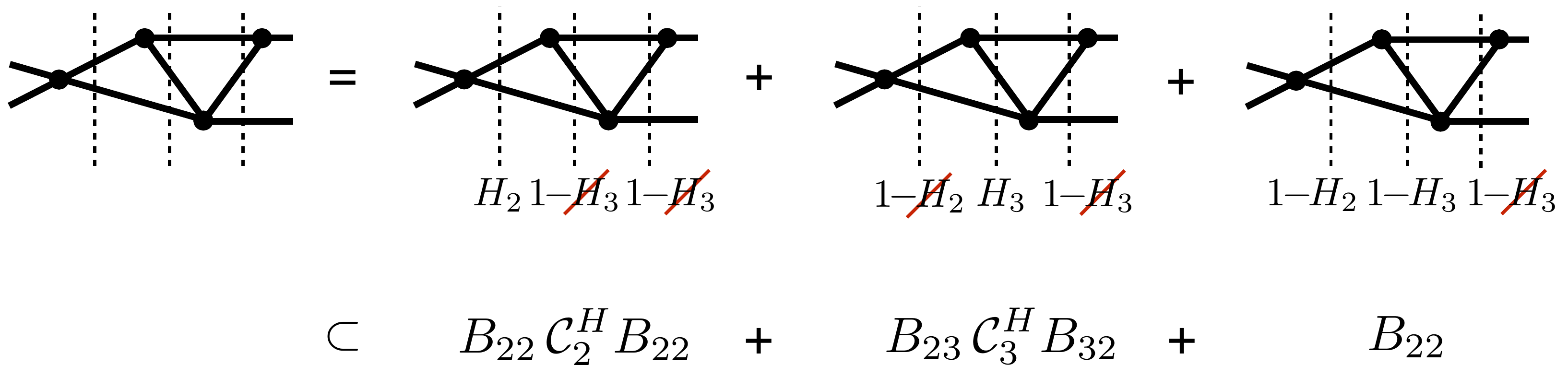}

\caption{The reduction procedure for one time ordering of the diagram 
of Fig.~\ref{fig:fish}. Notation as in Fig.~\ref{fig:loopreduce_all}.
See text for detailed discussion. 
\label{fig:fishreduce}}
\end{center}
\end{figure}

To see how this works we consider three examples, given in 
Figs.~\ref{fig:loopreduce_all} and \ref{fig:fishreduce}. 
These show how a particular time ordering is reduced 
to a product of smooth kernels and regulated cut factors, $\cC_2^H$ and $\cC_3^H$.
Figure~\ref{fig:loopreduce_all}(a) shows a diagram containing a class 2 loop.
We recall that, although two-cuts appear in the TOPT expression, the factor of $1\!-\!H_2$ cancels the
poles.\footnote{%
A single factor of $1\!-\!H_2$ can cancel any number of poles 
since it has an essential zero at the pole.}
Now we insert the identity (\ref{eq:H3iden}) on the three-cut, 
leading to the two diagrams on the right-hand side of the equality.
For that containing $H_3$, we use $H_2(\vec p) H_3(\vec p,\vec a)=0$ 
to drop the factor of $H_2$, as shown.\footnote{%
The fact that the $H_2$ can be dropped means that we do not have to worry
about distributing the $1\!-\!H_2$ factor between the kernels $\tM_{23}$ and $\tM_{32}$ 
on either side of the three-cut. This is important since we want to
treat all such kernels in a consistent manner.}
In other words, the presence of the $H_3$ in $\cC_3^H$ is sufficient to
ensure that there are no on-shell two-cuts. Thus we can decompose this diagram in the
form shown on the second line, with two smooth kernels and a single pole factor.

The diagram containing $1\!-\!H_3$ is simpler to analyze. Since both two- and three-particle poles are
canceled, the two loop sums have smooth summands, and can be converted into integrals.
Thus this contribution has no pole, and gives only a smooth kernel.
It is important to note that the $1\!-\!H_2$ factor, which remains for this time ordering,
is not associated with the left-hand cut, but rather with the entire outer loop.

We now turn to Fig.~\ref{fig:loopreduce_all}(b), which is a different time ordering
of the diagram in Fig.~\ref{fig:loopreduce_all}. In this case there are no cuts that require
the use of the identities in Eqs.~(\ref{eq:H2iden}) and (\ref{eq:H3iden}). All cuts are
nonsingular in our kinematic region (the two-cuts due to the factor of $1\!-\!H_2$, and the
5-cut due to the kinematic constraints), and so both loop sums can be replaced by
integrals, leading to a contribution to the kernel $\tM_{22}$.

Finally, we consider Fig.~\ref{fig:fishreduce}, which is one time ordering of the diagram with
overlapping class 1 and class 3 loops shown in Fig.~\ref{fig:fish}.
It thus comes with no explicit factors of $H_i$, and we must insert the
identities of Eqs.~(\ref{eq:H2iden}) and (\ref{eq:H3iden}) on all three cuts.
This leads to $2^3$ terms, but only the three shown survive.

To see this note that, because the rightmost two-particle state is on shell, it follows that the three particles present in the adjacent three-cut cannot all simultaneously go on shell, as they share an unscattered particle.
This already tells us that only $2^2$ terms will be nonzero. 
In other words, the right-hand cut cannot have a factor of $H_3$,
so only the $1\!-\!H_3$ factor survives for this cut, and furthermore we can set $1\!-\!H_3\to 1$.

A further reduction occurs if we choose $H_2$ for the left-hand cut,
for then the middle cut cannot have a factor of $H_3$.
If the left-hand cut has a factor of $1\!-\!H_2$, however, then the
middle cut can contain either $H_3$ or $1\!-\!H_3$, as shown.
In the former case, the $H_2$ in the left-hand cut can be dropped.
The net result is that there are only three diagrams.
These give the kernel and cut-factors shown in the figure,
where all momentum sums within the kernels can be replaced by integrals.

We can make several important general observations from these examples.
First, the off-diagonal kernels $\tM_{23}$ and $\tM_{32}$ 
produced by this reduction {\em do not have disconnected contributions}.
This is simply because such contributions necessarily come with a factor
of $\cC_2^H \cC_3^H\propto H_2 H_3$ which vanishes when one propagator is unscattered.
Thus, unlike in the na\"ive approach of Sec.~\ref{sec:naive}, where 
$\tA_{23}$ and $\tA_{32}$ had disconnected contributions that could be dropped,
here the corresponding kernels simply do not have such contributions.

The second observation is that 
{\em there are no disconnected contributions to $\tM_{22}$.} 
Such contributions arise in the na\"ive method of Sec.~\ref{sec:naive} 
from diagrams involving self-energy insertions 
such as Fig.~\ref{fig:loopreduce_all}.
For example, in Fig.~\ref{fig:loopreduce_all}(b), the loop lying between the two-cuts gives
a disconnected contribution to $A_{22}$. Here, however, all such contributions
are avoided because of the presence of the factor of $1\!-\!H_2$
(and the renormalization scheme chosen), which cancels the poles in the two-cuts. 

The third observation is that {\em the kernel $\tM_{33}$, unlike the other components of $\tM$,
can have disconnected parts.} An example where this arises is shown in 
Fig.~\ref{fig:loopreduce2}.
A disconnected contribution occurs in the first diagram on the right-hand side of the
equality, arising from a $2\leftrightarrow 2$ scattering.
The explicit form of the disconnected part is shown in Eq.~(\ref{eq:tM33disc}) below.
Note that completely disconnected parts cannot occur because there must be a vertex
between the two cuts, and self-energy insertions are not allowed on fully dressed propagators.

The final observation is more technical, but nevertheless important for the following
development.
This is that {\em all factors of $1 \!-\! H_2$ remaining after reduction
lie within loops that are integrated.}\footnote{%
The same is not true of factors of $1\!-\! H_3$, which can appear in
tree-level contributions to $\tM_{33}$.}
The observation can be demonstrated simply by noting that the
loop momentum running through the $1\!-\!H_2$ cannot be shared with either a
$\cC_2^H$ or a $\cC_3^H$ cut. 
The former possibility is ruled out by
the construction of Appendix~\ref{app:partialreduction}, in which
only a single regulator function was applied to each two-particle loop.
The latter is ruled out because, if a momentum is shared, then one can use 
the $H_2 H_3=0$ identity to replace $1\!-\! H_2$ with $1$.
The importance of this observation can be seen most easily from
the middle diagrams on the right-hand side of Fig.~\ref{fig:loopreduce2}.
Here the $1\!-\!H_2$ is not in an integrated loop, so there would be an ambiguity as to
which two-cut it is attached. In fact, since $1\!-\!H_2$ can be replaced by $1$, this
problem is absent. In the right-hand diagram, where the $1\!-\!H_2$ remains, it
can be unambiguously attached to the integrated loop as a whole.
This means that there is a well-defined set of rules for assigning factors of $1\!-\!H_i$ to
the diagrams contributing to the kernels.

\subsection{Final summation}
\label{app:stepseven}

After following the steps described above we have decomposed $\MLL$ into the
following sum of terms
\begin{align}
\MLL &= \sum_{n=1}^\infty \mathcal M_L^{(n)},
\label{eq:MLsum}
\\
\label{eq:MLgenterm}
\mathcal M_{L}^{(n)} &= \sum_{i\in {\rm diagrams}}
\tM^{(n,i;1)} \cC^H \tM^{(n,i;2)} \cC^H \cdots \cC^H \tM^{(n,i;n-1)}\cC^H \tM^{(n,i;n)}  \,.
\end{align}
Here we have reverted to the $2\times2$ matrix notation.
The sum over $i$ runs over all contributions (coming from the different time orderings
of all Feynman diagrams with all possible appearances of regulator factors after the
reduction described above) containing $n\!-\!1$ factors of $\cC^H$.
From the previous section we know that the kernels 
$\tM^{(n,i;j)}_{22}$, $\tM^{(n,i;j)}_{23}$ and $\tM^{(n,i;j)}_{32}$
are connected, smooth, infinite-volume ($L$-independent) functions. The $\tM^{(n,i;j)}_{33}$, however, consist of a connected, smooth, infinite-volume part plus
a term involving a Kronecker delta and factor of $L^3$ multiplying a two-to-two smooth,
infinite-volume kernel [as in Eq.~(\ref{eq:tM33disc})].

The construction of the $\tM^{(n,i;j)}$ follows the rather involved steps described in the previous
sections of this Appendix. What we show in this final section is that the sum over $i$ in
Eq.~(\ref{eq:MLgenterm}) leads to the simple form\footnote{%
As discussed in the main text, this is a slight oversimplification, in that the
matrix indices at the end of the chain are slightly different from those in the middle.
As reiterated below, however, all the kernels $\tM$ can be obtained from a single
master function, analogous to that in Eq.~(\ref{eq:masterA}).}
\begin{equation}
\mathcal M_{L}^{(n)} + I^{(n)} =
\underbrace{\tM \;\cC^H \tM \;\cC^H \cdots \cC^H \tM\;\cC^H \tM}_{n \ {\rm kernels}}
\label{eq:MLnres}\,.
\end{equation}
Here $I^{(n)}$ contains only disconnected contributions.
The key claim in this result is that the same kernels appear
in all positions and for all values of $n$.
Summing over $n$ then leads to the claimed result, Eq.~(\ref{eq:CHdecom}),
with the full subtraction given by $I = \sum_{n=1}^\infty I^{(n)}$.

\begin{figure}
\begin{center}
\includegraphics[width = \textwidth]{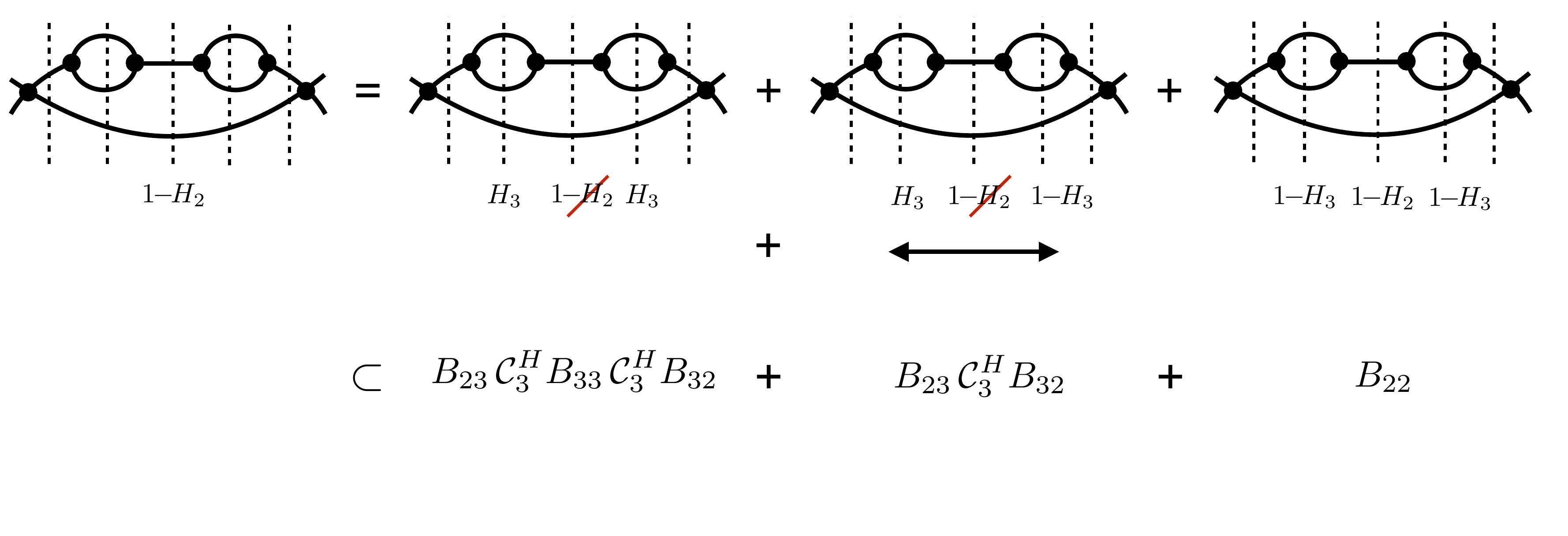}

\vspace{-40pt}

\caption{The reduction procedure for one time ordering
of a diagram with a class 2 loop containing two self-energy insertions. 
Notation as in Fig.~\ref{fig:loopreduce_all}.
There are four diagrams on the right-hand side of the equality, with the middle two related
by a horizontal reflection. Both contribute to $\tM_{23} \;\cC_3^H \tM_{32}$.
\label{fig:loopreduce2}}
\end{center}
\end{figure}

Before demonstrating Eq.~(\ref{eq:MLnres}) we recall the need for the subtraction term $I^{(n)}$.
We know from diagrams such as Fig.~\ref{fig:loopreduce2} that the kernel $B$ must
contain disconnected parts in the $33$ component.
If there were no subtraction in Eq.~(\ref{eq:MLnres}), then $\mathcal M_L^{(1)}$ would equal $B$,
and thus contain a disconnected part, which is inconsistent with its definition.
In other words, in order for the same kernel $B$ to appear in $\mathcal M_L^{(n)}$ for all $n$,
a subtraction is required.

Before demonstrating Eq.~(\ref{eq:MLnres}) we recall the need for the subtraction.
This arises from a mismatch between the kernels appearing in $\mathcal M_L^{(1)}$
and those in the higher-order terms. The former must be connected (since $\MLL$ is) 
while those appearing in higher order terms must contain disconnected parts in the
$33$ component (in order to accommodate diagrams such as that in Fig.~\ref{fig:loopreduce2}).
In order to have a uniform definition of the kernel a subtraction is required. 

To proceed we next give a precise definition of the kernel $B$. This is done by following
exactly the same steps as described in the preceding subsections, but instead of
starting with the fully connected $\MLL$, we allow also diagrams with 
$2\to 2$ scattering and a single disconnected propagator in $\ML{33}$. 
Fully disconnected diagrams are not included,
nor are those involving a $1\leftrightarrow 2$ subprocess in the $32$ or $23$ components.
We call this extended quantity $\ML{\rm ext}$.
It can be expanded in powers of the number of pole factors $\cC^H$,
just as in Eq.~(\ref{eq:MLsum}).
By construction, we then have that
\begin{equation}
\ML{\rm ext}^{(n)} = \mathcal M_L^{(n)} + I^{(n)} 
\,,
\label{eq:MLext}
\end{equation}
where $I^{(n)}$ is simply the disconnected part
of the left-hand side (which can be unambiguously identified).
$B$ is simply defined as the part of $\ML{\rm ext}$ without factors of $\cC^H$:
\begin{equation}
B \equiv \ML{\rm ext}^{(1)}\,.
\label{eq:firstorder}
\end{equation}

Using the new extended $\MLL$, we can reformulate the result 
Eq.~(\ref{eq:MLnres}) in the simpler form
\begin{equation}
\ML{\rm ext}^{(n)} = B \left(\cC^H B\right)^{n-1}
\,.
\label{eq:nthorder}
\end{equation}
We now recall that, when we say that all factors of $B$ are equal in (\ref{eq:nthorder}), 
we mean aside from the different momenta at which they are sampled.
In particular, we define a master kernel
\begin{equation}
\tM(\vec p \, ', \vec k', \vec a'; \vec p, \vec k, \vec a) = 
 \begin{pmatrix}
  \tM_{22}(\vec p \, '; \vec p) & \tM_{23}(\vec p \, '; \vec k, \vec a) \\
 \tM_{32}(\vec k', \vec a'; \vec p) & \tM_{33}(\vec k', \vec a'; \vec k, \vec a) 
  \end{pmatrix} \,,
 \end{equation}
 by extending the on-shell definition of $\tM $ to 
 general momenta $\vec p \, ', \vec k', \vec a'; \vec p, \vec k, \vec a$. 
 Then the kernel in Eq.~(\ref{eq:nthorder}) is given by restricting the
 momenta in the master kernel appropriately: External momenta are set on shell, while
 internal coordinates (those contracted with $\cC^H$) are restricted to the finite-volume set.
 This is identical to the description given for the na\"ive kernel $A$ in the main text following
 Eq.~(\ref{eq:masterA}).
 
By definition, Eq.~(\ref{eq:nthorder}) holds true for $n=1$, so we begin by considering the
$n=2$ case. We know that, using the procedure of previous subsections, we can
bring the contributions to $\ML{\rm ext}$ with a single $\cC^H$ into the form
\begin{equation}
\ML{\rm ext}^{(2)} = \tM'\; \cC^H \tM'\,,
\end{equation}
with $\tM'$ a matrix of kernels having the same properties as $B$ (smooth
and connected except for $\tM'_{33}$).
These kernels are constructed of all possible time orderings of the allowed
Feynman diagrams lying between the external states and the cut $\cC^H$, 
with appropriate factors of $1\!-\!H_i$ inserted, and all loops integrated.
Since the same set of orderings can occur on both sides of the $\cC^H$, the
two kernels are equal.\footnote{%
This relies on the fact that the cut factors $\cC^H$ act just like amputation on the
external legs: Removing the factors associated with the cut propagators from the kernels,
and only allowing time orderings in which the vertices lie between the external states.}
What we need to show is that $B'=B$, i.e. that all contributions to $B'$ are contained in $B$
and {\em vice versa}. The former property is clear---any diagram connecting
an external state to a cut $\cC^H$ can also serve to connect two external states
(or, as needed below, two cut factors). The latter property follows because every
contribution contained in $\tM \cC^H \tM$ will occur in $\ML{\rm ext}^{(2)}$,
simply by gluing the two halves together and inserting the cut factor.

This argument extends straightforwardly to arbitrary $n$, and 
completes the demonstration of Eq.~(\ref{eq:nthorder}).

\section{Finite-volume dependence from the TOPT results}
\label{app:FVTOPT}

In this appendix we sketch the derivations of various results quoted in the main text.
We first discuss quantities involving only two-cuts, and then we consider those containing three-cuts.

\subsection{Derivation of the result for $X_{22}$}
\label{app:X22}

The analysis of Refs.~\cite{Kim:2005gf,Hansen:2014eka, Hansen:2015zga} uses a skeleton expansion applied
to standard relativistic Feynman diagrams.
This is in contrast to the analysis in the main text, which uses TOPT,
leading to the expression Eq.~(\ref{eq:X22}) for $\MLL$.
While the two approaches lead to the same poles, as they must, they
differ in the way that various nonpole parts are allocated to nonsingular kernels.
For example, the quantity $B_{22}$ in Eq.~(\ref{eq:X22}) 
differs from the Bethe-Salpeter kernel $\widetilde B_2$ that appears
in the analogous expression from the Feynman diagram analysis
(as discussed further in Appendix~\ref{app:X33} below).
Because of this, there is no simple way to recast the TOPT expression 
for $X_{22}$ back into a Feynman-diagram form.
Thus we cannot directly apply the results obtained in Refs.~\cite{Kim:2005gf, Hansen:2014eka, Hansen:2015zga}.
Instead, we apply the methodology developed in those references directly to the TOPT expression.

Starting from Eq.~(\ref{eq:X22}), we focus on one of the two-cuts, and make
the  matrix multiplications explicit, leading to 
\begin{align}
\left[B_{22}\, \CH_2 \!B_{22} \right]_{p'';p'}
&=
\sum_{\vec p,\vec r} B_{22;p'';p} \, \mathcal C_{2;p;r}\, H_2(\vec r)\; B_{22;r;p'} \,, 
\\
&=
-\frac1{L^3} \sum_{\vec p}
B_{22;p'';p} \frac{1}{2} \frac1{2\omega_p 2\omega_{Pp} (E-\omega_p-\omega_{Pp})} 
H_2(\vec p)\; B_{22;p;p'} 
\,.
\label{eq:genericX2}
\end{align}
The factor of $-1$ coming with $\CH$ arises from the product of the
$i$ associated with the energy denominator and that associated with one of the adjacent vertices. 
We now recall that the key property of $B_{22}$ for our purposes is that it is a smooth
function of its momentum arguments. Thus the only singularity in the summand is that
from the explicit pole in $\CH$.

We now write the sum over $\vec p$ as an integral plus a sum-integral difference to reach
\begin{multline}
\left[B_{22} \,\CH_2 \!B_{22} \right]_{p'';p'} = 
- {\rm PV} \int_{\vec p}
B_{22;p'';p}  \frac{H_2(\vec p)}{8\omega_p \omega_{Pp} (E-\omega_p-\omega_{Pp})} 
\; B_{22;p;p'}  \\
-\bigg [\frac1{L^3} \sum_{\vec p} - {\rm PV} \int_{\vec p} \bigg ]
B_{22;p'';p} \frac{1}{2} \frac{h(\vec p)}{2\omega_p 2\omega_{Pp} (E-\omega_p-\omega_{Pp})} 
\; B_{22;p;p'} 
\,.
\label{eq:genericX2_v2}
\end{multline}
Here we have also replaced $H_2(\vec p)$ with $h(\vec p)$ in the sum-integral difference, with $h(\vec p)$ the UV regulator introduced in Eq.~(\ref{eq:F2def}) above. This substitution is justified because $H_2(\vec p) - h(\vec p)$ vanishes at the pole so that the replacement is equivalent to dropping the sum-integral difference of a function that is smooth for all real $\vec p$, i.e.~dropping a contribution that is exponentially suppressed.  Here and below we keep implicit the fact that we are dropping exponentially suppressed terms.

From here we follow the steps outlined in Ref.~\cite{Kim:2005gf} to rewrite the sum-integral difference in terms of the zeta function $F_2$, defined in Eq.~(\ref{eq:F2def}).
Given that $B_{22}$ is a smooth function, the dominant finite-volume corrections from the second term above are due to the explicit propagator pole. As a result, one can replace $B_{22}$ with its value when the internal momentum $p$ is projected on shell. This is effected by setting the CM frame magnitude to equal $q^*$. This fixes the magnitude but not the direction and this remaining degree of freedom motivates us to decompose $B_{22}$ in spherical harmonics
\begin{align}
B_{22;p'';p} \bigg \vert_{p^* = q^*} = \sqrt{4\pi} \,Y_{\ell' m'}(\hat p^*) B_{22;p'';\ell' m'},\hspace{1cm}   
B_{22;p;p'} \bigg \vert_{p^* = q^*}
= \sqrt{4\pi}\,Y^*_{\ell,m}(\hat p^*)B_{22;\ell m;p'}\,.
\end{align}
Using the sum-integral-difference identity of Ref.~\cite{Kim:2005gf}, as expressed in Appendix A of
Ref.~\cite{Hansen:2014eka}, we find
\begin{equation}
\left[B_{22}\, \CH_2 \!B_{22} \right]_{p'';p'} = -{\rm PV} \int_q
B_{22;p'';q}  \frac{H_2(\vec q)}{8\omega_q \omega_{Pq} (E-\omega_q-\omega_{Pq})} 
\; B_{22;q;p'} 
-
B_{22;p';\ell' m'} F_{2;\ell'm';\ell m} B_{22;\ell m;p'}
\,.
\label{eq:intplusF2}
\end{equation}
We summarize this result in shorthand notation as
\begin{equation}
B_{22} \,\CH_2\! B_{22} = - B_{22} I_C B_{22} - B_{22} F_2 B_{22}
\,,
\label{eq:identity}
\end{equation}
with $I_C$ an integral operator. 
We note that this identity holds for any choice of kernels on the left- and right-hand sides,
as long as they are smooth functions of momenta.
We can thus condense the notation even further and write
\begin{equation}
\CH_2 = - I_C - F_2\,.
\label{eq:identity2}
\end{equation}

Using this identity, we can reorganize the sum in
Eq.~(\ref{eq:X22}) into a series in powers of $F_2$ (following the method of Ref.~\cite{Kim:2005gf})
\begin{align}
X_{22} &= B_{22} \sum_{n=0}^\infty \left[(-I_C - F_2)B_{22}\right]^n \,,
\\
&= \cK_{22,D} \sum_{n=0}^\infty \left[ -F_2 \cK_{22,D}\right]^n
\,,
\label{eq:X22resint}
\end{align}
where 
\begin{equation}
\cK_{22,D} = \sum_{n=0}^\infty B_{22} [-I_C B_{22}]^n \,.
\label{eq:K2Dres}
\end{equation}
Summing the geometric series in Eq.~(\ref{eq:X22resint}) leads to the result quoted in the main text,
Eq.~(\ref{eq:X22res}).

\subsection{Derivation of the results for $Y_{22}$ and $Z_{23}$}
\label{app:Y22}

The determination of the volume dependence of  $Y_{22}$, defined in Eq.~(\ref{eq:Y22def}),
follows similar steps to those described in Appendix~\ref{app:X22} for $X_{22}$.
We can use the identity (\ref{eq:identity2}) for all two-cuts, since the kernels on either side
of the cut involve the smooth functions $B_{22}$, $B_{23}$ or $B_{32}$.
Collecting terms according to the number of factors of $F_2$, we find
\begin{align}
Y_{22} &= B_{32}\left[ \CH_2 + \CH_2 B_{22} \CH_2 + \cdots \right] B_{23} \,,
\\
\begin{split}
&= B_{32} \left[ - I_C + I_C B_{22} I_C - \cdots \right] B_{23}
- B_{32} \left[1 - I_C B_{22} + \cdots \right] F_2 \left[1 - B_{22} I_C + \cdots \right] B_{23}
\\
&\quad +
B_{32} \left[1 - I_C B_{22} + \cdots \right] F_2 
\left[B_{22} - B_{22} I_C B_{22} + \cdots\right] F_2
\left[1 - B_{22} I_C + \cdots \right] B_{23}
+ \cdots \,,
\end{split}\\
&= 
B_{32} \cD_{C,2} B_{23} - B_{32}\cD_{A',2} F_2 \cD_{A,2} B_{23}
+ B_{32}\cD_{A',2} F_2 \cK_{22,D} F_2  \cD_{A,2} B_{23} - \cdots \,,
\label{eq:Y22int}
\end{align}
where in the last step we have used Eq.~(\ref{eq:K2Dres}) and defined the integral operators
\begin{align}
\cD_{C,2} &=  \left[ - I_C + I_C B_{22} I_C - \cdots \right] \,,
\label{eq:DC2def}
\\
\cD_{A',2} &=  \left[1 - I_C B_{22} + \cdots \right]\,,
\label{eq:DAp2def}
\\
\cD_{A,2} &= \left[1 - B_{22} I_C + \cdots \right] \,.
\label{eq:DA2def}
\end{align}
Summing the geometric series in Eq.~(\ref{eq:Y22int}) leads
to the result quoted in the main text, Eq.~(\ref{eq:Y22res}).

This derivation applies also for $Z_{23}$, the only change being the replacement
of $B_{32}$ on the left with $B_{22}$. Thus from Eq.~(\ref{eq:Y22res}) we obtain
\begin{equation}
Z_{23} = B_{22}
\left[ \cD_{C,2} 
- \cD_{A',2} F_2 \frac1{1+ \cK_{22,D} F_2} \cD_{A,2}\right]B_{23}\,.
\label{eq:Z23res1}
\end{equation}
This can be simplified using the identities
\begin{align}
B_{22} \cD_{A',2} &=  \cK_{22,D}\,,
\\
B_{22} \cD_{C,2} &= \cD_{A,2}-1\,,
\end{align}
leading to
\begin{equation}
Z_{23} = \left[ \cD_{A,2} - 1 
-\cK_{22,D} F_2 \frac1{1+ \cK_{22,D} F_2} \cD_{A,2}\right]B_{23}\,.
\label{eq:Z23res2}
\end{equation}
The result for $Z_{23}$ in the main text, Eq.~(\ref{eq:Z23res}), follows immediately.

\subsection{Comments on the derivation of the result for $X_{33}$}
\label{app:X33}

As explained in the main text, to determine $X_{33}$ we must repeat the
analysis of Refs.~\cite{Hansen:2014eka, Hansen:2015zga} starting from the TOPT decomposition of Eq.~(\ref{eq:X33})
instead of the skeleton expansion of Feynman diagrams.
To do so, we use the decomposition of $B_{33}$ into connected and disconnected parts,
Eq.~(\ref{eq:tM33decomp}).
$B_{33}^{\rm conn}$ is the analog in the present analysis of the three-particle
Bethe-Salpeter amplitude $B_{3}$ in the analysis of Refs.~\cite{Hansen:2014eka, Hansen:2015zga}. 
The disconnected part can be written
\begin{equation}
B_{33;k'a';ka}^{\rm disc} = 2 \omega_k L^3 \delta_{k' k}  \widetilde B_{2}(\vec k)_{a';a}
+ \textrm{permutations}
\,,
\label{eq:tM33disc}
\end{equation}
where $\widetilde B_{2}$ plays the role here
of the two-to-two Bethe-Salpeter kernel $B_2$ appearing in Ref.~\cite{Hansen:2014eka}, 
with some important distinctions that we discuss below. 
``Permutations'' refers to the inclusion of all possible choices 
 of incoming and outgoing spectator momenta. 
 There are nine terms in total,
corresponding  to the three different choices of the momentum of the
spectator particle in both initial and final states
(e.g. $\vec k$, $\vec a$ or $\vec P-\vec k -\vec a$ in the initial state).
Thus we can rewrite the result 
using the symmetrization operators introduced in the  main text:
\begin{equation}
B_{33;k'a';ka}^{\rm disc} = \cS_{\cL}\left\{
2 \omega_k L^3  \delta_{k' k}  \widetilde B_{2}(\vec k)_{a';a}
\right\}\cS_{\cR}
\,.
\end{equation}
The factor of $2\omega_k$ is needed to cancel the $1/(2\omega_k)$ contained
in the adjacent three-cut, $\CH_3$, since each disconnected propagator should come with only
one overall factor of $1/(2\omega_k)$, and this factor is provided by the first $\CH_3$. Similarly, the factor of $L^3$ is introduced to assure that diagrams with insertions of $B_{33}^{\rm disc}$ have the correct powers of $L$. 

\begin{figure}[t]
\begin{center}
\includegraphics[width = \textwidth]{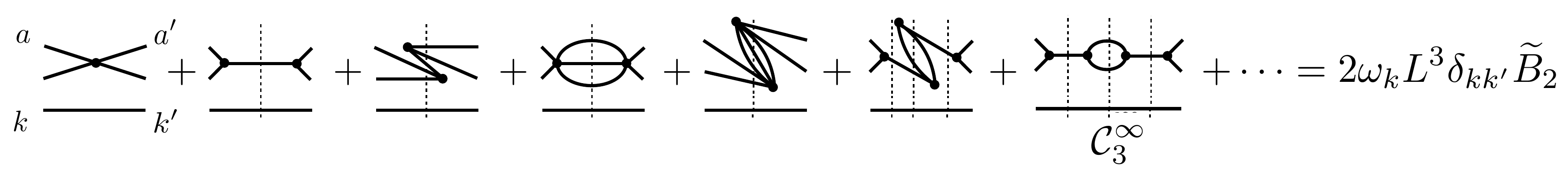}

\caption{Examples of TOPT diagrams contributing to $B^{\rm disc}_{33}$, 
and thus to $\widetilde B_2$,  in a general EFT. 
The external three-particle state is either on shell or has a factor of $\CH_3$.
The vertical dashed lines indicate intermediate states, which
come with factors of $1/(E- \sum_i \omega_i)$. Three-particle intermediate states also
include a factor of $(1\!-\!H_3)$, as indicated by the $\mathcal C_3^\infty$ in the last diagram.
Two-particle intermediate states do not contain factors of $H_2$, and there are no
$H_i$ factors in intermediate states containing four or more particles.
No vertices are allowed before the initial time or after the final time.
All loop momenta are integrated rather than summed (since there are no physical cuts). }
\label{fig:appCfig1}
\end{center}
\end{figure}

It is important to understand in some detail the differences between 
the Bethe-Salpeter kernel, $B_2$, and the quantity appearing here, $\widetilde B_2$.
$B_2$ consists of all amputated two-to-two Feynman diagrams that are two-particle
irreducible in the $s$ channel. 
$\widetilde B_{2}$ contains all the time orderings arising from these Feynman diagrams,
except those in which any vertex lies before the initial three-cut or after the final three-cut.
In addition, because of the definition of $B$ described in Sec.~\ref{sec:naive},
$\widetilde B_{2}$ includes time orderings (constrained as above)  
from two-to-two diagrams that are two-particle {\em reducible} in the $s$ channel.
These, however, are weighted by a factor of $1\!-\!H_3$, so that there is no physical cut.
(The weight involves $H_3$ and not $H_2$ because this is part of a three-particle kernel.)
These features are illustrated in Fig.~\ref{fig:appCfig1}. Because of the appearance of
$1\!-\!H_3$ in some intermediate states, $\widetilde B_{2}$ is an unconventional quantity.  

\begin{figure}[t]
\begin{center}
\includegraphics[width = \textwidth]{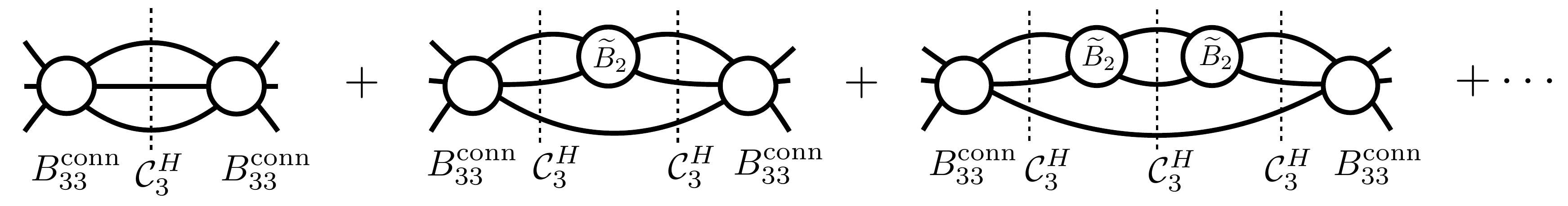}

\caption{Contributions to $X_{33}$ in TOPT that correspond to the ``no switch" diagrams
considered in Ref.~\cite{Hansen:2014eka}.}
\label{fig:appCfig2}
\end{center}
\end{figure}

We now proceed through the steps of the derivation in Refs.~\cite{Hansen:2014eka, Hansen:2015zga}.
We recall that Ref.~\cite{Hansen:2015zga} studied the quantity of interest, $\MLL$, but
made heavy use of the work in Ref.~\cite{Hansen:2014eka}, so we need to repeat
the steps from both references. We stress that the steps we need to take using
the TOPT decomposition are in one-to-one correspondence
with those using the skeleton expansion. To illustrate this correspondence
we consider the following contributions to $X_{33}$:
\begin{equation}
X_{33} \supset 
B_{33}^{\rm conn} 
\left[\CH_3 + \CH_3 B_{33}^{\rm disc} \CH_3 
+ \CH_3 B_{33}^{\rm disc} \CH_3 B_{33}^{\rm disc} \CH_3+ \cdots \right] 
B_{33}^{\rm conn}
\,.
\end{equation}
If we keep the subset of these contributions in which the spectator meson
remains the same for all factors of $B_{33}$ 
then we obtain the diagrams shown in Fig.~\ref{fig:appCfig2}.
These correspond to the
``no switch" diagrams considered in Sec. IVA of Ref.~\cite{Hansen:2014eka},
and shown in Fig.~7 of that work.
The differences between the expressions represented by the diagrams
are as follows:
First, while here the ``end caps" are provided by factors
of $B_{33}^{\rm conn}$, in Ref.~\cite{Hansen:2014eka} they are given by the
external operators $\sigma$ and $\sigma^\dagger$.
As noted in Ref.~\cite{Hansen:2014eka}, however, as long as they are nonsingular, 
the choice of end caps has no impact on the form of the result.
Second, as already described, $\widetilde B_{2}$ here is replaced by
$B_2$ in Ref.~\cite{Hansen:2014eka}.
Last, the expression for $\CH_3$ differs from the ``cut" that arises in
Ref.~\cite{Hansen:2014eka}.
The key point, however, is that the residue of the pole is the same in both cases,
with the differences appearing in nonsingular terms.
This can be seen, for example, from Eq.~(56) of Ref.~\cite{Hansen:2014eka}, which is
proportional to $\CH_3$.
Indeed, the essential difference between the TOPT analysis and that using
Feynman diagrams is that nonsingular terms are reshuffled between the kernels.

In the expression represented by the diagrams of Fig.~\ref{fig:appCfig2},
the three-momentum sums associated with each $\CH_3$ factor are replaced 
by integrals and a zeta function, 
using a generalization of the identity given in Eq.~(\ref{eq:identity}).
Following the steps of Ref.~\cite{Hansen:2014eka}, we find that this class of diagrams
leads to the following volume-dependent terms
\begin{equation}
\MLL \supset -B_{33}^{\rm conn} (1 + \cD_{A',3}^{(1,u)}) \frac{F}{2\omega L^3}
\frac1{1 + {\mathcal K}_{22} F} (1 + \cD_{A,3}^{(1,u)} ) B_{33}^{\rm conn}
+ \frac{2}{3} B_{33}^{\rm conn} \frac{F}{2\omega L^3} B_{33}^{\rm conn} \,.
\label{eq:oneswitch}
\end{equation} 
Here $F$ is defined in Eq.~(\ref{eq:Fdef}), $ {\mathcal K}_{22}$ is given by
\begin{equation}
{\mathcal K}_{22;k'\ell'm';k\ell m} = \delta_{k' k}\left[
\widetilde B_{2}(\vec k) + {\rm PV}\!\! \int\widetilde B_{2}(\vec k) 6 \omega_k L^6\, \CH_3 
\widetilde B_{2} (\vec k) + \cdots \right]_{\ell'm';\ell m}  \,,
\label{eq:tK2def}
\end{equation}
(where the integral runs over the implicit $\vec a$ dependence of the two $\widetilde B_{2}$
factors and of $\CH_3$),
and $\cD_{A',3}^{(1,u)}$ and $\cD_{A,3}^{(1,u)}$ are the first contributions to
the decoration operators $\cD_{A',3}$ and $\cD_{A,3}$ discussed in the main text.
The result (\ref{eq:oneswitch}) has the same form as Eq.~(92) of Ref.~\cite{Hansen:2014eka}.

We have checked that all subsequent steps in the lengthy derivations
of Refs.~\cite{Hansen:2014eka, Hansen:2015zga} go through, and we do not present further details.
The conclusion is that we can read off the final result for $X_{33}$ from that
for $\MLL$ given in Eq.~(68) of Ref.~\cite{Hansen:2015zga}, as long as we change the
meaning of the symbols appropriately. This is what we have done in
Eqs.~(\ref{eq:X33res})-(\ref{eq:F3res}).

\begin{figure}[t]
\begin{center}
\includegraphics[scale=0.5]{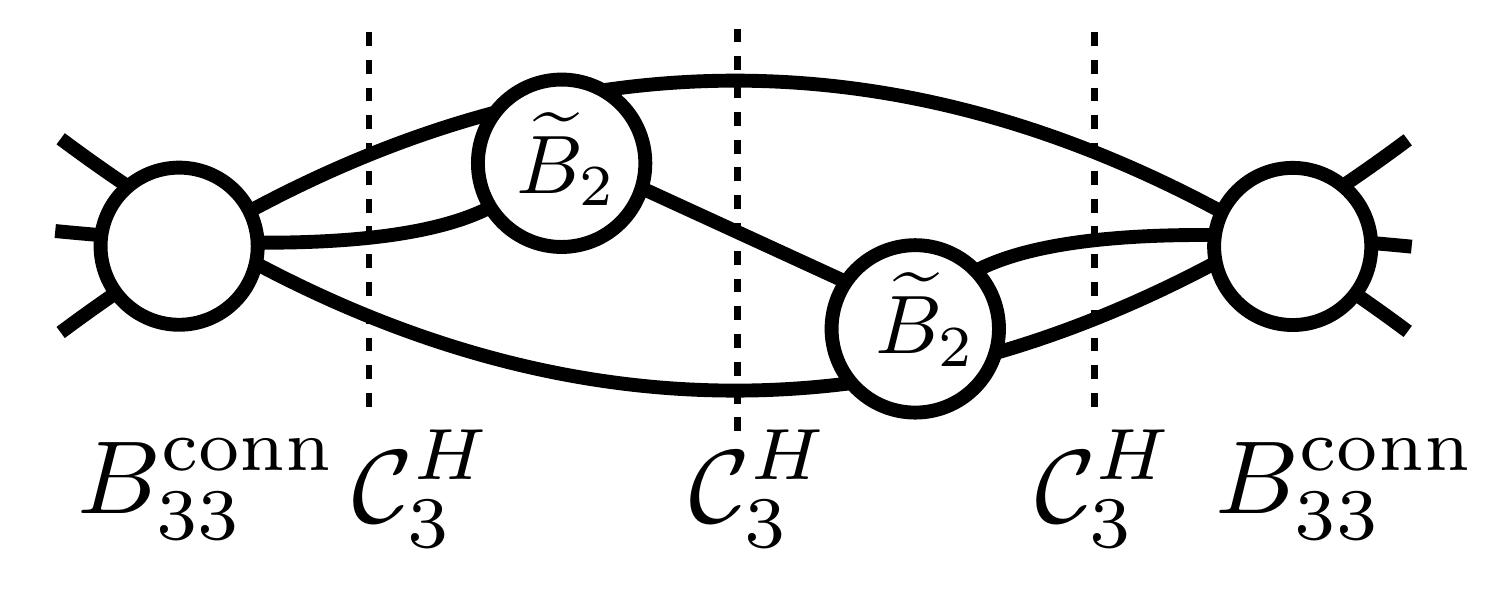}
\caption{Example of one-switch diagram contributing to $X_{33}$ in TOPT.}
\label{fig:appCfig3}
\end{center}
\end{figure}

There are, however, two features of the result that deserve further mention.
The first concerns the matrix $G^H$. This arises from diagrams involving switches,
the simplest of which is shown in Fig.~\ref{fig:appCfig3}.
The corresponding diagram is analyzed in Sec.~IVB of Ref.~\cite{Hansen:2014eka}.
In one of the volume-dependent contributions, the two outer $\CH_3$ factors
are replaced by $F$ factors, while the central factor gives rise to a switch matrix $G^H$:
\begin{equation}
G^H_{p\ell' m';k\ell m} = \left(\frac{k^*}{q_p^*}\right)^{\ell'}
\frac{4\pi Y_{\ell'm'}(\hat k^*) H_3(\vec p,\vec k) Y^*_{\ell m}(\hat p^*)}
{2\omega_{Pkp}(E-\omega_k-\omega_p-\omega_{Pkp})}
\left(\frac{p^*}{q_k^*}\right)^{\ell} \frac1{2\omega_k L^3}\,.
\label{eq:GHdef}
\end{equation}
This switches the interacting pair from the upper two to the lower two particles.
The key point here is that $G^H$ inherits the cutoff 
$H_3 = H(\vec p) H(\vec k) H(\vec b_{kp})$ from $\CH_3$.
By contrast, in Ref.~\cite{Hansen:2014eka}, where the switch matrix is first introduced in
Eq.~(116), there is some freedom in the choice of the cutoff function, and the
choice made there is $H(\vec p)H(\vec k)$. Thus $G^H$ and $G$ differ by
a factor of $H(\vec b_{kp})$. We note, however, that in Ref.~\cite{Hansen:2014eka}
one could equally well have included the full $H_3$ in the definition of $G$ without
changing the derivation.
In other words, the form of $G$ that is forced on us here is a completely viable
option in Ref.~\cite{Hansen:2014eka} as well.

The second feature of the result for $X_{33}$ concerns $ {\mathcal K}_{22}$, defined in Eq.~(\ref{eq:tK2def}). 
We find that
\begin{equation}
 {\mathcal K}_{22;k'\ell' m';k \ell m} = \delta_{k' k}  {\mathcal K}_{2;\ell' m'; \ell m}(E-\omega_k,\vec P-\vec k)
\,,
\label{eq:tK2physical}
\end{equation}
i.e. $ {\mathcal K}_{22}$ in fact contains the {\em physical} two-particle $K$ matrix.
To show this requires two further results: The unphysical dependence of $\widetilde B_{2}$ on $H_3$ 
must cancel, and the missing time orderings in $\widetilde B_{2}$ must become irrelevant.
To explain the cancellation of $H_3$ dependence, we rewrite $B_{22}$ to
make its dependence on $H_3$ explicit:
\begin{equation}
\widetilde B_{2} = \overline B_{2} + {\rm PV} \int \overline B_{2} 6\omega_k L^6 \cC_3^\infty
\overline B_{2} + \cdots \,.
\label{eq:B2bardef}
\end{equation}
Here $\overline B_{2}$ 
is the result obtained when all diagrams containing $\cC_3^\infty$ are dropped,
and thus is independent of $H_3$.
For example, in Fig.~\ref{fig:appCfig1}, the last diagram would be dropped.
Thus $\overline B_{2}$ differs from the Bethe-Salpeter amplitude $B_2$ only in that 
certain time orderings are not included in the former. 
The $H_3$ dependence of $\widetilde B_{2}$ is then reintroduced by the terms involving
integrals in Eq.~(\ref{eq:B2bardef}), corresponding to adding back in diagrams like the
last one in Fig.~\ref{fig:appCfig1}.
Substituting this result into Eq.~(\ref{eq:tK2def}), and rearranging terms, we find that
\begin{equation}
{\mathcal K}_{22;k'\ell'm';k\ell m} = \delta_{k' k}\left[
\overline B_{2}(\vec k) + {\rm PV}\!\! \int \overline B_{2}(\vec k) 6\omega_k L^6\, \cC_3 
\overline B_{2} (\vec k) + \cdots \right]_{\ell'm';\ell m} \,.
\label{eq:tK2res2}
\end{equation}
The $H_3$ dependence has canceled because $\CH_3+\cC_3^\infty=\cC_3$.
Thus $ {\mathcal K}_{22}$ receives contributions from all amputated two-to-two TOPT diagrams,
except that no time orderings are allowed in which vertices lie before the initial cut
or after the final cut. However, as indicated by the spherical harmonic indices in 
Eq.~(\ref{eq:tK2res2}), these diagrams are evaluated on shell assuring that 
diagrams with the missing time orderings vanish. Thus we find the result (\ref{eq:tK2physical}).

\subsection{Derivation of the result for $Z_{32}$}
\label{app:Z32}

The final quantity we consider in this appendix is $Z_{32}=B_{33}\Xi_{33}B_{32}$.
As noted in the main text, this is not a quantity for which a result can simply be
read off from Refs.~\cite{Hansen:2014eka, Hansen:2015zga}, since it has disconnected parts on one end
but not the other. Nevertheless, by a small extension of Eq.~(64) in Ref.~\cite{Hansen:2015zga},
the relevant result can be found. This equation gives a result for $\cM_{3,L}^{(u,u)}$,
the unsymmetrized three-particle finite-volume amplitude, with all factors of
$B_3$ (the fully connected three-particle Bethe-Salpeter amplitude) explicit.
To obtain $Z_{32}$ we must (a) drop any contribution in which there is no $B_3$,
(b) replace the rightmost $ B_3$ with $B_{32}$, (c) replace all other
factors of $B_3$ with $B_{33}^{\rm conn}$, and (d) symmetrize on the left. 
The result is
\begin{equation}
Z_{32} = \cS_{\cL,3}\left\{\cL^{(u,u)}_{L,3} \widetilde \cZ
\cD_{A,3}^{[B_2,\rho]} 
\sum_{n=0}^\infty \left(B_{33}^{\rm conn} M^{[B_2,\rho]}\right)^n B_{32}\right\}
-B_{32}\,,
\label{eq:Z32step1}
\end{equation}
where $\cL^{(u,u)}_{L,3}$ is defined in Eq.~(\ref{eq:LL3res}), while
\begin{align}
\widetilde \cZ &= \frac1{1 + \cK_{{\rm df},33,D}^{[B_2,\rho]} F_3}\,,
\label{eq:tZdef}
\\
M^{[B_2,\rho]} &= \cD_{C,3}^{[B_2,\rho]}
- \cD_{A',3}^{[B_2,\rho]} F_3 \widetilde \cZ \cD_{A,3}^{[B_2,\rho]}
\,.
\label{eq:MB2rhodef}
\end{align}
The superscript $[B_2,\rho]$, which is defined in Ref.~\cite{Hansen:2014eka}, indicates
the parts of the integral operators that do not contain factors of
$B_{33}^{\rm conn}$. The relation between these parts and the full integral operators 
can be read off from Eqs.~(247)-(249) of Ref.~\cite{Hansen:2014eka}, and is\footnote{%
We comment that the decoration operator $\mathcal D_{C,3}$ used here and the analog used in Ref.~\cite{Hansen:2014eka}, denoted $D_C$, differ by a trivial relative phase. In particular, in the limit where the two-to-three coupling is set to zero, the operators are related by $\mathcal D_{C,3} = i D_C$.}
\begin{align}
\cD_{C,3} &= \cD_{C,3}^{[B_2,\rho]}
 \sum_{n=0}^\infty \left( B_{33}^{\rm conn} \cD_{C,3}^{[B_2,\rho]}\right)^n \,,
 \\
 \cD_{A,3} &= \cD_{A,3}^{[B_2,\rho]}
 \sum_{n=0}^\infty \left( B_{33}^{\rm conn} \cD_{C,3}^{[B_2,\rho]}\right)^n \,,
\\
\cD_{A',3} &= 
 \sum_{n=0}^\infty \left( \cD_{C,3}^{[B_2,\rho]}B_{33}^{\rm conn} \right)^n 
\cD_{A',3}^{[B_2,\rho]}\,.
\end{align}
These operators appear in the expression for $Y_{33}$, Eq.~(\ref{eq:Y33res}).

Our final comment about Eq.~(\ref{eq:Z32step1}) concerns the subtraction of $B_{32}$ on the 
right-hand side. This is required to cancel the leading contribution from the first term on the right-hand side, which comes from the symmetrization of the product of the
$1/3$ term in $\cL_{L,3}^{(u,u)}$ [Eq.~(\ref{eq:LL3res})], the $1$ in $\widetilde \cZ$, the $1$ in $\cD_{A,3}^{[B_2,\rho]}$, and the $n=0$ term in the sum.
This $B_{32}$ term is absent in $Z_{32}$.

The next step is to substitute the result (\ref{eq:MB2rhodef}) into Eq.~(\ref{eq:Z32step1})
and collect terms according to the number of $F_3$ factors. This leads to
\begin{equation}
Z_{32}+B_{32} = \cS_{\cL,3}\left\{\cL^{(u,u)}_{L,3} \widetilde \cZ
\left[1 - B_{33}^{{\rm conn}, CA} F_3 \widetilde \cZ
+ \left(B_{33}^{{\rm conn}, CA} F_3 \widetilde \cZ \right)^2 + \cdots
\right]
\cD_{A,3} B_{32}\right\}\,,
\label{eq:Z32step2}
\end{equation}
where
\begin{equation}
B_{33}^{{\rm conn},CA} = \cD_{A,3}^{[B_2,\rho]}
\sum_{n=0}^\infty \left( B_{33}^{\rm conn} \cD_{C,3}^{[B_2,\rho]}\right)^n B_{33}^{\rm conn}
\cD_{A',3}^{[B_2,\rho]} \,,
\end{equation}
is the analog here of the quantity $B_3^{[B_2,\rho]}$ in Ref.~\cite{Hansen:2015zga}.
Finally, summing the geometric series in Eq.~(\ref{eq:Z32step2}), performing some
algebraic manipulations, and using 
\begin{equation}
\mathcal K_{\rm df,33,D} = \cK_{\rm df,33}^{[B_2,\rho]} + B_{33}^{{\rm conn},CA} \,,
\label{eq:Kdf3Ddef}
\end{equation}
[the analog of Eq.~(65) of Ref.~\cite{Hansen:2015zga}],
leads to the claimed result, Eq.~(\ref{eq:Z32res}).

\section{Time-reversal and parity invariance \label{app:time_rev}}

In this section we investigate the implications for $\mathcal K_{\rm df}$ of assuming
that time-reversal and parity invariance hold in the underlying theory.
We first discuss the consequences of time-reversal invariance; the consequences of parity invariance can then be inferred by a straightforward modification.

Na\"ively, one might expect that, since $\mathcal{K}_{\rm df}$ is an infinite-volume
scattering quantity, it should transform under time reversal in the same way as $\mathcal M$.
However, upon closer inspection, this result is far from obvious.
For example, the definition of $\mathcal K_{{\rm df},33}$, the three-to-three component of $\mathcal{K}_{\rm df}$, 
involves a choice of ordering of loop integrals that is not manifestly time-reversal 
invariant~\cite{Hansen:2014eka}. 
Nevertheless, as we show in this appendix, given the relations 
between $\mathcal K_{\rm df}$ and $\mathcal M$ derived in Sec.~\ref{sec:MtoK},
the transformation properties of $\mathcal M$ are indeed inherited by $\mathcal K_{\rm df}$.

Time-reversal invariance implies that the components of the scattering amplitude satisfy
\begin{align}
{\cal M}_{22;\vec P}(\hat p'^*; \hat p^*) &= {\cal M}_{22;-\vec P}(-\hat p^*; -\hat p'^*) \,,
\label{eq:M22_Trev}
\\
{\cal M}_{23;\vec P}(\hat p'^*; \vec k, \hat a^*) &= {\cal M}_{32;-\vec P}(-\vec k, -\hat a^*; -\hat p'^*) \,,
\label{eq:M23_Trev}\\
{\cal M}_{{\rm df},33;\vec P} (\vec k', \hat a'^*; \vec k, \hat a^*) &= {\cal M}_{{\rm df,33};-\vec P} (-\vec k,-\hat a^*;-\vec k', -\hat a'^*) \,,
\label{eq:M33_Trev}
\end{align}
where we have denoted dependence on the total momentum, $\vec P$, as a subscript.\footnote{%
Previously the dependence on $\vec P$ has been implicit. We make it explicit throughout
this appendix.
}
Decomposing 
using spherical harmonics, one finds that the various components satisfy
\begin{align}
\label{eq:M22_Trev2}
{\cal M}_{22; \ell m; \ell'm';\vec P} &= (-1)^{\ell+m+\ell'+m'}\,{\cal M}_{22; \ell' -m'; \ell -m;-\vec P} \,,\\
\label{eq:M23_Trev2}
{\cal M}_{23;\ell m; \ell' m';\vec P}( \vec k) &= (-1)^{\ell+m+\ell'+m'}\,{\cal M}_{32;\ell' -m'; \ell -m;-\vec P}(-\vec k)\,,\\
\label{eq:M33_Trev2}
{\cal M}_{{\rm df,33};\ell m; \ell' m';\vec P}(\vec k'; \vec k) &= (-1)^{\ell+m+\ell'+m'}\, {\cal M}_{{\rm df,33};\ell' -m'; \ell -m;-\vec P}(-\vec k; -\vec k') \,.
\end{align}
To obtain these results
we have used standard properties of the spherical harmonics under complex conjugation and parity 
transformation. Note that, since we are considering the divergence-free form
of ${\cal M}_{33}$, we can decompose in spherical harmonics. 
From these results we conclude that it is sufficient to determine 
${\cal M}_{22}$, ${\cal M}_{23}$, and ${\cal M}_{\rm df, 33}$, 
since $\mathcal M_{32}$ then follows trivially from Eq.~(\ref{eq:M23_Trev2}).
 
In the following, we will say that a quantity has ``standard time-reversal transformation properties"
if Eqs.~(\ref{eq:M22_Trev2})-(\ref{eq:M33_Trev2}) hold with the quantity substituted for 
$\mathcal M$.
We recall from Sec.~\ref{sec:MtoK} that $\mathcal K_{\rm df}$ is obtained
from $\mathcal M$ in two steps. First, the intermediate quantity $\T$ is obtained from $\mathcal M$
using Eqs.~(\ref{eq:T23_M23})-(\ref{eq:T33_M33}),
and, second, $\mathcal K_{\rm df}$ is obtained from $\T$ using 
Eqs.~(\ref{eq:K22_T22})-(\ref{eq:K33_T33}).
In what follows we first show that $\T$ has standard time-reversal transformation properties
and then show that the same holds for $\cK_{\rm df}$.

$\T$ is obtained from $\mathcal M$ by integrating with the kernels $I_{\cR}$ and $I_{\cL}$,
which are themselves obtained from $\Delta_{\cL}$ and $\Delta_{\cR}$ by solving
the integral equations (\ref{eq:ILdef}) and (\ref{eq:IRdef}), respectively.
The latter kernels are essentially the symmetrized forms of $\cL_3^{(u,u)}$
and $\cR_3^{(u,u)}$, as shown by Eqs.~(\ref{eq:M33dfwithLR}) and (\ref{eq:M33df_final}).
Thus, to proceed, we need to understand the time-reversal transformation properties of $\cL_3^{(u,u)}$ and $\cR_3^{(u,u)}$, defined in Eqs.~(\ref{eq:Luures}) and (\ref{eq:Ruures}), respectively. 
These are built using $\cD_3^{(u,u)}$, which, as shown in Eq.~(\ref{eq:Duuinteq}),
involves the kernel $G^\infty$ of Eq.~(\ref{eq:Ginfdef}).

Thus we begin by studying the transformation properties of $G^\infty$.
It follows from its definition that
\begin{align}
\label{eq:Ginf_Trev}
G^{\infty}_{\ell m; \ell' m';\vec P}(\vec k'; \vec k) = (-1)^{\ell+m+\ell'+m'}\,G^{\infty}_{\ell' -m'; \ell -m;-\vec P}(-\vec k; -\vec k')\,,
\end{align}
where we have used
$H_{3;\vec P}(\vec k', \vec k) = H_{3; - \vec P}(- \vec k, - \vec k') $.
Using the definition of $\mathcal D_3^{(u,u)}$, Eq.~(\ref{eq:Duuinteq}), and substituting 
the symmetry relations for $\mathcal M_{22}$, Eq.~(\ref{eq:M22_Trev2}), and $G^\infty$, Eq.~(\ref{eq:Ginf_Trev}), we find
\begin{equation}
 \mathcal{D}^{(u,u)}_{3;\ell m; \ell' m';\vec P}(\vec k'; \vec k) =     (-1)^{\ell+m+\ell'+m'} \mathcal{D}^{(u,u)}_{3;\ell' -m';\ell -m;-\vec P}(-\vec k;-\vec k') \,,
\end{equation}
i.e. $\mathcal D_3^{(u,u)}$ transforms in the same way as $G^\infty$.
It is now straightforward to use the definitions, Eqs.~(\ref{eq:Luures}) and (\ref{eq:Ruures}),
 to show that the components of $ \mathcal L_3^{(u,u)}$ and $ \mathcal R_3^{(u,u)}$ satisfy 
 \begin{align}
 \mathcal L^{(u,u)}_{3;\ell m; \ell' m';\vec P}(\vec k', \vec k)
 =
 (-1)^{\ell +m+ \ell' +m'} \mathcal R^{(u,u)}_{3;\ell' -m'; \ell -m;-\vec P}(-\vec k,-\vec k') \,.
\end{align}
We further note that $\mathcal L_3^{(u,u)}$ and $\mathcal R_3^{(u,u)}$ satisfy
\begin{align}
 \frac{  \rho_3(\vec k ' )}{2 \omega_{k'}}\,
 {\cal L}_{3}^{(u,u)}(\vec k ' ,\vec k )  
 =
 {\cal R}_{3}^{(u,u)}(\vec k ' ,\vec k )  
 \,
  \frac{  \rho_3(\vec k )}{2 \omega_{k}} \,,
 \end{align}
and from this and Eq.~(\ref{eq:M33df_final}), we deduce
\begin{align}
\Delta_{\mathcal L;\ell m; \ell' m';\vec P}(\vec p, \vec k)
 =
 (-1)^{\ell +m+ \ell' +m'}\,\Delta_{\mathcal R;\ell' -m'; \ell -m;-\vec P}(-\vec k,-\vec p)
\,.
\end{align}
Inserting this into Eq.~(\ref{eq:ILdef}) and solving for $ I_{\cal L}$ iteratively then gives
\begin{align}
I_{\mathcal L;\ell m; \ell' m';\vec P}(\vec p, \vec k)
 =
 (-1)^{\ell +m+ \ell' +m'}\,I_{\mathcal R;\ell' -m'; \ell -m;-\vec P}(-\vec k,-\vec p)
\,.
\end{align}

Substituting these properties of $I_{\mathcal L}$ and $I_{\mathcal R}$ along with 
the standard time-reversal transformation properties of $\mathcal M$ into
Eqs.~(\ref{eq:T23_M23})-(\ref{eq:T33_M33}), it then follows immediately that
$\T$ has standard transformation properties.
Using this result in Eqs.~(\ref{eq:K22_T22})-(\ref{eq:K33_T33}),
we find the claimed result that $\mathcal{K}_{\rm df}$ also has standard time-reversal transformation properties, i.e.
\begin{align}
{\cal K}_{22; \ell m; \ell'm';\vec P} &= (-1)^{\ell+m+\ell'+m'}\,{\cal K}_{22; \ell' -m'; \ell -m;-\vec P} \,,
\label{eq:K22_Trev}\\
{\cal K}_{23;\ell m; \ell' m';\vec P}( \vec k) &= (-1)^{\ell+m+\ell'+m'}\,{\cal K}_{32;\ell' -m'; \ell -m;-\vec P}(-\vec k) \,, \\
{\cal K}_{{\rm df,33};\ell m; \ell' m';\vec P}(\vec k'; \vec k) &=(-1)^{\ell+m+\ell'+m'}\, {\cal K}_{{\rm df,33};\ell' -m'; \ell -m;-\vec P}(-\vec k; -\vec k') \,.
\end{align}
We conclude that the $K$ matrix appearing in the quantization condition, Eq.~(\ref{eq:QC}), satisfies the same time-reversal transformation properties as a standard $K$ matrix. 
This implies that only three of the four components of the $K$ matrix 
must be determined from the finite-volume spectrum.

We can extend this result if we also assume parity invariance. Since there is nothing in the
construction of $\mathcal K_{\rm df}$ that violates parity, it transforms in the same way
as $\mathcal M$ under parity, namely by flipping the sign of all vectors and multiplying
spherical harmonics by $(-1)^\ell$. We thus arrive at the following relations in a theory
that is invariant under both time-reversal and parity transformations:
\begin{align}
{\cal K}_{22; \ell m; \ell'm';\vec P} &= (-1)^{m+m'}\,{\cal K}_{22; \ell' -m'; \ell -m;\vec P} \,,\label{eq:K22_TPrev}
\\
{\cal K}_{23;\ell m; \ell' m';\vec P}( \vec k) &= (-1)^{m+m'}\,{\cal K}_{32;\ell' -m'; \ell -m;\vec P}(\vec k)\,, \label{eq:TPsymmK23} \\
{\cal K}_{{\rm df,33};\ell m; \ell' m';\vec P}(\vec k'; \vec k) &=(-1)^{m+m'}\, {\cal K}_{{\rm df,33};\ell' -m'; \ell -m;\vec P}(\vec k; \vec k')\,.
\label{eq:TPsymmK33}
\end{align}
These relations are more useful since the same value of the total three-momentum appears
on both sides. In particular, the second relation shows that
$\mathcal K_{23}$ is not independent of $\mathcal K_{32}$.

\bibliographystyle{apsrev4-1} 
\bibliography{bibi} 

\end{document}